\documentclass[hyper]{JHEP3}

\input{epsf}
\usepackage{epsfig}
\usepackage{amssymb}
\usepackage{amsfonts}
\usepackage{amsbsy}
\usepackage{amsmath}

\usepackage{epstopdf}

\def\ch{{\rm ch}}

\def\im{\mbox{Im }}

\def\IC{\mathbb{C}}

\def\IZ{{\mathbb{Z}}}
\def\IR{{\mathbb{R}}}
\def\IP{\mathbb{P}}
\def\IQ{\mathbb{Q}}

\def\CC {{\cal C}}
\def\CM {{\cal M}}

\def\CN {{\cal N}}

\def\CF {{\cal F}}

\def\CP {{\cal P }}
\def\CL {{\cal L}}
\def\CV {{\cal V}}
\def\CW {{\cal W}}

\def\CO {{\cal O}}
\def\CZ {{\cal Z}}

\def\CH {{\cal H}}

\def\CS {{\cal S}}

\def\CU{{\cal U}}
\def\CZ{{\cal Z}}

\def\half{\frac{1}{2}}

\renewcommand{\Im}{{\rm Im }}
\renewcommand{\Re}{{\rm Re }}

\def\one{{\hbox{ 1\kern-.8mm l}}}

\def\p{\partial}

\def\half{\frac{1}{2}}

\def\Im{{\rm Im}}

\newcommand\fro{{\overline{\underline{\Omega}}}}

\title{Bound state transformation walls}

\author{Evgeny~Andriyash$^1$, Frederik~Denef$^{2,3}$, Daniel~L.~Jafferis$^4$ and Gregory~W.~Moore$^1$\\
\\
${}^1$ NHETC and Department of Physics and Astronomy,
Rutgers University,\\
$^2$ Center for the Fundamental Laws of Nature,
Harvard University,\\
$^3$ Institute for Theoretical Physics,
University of Leuven, \\
$^4$ School of Natural Sciences, Institute for Advanced Study, Princeton.\\
\\
{\tt andriyas@physics.rutgers.edu, denef@physics.harvard.edu,\\
jafferis@ias.edu, gmoore@physics.rutgers.edu} }

\abstract{ In four dimensional N=2 supergravity theories, BPS bound states near marginal stability are described by configurations of widely separated constituents with nearly parallel central charges. When the vacuum moduli can be dialed adiabatically until the central charges become \emph{anti}-parallel, a paradox arises. We show that this paradox is always resolved by the existence of ``bound state transformation walls'' across which the nature of the bound state changes, although the index does not jump. We find that there are two distinct phenomena that can take place on these walls, which we call recombination and conjugation. The latter is associated to the presence of singularities at finite distance in moduli space. Consistency of conjugation and wall-crossing rules near these singularities leads to new constraints on the BPS spectrum. Singular loci supporting massless vector bosons are particularly subtle in this respect. We argue that the spectrum at such loci necessarily contains massless magnetic monopoles, and that bound states around them transform by intricate hybrids of conjugation and recombination.

}

\preprint{RUNHETC-2010-05}

\begin{document}

\section{Introduction and qualitative discussion of basic ideas}

The spectrum of BPS states in four dimensional ${\cal N}=2$ supersymmetric theories shows interesting behavior when the vacuum moduli are varied. Well known are jumps at walls of marginal stability, where BPS bound states can decay into, or be assembled from, mutually supersymmetric constituents. In supergravity these bound states are described by multicentered black hole or particle configurations \cite{Denef:2000nb,LopesCardoso:2000qm}, providing an intuitive ``molecular'' picture of such bound states and their wall crossing behavior \cite{Denef:2007vg} (for recent reviews, see \cite{DimofteThesis,PiTPLectures}).  However as pointed out e.g.\ in \cite{Jafferis:2008uf}, this picture leads to an apparent paradox, reviewed below. In this paper we explain how this paradox is resolved in the most general case. Doing so will lead us to define certain walls in moduli space, across which in a sense the structure of the bound state changes, or more accurately, the attractor flow tree description of the state changes. We will call these walls bound state transformation (BST) walls, and distinguish between recombination and conjugation walls, defined below. These walls are not marginal stability walls, and hence the BPS index must remain constant when crossing them. This leads to various consistency conditions on the spectrum, which we study in some detail. In fact, one of those consistency conditions is nothing but the Kontsevich-Soibelman wall crossing formula, leading to a simple universal derivation of it. This is the subject of a companion paper \cite{Andriyash:2010qv}.   In the present paper on the other hand we mostly focus on consistency conditions on the massless spectrum at singularities.

  The BPS spectrum near singularities, including constraints from monodromy and stability, is of course a well-studied subject, going back to the original works \cite{Cecotti:1992qh,Cecotti:1992rm,Seiberg:1994rs}. Scattered examples of supergravity bound state transformation phenomena have appeared before in the literature \cite{Denef:2000nb,Denef:2001xn,Denef:2007vg,Manschot:2010xp}. Related phenomena have been exhibited in other pictures of BPS bound states; for example the conjugation phenomenon which we describe is related to quiver mutations or Seiberg dualities in cases where the quiver description of BPS bound states holds \cite{Berenstein:2002fi,Aganagic:2010qr}, and to the string - string junction transition in the D-string description of BPS states in brane engineered field theories \cite{Gaberdiel:1998mv,Bergman:1998br,Mikhailov:1998bx}. The goal of this article is to study bound state transformations in full generality in the supergravity attractor flow tree picture of BPS states, and to determine how they constrain the BPS spectrum. We believe our constraints on the spectrum of massless states discussed in Sections \ref{subsec:BPS-Indices} and \ref{subsec:IRfree} are new.

\subsection{Puzzle }\label{sec:puzzle}

The simplest example of a BPS bound state in supergravity is a 2-centered bound state of charges
$\Gamma_1$ and $\Gamma_2$. The equilibrium distance between the two centers is given by \cite{Denef:2000nb}
\begin{equation} \label{Req}
 R = \frac{\langle \Gamma_1,\Gamma_2 \rangle}{2} \, \frac{|Z_1+Z_2|}{{\rm Im}(Z_1 \overline{Z}_2)}
\end{equation}
where the central charges $Z_i$ of $\Gamma_i$ are evaluated at spatial infinity. Existence of the bound state requires $R>0$. When one dials the moduli at infinity through a marginal stability wall, the equilibrium distance $R$ diverges and the BPS state decays. The same is true when the two centers themselves are replaced by clusters of black holes or particles, or by a multi-particle `halo' \cite{Denef:2002ru}. This simple physical picture has led to a number of notable successes, including the derivation of universal wall crossing formulae
\cite{Denef:2007vg,Andriyash:2010qv}.

\begin{figure}[htp]
\centering
\includegraphics[scale=0.5,angle=0,trim=0 0 0 0]{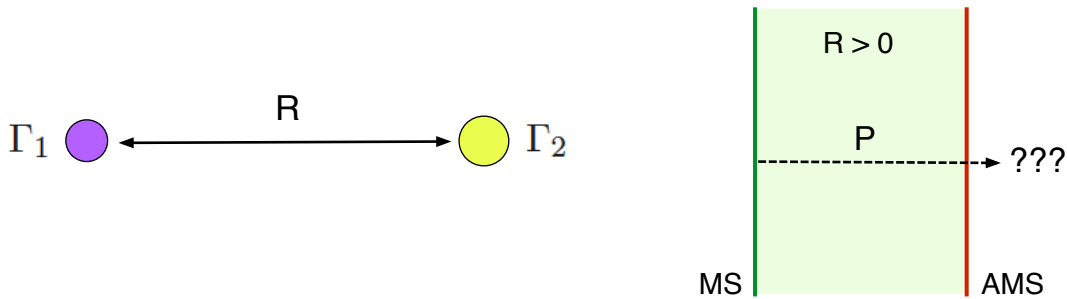}
\caption{BPS bound states appear to be adiabatically transportable from marginal stability to anti-marginal stability keeping $R>0$, violating conservation of energy.}
\label{fig:puzzle}
\end{figure}

These successes notwithstanding, it does not take much effort to arrive at the following disturbing observation. It is often possible \cite{Denef:2000nb,Denef:2001xn,Jafferis:2008uf} to dial the moduli \emph{while keeping $R$ positive
and finite}, from a marginal to an \emph{anti-}marginal stability wall, as illustrated in fig.\ \ref{fig:puzzle}. At an anti-marginal stability wall, the phases of $Z_1$ and $Z_2$ anti-align. It would appear from (\ref{Req}) that this simply leads to $R \to \infty$ again and a decay $\Gamma \to \Gamma_1 + \Gamma_2$. However, this obviously violates conservation of energy: the energy of the BPS bound state at the anti-marginal stability wall is $|Z| =\left||Z_1|-|Z_2|\right|$, while the total energy of the decay products equals $|Z_1|+|Z_2|$! Either we have created a perpetuum mobile, or something dramatic must have happened to the bound state along the way.

The puzzle is not tied to the supergravity approximation; it similarly arises, perhaps even more sharply, when thinking of BPS bound states as characterized by attractor flow trees \cite{Denef:2000nb,Denef:2000ar,Denef:2001xn,Denef:2007vg}. The branches of such trees are attractor flows, splitting on walls of marginal stability, terminating on the attractor points of the constituents, and rooted at the vacuum value of the moduli. A flow tree can be thought of as the ``skeleton'' of a supergravity solution,\footnote{In \cite{Denef:2007vg} a conjecture was formulated (the ``split attractor flow conjecture'') stating that there is a one to one correspondence between connected components of the supergravity solution space and attractor flow trees. In the course of this work we noticed this is not quite correct: the loci in moduli space where solution spaces split and join do not exactly coincide with loci in moduli space where flow trees split and join; see section \ref{sec:sfconj} for details. None of the results on indices and wall crossing in the literature are affected by this, as those required only the interpretation of flow trees as canonical procedures to assemble or disassemble BPS states. The same holds for the results in this paper, but for this reason we are careful to phrase definitions of bound state transformation walls in terms of trees, not solutions.} but can be given a meaning independent of supergravity \cite{Denef:2001ix,Denef:2007vg}, as a canonical recipe to assemble or disassemble a BPS state. What makes an attractor flow special compared to other paths in moduli space is that for any pair of constituent charges it either crosses a marginal stability wall once, or an anti-marginal stability wall once, or it crosses neither. In particular it is not possible to cross both a marginal and an anti-marginal stability wall. Our puzzle is then how to reconcile this with the fact that the root point of the tree can be moved along the path $\CP$ shown in fig.\ \ref{fig:puzzle}, seemingly forcing the trunk to cross both the AMS and the MS wall. A related puzzle has been discussed in \cite{Chalmers:1996ya}.

In the following subsection we will give a qualitative resolution of the puzzle.
A more precise  description together with several examples and detailed arguments will be
given in subsequent sections of the paper.

\subsection{Resolution}

The most straightforward resolution of the puzzle would appear to be that somewhere along the path, the BPS state simply gets lifted. Indeed for classical solutions this ``elevation'' phenomenon was noticed some time ago already \cite{Denef:2001xn} (fig.\ 18); see also fig.\ \ref{fig:elevation} below. However, this is only possible at the quantum level if the BPS index was zero to begin with. If the index is nonzero, as in the situation raised in \cite{Jafferis:2008uf}, something more dramatic needs to happen to prevent the paradox.

In the flow tree picture something dramatic can only happen when the flow tree degenerates, i.e.\ when an edge shrinks to zero size. This edge can be the trunk, an internal edge or a terminal edge. The first case corresponds to crossing a marginal stability wall, which we have excluded from the start. The second case is associated to constituents rearranging themselves, and the third case to constituents becoming massless and charge conjugate particles being created. They will be referred to as recombination resp.\ conjugation walls.

\begin{figure}[htp]
\centering
\includegraphics[scale=0.5,angle=0,trim=0 0 0 0]{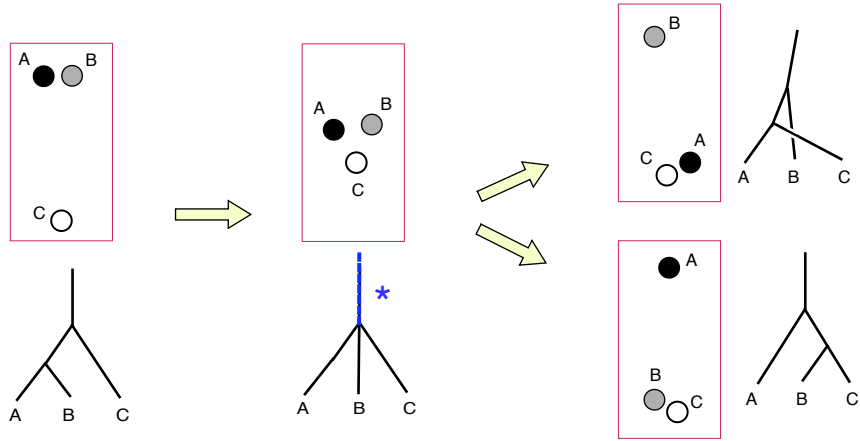}
\caption{
{\bf Recombination:} Constituents rearrange themselves into different clusters. The example represents a family of configurations with $A$ tightly bound to $B$ evolving into a family with $A$ tightly bound to $C$, and a family with $C$ tightly bound to $B$. The corresponding attractor flow tree evolves from an $((A,B),C)$ tree to a $((C,A),B)$ tree plus a $((B,C),A)$ tree. At the transition point, the flow tree has two 3-valent vertices coalescing into a 4-valent vertex. The recombination wall is the blue line with the asterisk next to it.}
\label{fig:recombination}
\end{figure}

\begin{enumerate}

\item {\bf Recombination wall} (fig.\ \ref{fig:recombination}): If the charges $\Gamma_1$ and $\Gamma_2$ themselves are composite bound states, it is possible that along the way the different constituents recombine into new clusters. This invalidates the hidden assumption in the formulation of the puzzle that the BPS state can at all times be viewed as a bound state with well separated clusters of charge $\Gamma_1$ and $\Gamma_2$. What happens instead is that before the troubling AMS wall is reached, the constituents rearrange themselves to make the AMS wall irrelevant. A sketch of possible recombination processes is shown in fig.\ \ref{fig:recombination}. In the corresponding flow trees we see two 3-valent vertices coalesce into a 4-valent vertex, which then again separates into 3-valent vertices, but with a different tree structure. The degenerate 4-valent vertex lies at the intersection of the marginal stability walls for the different partitions of the constituents. The union of critical ingoing flows, i.e.\ the set of all moduli values flowing into the degenerate vertex, forms a codimension 1 wall in moduli space, the recombination wall. We will check that both index and spin character remain constant across a recombination wall, provided we sum over all trees of the given charge. An example of such a recombination process appeared in fig.\ 14 of \cite{Denef:2007vg}. More recently it was also discussed in \cite{Manschot:2010xp}.

\begin{figure}[htp]
\centering
\includegraphics[scale=0.42,angle=0,trim=0 0 0 0]{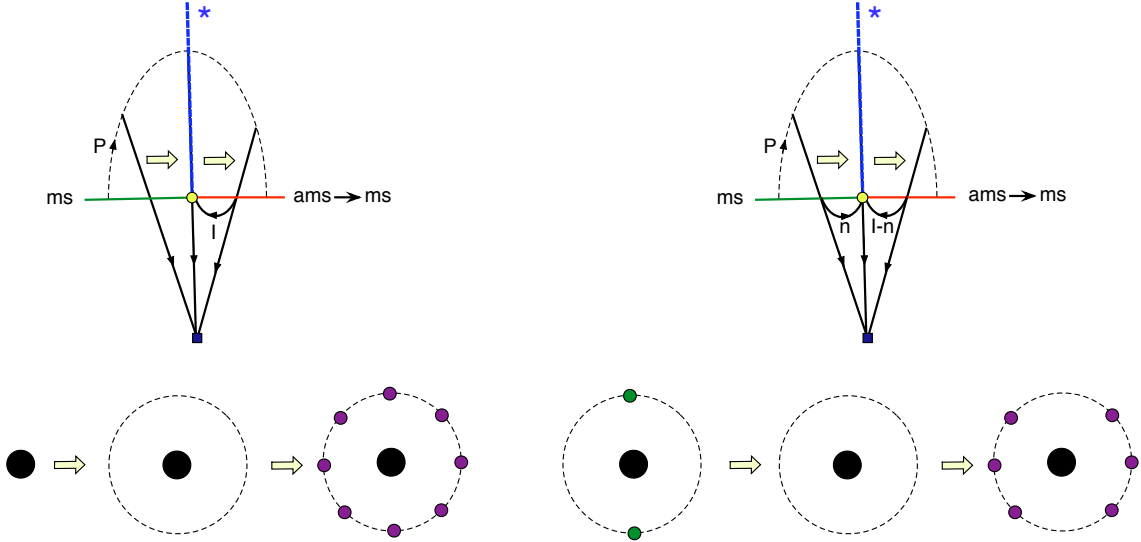}
\caption{
{\bf Conjugation:} Left: a single centered $\Gamma_2$ flow turns into a $(\Gamma_2,-I \Gamma_1)$ split flow when pulled through the locus where the hyper $\Gamma_1$ becomes massless and around which we have a monodromy $\Gamma_2 \to \Gamma_2 + I \Gamma_1$. The corresponding spacetime picture is the creation of a fully filled fermi shell of $I$ particles of charge $-\Gamma_1$. Right: a $(\Gamma_2,n \Gamma_1)$ split flow turns similarly into a $(\Gamma_2,(I-n)(-\Gamma_1))$ split flow.
\label{fig:conjugation}}
\end{figure}

\item {\bf Conjugation wall}: The second possibility is more subtle, and is the one that solves the particular instance of the puzzle raised in \cite{Jafferis:2008uf}. It is associated to a vanishing terminal edge of the flow tree. This is only possible if the end point of this edge is a singularity, since regular attractor points can never lie on marginal stability walls. Thus, for example, it occurs when one of the constituents, say $\Gamma_1$, is a particle in a hypermultiplet which becomes massless at a singular locus, where the MS and AMS walls meet. In such cases there is a log-monodromy around the massless locus: $\Gamma_2 \to \Gamma_2 + I \Gamma_1$, where $I=|\langle \Gamma_1,\Gamma_2 \rangle|$. As was pointed out in \cite{Denef:2000nb}, trying to pull a single $\Gamma_2$ attractor flow through such a massless locus going from MS($\Gamma_1,\Gamma_2$) to AMS($\Gamma_1,\Gamma_2$) will cause the creation of a new tree branch, corresponding to charge $-I \Gamma_1$, terminating on the massless locus, as shown in fig.\ \ref{fig:conjugation} on the left. This is required by charge conservation. The spacetime picture of this is that a shell of $I$ particles of charge $-\Gamma_1$ gets created as a halo around a core of charge $\Gamma_2$ at the radius where the moduli pass through the massless locus. It was shown in \cite{Denef:2002ru} that these newly created particles form a completely filled fermi shell of spin 1/2 fermions. If we start off with a bound state of $\Gamma_1$ and $\Gamma_2$, the fermi shell of $I$ particles of charge $-\Gamma_1$ will again be generated, but now one $-\Gamma_1$-particle will annihilate with the $\Gamma_1$ particle already present, leaving behind a hole in the fermi shell, i.e.\ $I-1$  particles
    of charge $-\Gamma_1$. The troublesome AMS wall for $\Gamma_1$ and $\Gamma_2$ is now reinterpreted as a trouble-free MS wall for $\Gamma_2$ and the remaining  particles of charge $-\Gamma_1$. In particular the state remains BPS: we go from a flow tree $(\Gamma_2,\Gamma_1)$ to a flow tree $(\Gamma_2,(I-1) (-\Gamma_1))$. Similarly, if $n \leq I$ $\Gamma_1$ particles were present, we end up with $n$ holes or $I-n$  particles
    of charge $-\Gamma_1$. In flow tree language we go from a $(\Gamma_2,\Gamma_1)$ tree to a $(\Gamma_2,(I-n) (-\Gamma_1))$ tree. This is shown in fig.\ \ref{fig:conjugation} on the right. We call this process the fermi flip.

When $n>I$ this does not work: we end up with particles of charge $\Gamma_1$ rather than $-\Gamma_1$, the flow tree ceases to exist since splits on AMS walls are not allowed, and the bound state goes from being classically BPS to being classically non-BPS (the minimum of the interaction potential is no longer at the BPS bound). At the classical level, this is a realization of the elevation phenomenon mentioned earlier as the most straightforward resolution of the puzzle. At the quantum level, what happened here wasn't quite elevation, because there were initially no quantum BPS states at all. This is because more fermions were present ($n$) than the number of available 1-particle states ($I$). In some cases\footnote{This may require fine tuning of hypermultiplet moduli, and requires the absence of quantum tunneling phenomena pairing up and lifting unprotected BPS states.} elevation processes may occur also at the quantum level. An example is a bound state of some magnetically charged particle and an electrically charged $\CN=4$ vector multiplet, illustrated in fig.\ \ref{fig:elevation}.

\end{enumerate}

The above list is exhaustive, since the only possible degenerations of flow trees are collapses of edges. In general we can also get recombination-conjugation hybrids, at singularities where mutually nonlocal BPS states become massless. But the basic building blocks are given by the above classification.


\begin{figure}[htp]
\centering
\includegraphics[scale=0.5,angle=0,trim=0 0 0 0]{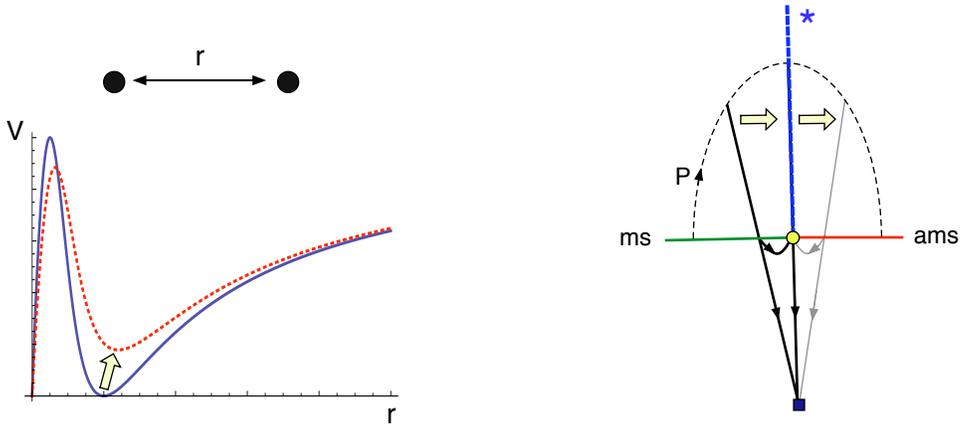}
\caption{
{\bf Elevation:} The initially BPS-saturated minimum of the interaction potential $V(r)$ gets lifted, and the bound state becomes classically non-BPS. The corresponding flow trees are shown on the right. The blue line with the asterisk is the elevation wall. It corresponds to a critical attractor flow hitting a locus in (a suitable finite cover of) moduli space where the mass of one of the constituent particles vanishes but there is no charge monodromy around it, as is the case for example if an $\CN=4$ vector multiplet becomes massless. The tree on the right is shown in grey because it does not represent an actual BPS flow tree, since the split occurs on an anti-marginal stability wall.
\label{fig:elevation}}
\end{figure}

The definitions we have given here will be made more precise in the following sections, and we will study more systematically under what conditions these phenomena occur. Besides solving the puzzle raised in \cite{Jafferis:2008uf}, these considerations will also lead to interesting constraints on the BPS spectrum. More precisely these follow from continuity of BPS indices across bound state transformation walls. For example, if only particles of charge proportional to $\gamma$ become massless at a certain locus at finite distance in moduli space, we find that the monodromy around this locus must be given by
\begin{equation}
 \Gamma \to \Gamma + I \, \gamma \, , \qquad
 I = \langle \gamma,\Gamma \rangle \sum_{k=1}^\infty k^2 \Omega(k \gamma) \, .
\end{equation}
Furthermore, if some $\Gamma_1 = k \gamma$ supports massless BPS vector particles with $\Omega(k \gamma) \neq 0$, we clearly run into trouble, since we could start with a BPS configuration $(\Gamma_2,n \Gamma_1)$ with $n > I$ vector-particles (as this is no longer forbidden by the exclusion principle), and after crossing the bound state transformation wall end up with $n-I>0$ particles of charge $\Gamma_1$, leading to a classically non-BPS configuration. By continuity of the index, this implies $\Omega(\Gamma_2 + n \Gamma_1) = 0$ on either side of the BST wall. The only way this is possible is if by some bizarre conspiracy the sum of all (nonzero) indices of individual configurations with total charge $\Gamma_2 + n \Gamma_1$ equals zero, for \emph{any} choice of $\Gamma_2$ and $n$. More plausibly, this situation simply cannot occur. A more precise version of this argument is
given in Sections \ref{subsec:BPS-Indices} and \ref{subsec:Rational-case} below.  Indeed the following independent argument corroborates this. We only expect massless vectors at the quantum level when the low energy gauge theory is IR free or conformal. For IR free gauge theories we can trust the smooth classical BPS monopole solutions that such theories have on their Coulomb branch. These monopoles have mass proportional to the W-boson mass, and so we will always get BPS states with mutually nonlocal charges becoming massless at the same locus as the vector, contradicting the assumption that only charges proportional to $\gamma$ become massless. See Section \ref{subsec:IRfree}
for further discussion.

The paper is organized as follows. In section \ref{subsec:Top-Arg} we give a general and precise description of BST walls. In section \ref{sec:BST-walls} we review multicentered halo solutions of supergravity. We investigate the conjugation wall and the way BPS Hilbert spaces change across it, detailing the physical conjugation process and its relation to monodromy around the singularity. Finally we discuss constraints on the spectrum of BPS states following from the continuity of BPS index. In section \ref{sec:conifold} we give the simplest example of a conjugation wall, associated to a single massless hypermultiplet, as in the example from \cite{Jafferis:2008uf}. In Section \ref{sec:TCwalls} the recombination walls are presented and we show how index and spin character are preserved. In addition we revisit the split attractor flow conjecture of \cite {Denef:2000nb} and add some important amendments to it.

In Section \ref{sec:Massless-VM} we describe the situation with a massless vector multiplet appearing at the singularity, and in section \ref{sec:MasslessVM-examples} we give a representative set of examples.
One important conclusion we draw from the example of Section \ref{subsec:ExtrTran}
is that the spectrum of low energy
particles in some models with extremal transitions has not been fully understood in the past. We
found this because published spectra disagreed with the general conclusions we had reached in
Section \ref{sec:BST-walls}. We have concluded that the published spectra were incomplete, and do
not constitute counterexamples to the prediction of Section \ref{subsec:BPS-Indices}.

In the companion paper \cite{Andriyash:2010qv} we explain how the halo picture in supergravity leads to a
very simple derivation of the Kontsevich-Soibelman wall-crossing formula (KSWCF). Moreover, in the presence
of the singularities studied in this paper, the derivation leads to a generalization of the
KSWCF which in fact constrains the spectrum of the theory. An example of this constraint
  is given in Section    \ref{sec:FHSV} below. Moreover, the examples of
Section \ref{sec:MasslessVM-examples} point to  several interesting directions for further
research.

\section{Walls from attractor flow trees}\label{subsec:Top-Arg}

One way to describe BPS boundstates in supergravity is via
attractor flow trees \cite{Denef:2000nb,Denef:2000ar,Denef:2001xn,
Denef:2007vg}.\footnote{In fact, the description of BPS boundstates via attractor flow trees is applicable in a far more general context than just the supergravity approximation.} Such trees describing a boundstate of
two subcomponents of charge $(\Gamma_1, \Gamma_2)$ (necessarily in
a stable region) begin with single-centered attractor flow for the
total charge $\Gamma:=\Gamma_1+\Gamma_2$. When describing
boundstates of two constituents $\Gamma_1,\Gamma_2$ the tree then
splits on a marginal stability wall $MS(\Gamma_1, \Gamma_2)$:

\begin{equation}
MS(\Gamma_1, \Gamma_2) :=\{ t\in \widetilde\CM \: | \: 0 <
Z(\Gamma_1; t) / Z(\Gamma_2; t) < + \infty \}
\end{equation}
where  $\widetilde\CM $ denotes the universal cover of vectormultiplet
moduli space. (In general our notation follows \cite{Denef:2007vg}.)

Let us now suppose we are in the situation of our puzzle.
The region of stability is defined by
\begin{equation}\label{eq:Stab-reg}
\langle \Gamma_1, \Gamma_2 \rangle {\rm Im}  Z(\Gamma_1;
t)\overline{Z(\Gamma_2; t)} > 0.
\end{equation}
Suppose that the   path $\CP$ is contained in the region of stability,  connecting a
point $t_{ms}$ on $MS(\Gamma_1,\Gamma_2)$ to a point $t_{ams}$ on
$AMS(\Gamma_1,\Gamma_2)$

\begin{equation}
AMS(\Gamma_1, \Gamma_2) :=\{ t\in \widetilde\CM \: | \: -\infty  <
Z(\Gamma_1; t) / Z(\Gamma_2; t) < 0 \}.
\end{equation}
The boundary of a region of stability is the set:
\begin{equation}
W(\Gamma_1, \Gamma_2) = \{ t\in \widetilde{\CM} \: | \: \Im \left[
Z(\Gamma_1; t) \bar Z(\Gamma_2; t) \right]=0 \},
\end{equation}
which can be decomposed as:
\begin{equation}
W(\Gamma_1, \Gamma_2)= MS(\Gamma_1,\Gamma_2)\amalg
AMS(\Gamma_1,\Gamma_2) \amalg \left(\CZ(\Gamma_1) \cup \CZ(\Gamma_2) \right),
\end{equation}
where
\begin{equation}
\CZ(\Gamma):= \{ t \in \widetilde{\CM} \vert Z(\Gamma;t)=0 \}.
\end{equation}


\begin{figure}[htp]
\centering
\includegraphics[scale=0.5,angle=90,trim=0 0 0 0]{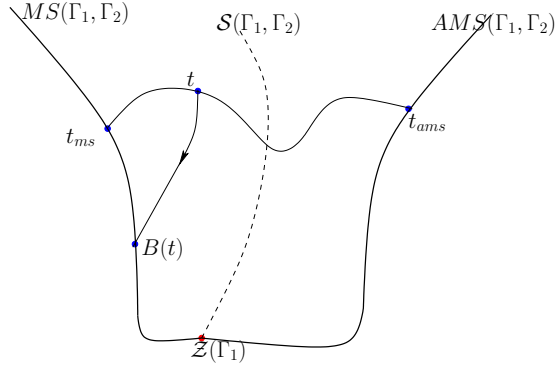}
\caption{Location of the $\CS(\Gamma_1, \Gamma_2)$ wall. }
\label{fig:BSTwall1}
\end{figure}

In figure \ref{fig:BSTwall1} we depict a caricature of the location of different components of $W(\Gamma_1, \Gamma_2)$ in a real dimension 2 surface in the moduli space. Denote by $t$ a point on
$\CP$ and consider the   behavior of the attractor flow tree as $t$
moves along $\CP$ from $t_{ms}$ towards $t_{ams}$. We want to prove the following

{\em \noindent \textit{ {\bf Statement:} There exists a point $t \in \CP$, such that the attractor flow for
$\Gamma_1+\Gamma_2$, starting at $t$, ends on either $\CZ(\Gamma_1)$ or
$\CZ(\Gamma_2)$.
}}

One can use the following simple argument. Notice that when $t$ is close to $t_{ms}$, the
attractor flow for $\Gamma_1+\Gamma_2$ will almost immediately hit
$MS(\Gamma_1, \Gamma_2)$. On the other hand, when $t$ is close to
$t_{ams}$, the flow will hit $AMS(\Gamma_1,\Gamma_2)$. Indeed, according to Property 3 from Appendix \ref{app:properties}, the attractor flow always has the direction from  stable to unstable side in the vicinity of an (anti)marginal stability locus. As
one can continuously get from $MS(\Gamma_1,\Gamma_2)$ to $AMS(\Gamma_1,\Gamma_2)$  only
through the loci $\CZ(\Gamma_1)$ or $\CZ(\Gamma_2)$, it is almost obvious that
for some $t \in \CP$ the attractor flow will crash on those zeros.
The only thing to check is that the attractor
flow for $\Gamma_1+ \Gamma_2$ does not run to a boundary of the
moduli space at infinite distance when we move $t$ from $t_{ms}$ to $t_{ams}$. Let  $B(t)$ denote the point where the flow hits $W(\Gamma_1, \Gamma_2)$. Property 2 from Appendix \ref{app:properties} says that $B(t)$ exists for all $t$ on path $\CP$ and Property 1 ensures that it is unique. Now define a real-valued function

\begin{equation}
\lambda(t): = \frac{Z(\Gamma_1; B(t))}{Z(\Gamma_2;B(t))} \in \IR
\cup \{ \pm \infty\}.
\end{equation}

%
%
%


We showed that it can be defined for every $t$. Given that
$\lambda(t) >0$ for $t$ near $t_{ms}$ and $\lambda(t)<0$ for $t$ near
$t_{ams}$, there must be some point where $\lambda(t)$ changes sign, going either to
zero or infinity. $\lambda(t)$ having zero
corresponds to crossing a wall $ \CS(\Gamma_1,\Gamma_2)$  and
$\lambda(t)$ having infinity corresponds to  crossing  a wall
$\CS(\Gamma_2,\Gamma_1)$ wall, where
\begin{equation}
 \CS(\Gamma_1,\Gamma_2) := \{t \: |\:   (\Gamma_1+\Gamma_2) \:
{\rm flow \: from} \: t \: {\rm crashes \: on} \: \CZ(\Gamma_1) \}.
\end{equation}
In principle, $\lambda(t)$ could have changed its sign more than once along the path $\CP$, crossing one or both of $ \CS(\Gamma_1,\Gamma_2)$, $ \CS(\Gamma_2,\Gamma_1)$ walls possibly several times. For resolving our
puzzle it will suffice to understand what happens when we cross just one wall.

Coming back to  the fate of the BPS bound state $(\Gamma_1+\Gamma_2)$, we suppose for definiteness that $\lambda(t)$ has a zero, the path $\CP$ crosses $\CS(\Gamma_1,\Gamma_2)$ and $Z(\Gamma_1)$ has a zero. The
physical discussion of BPS states depends on the following
dichotomy:

\begin{enumerate}

\item
Near the locus $\CZ(\Gamma_1)$, BPS states with charge $\gamma_1$ parallel to $\Gamma_1$
exist. That is, there is a positive rational number  and a charge $\gamma_1 = r \Gamma_1$
so that
\begin{equation}\label{eq:BSTwallHilbSpace}
    \CH(\gamma_1, t)|_{t \in \CZ(\Gamma_1)} \ne \emptyset,
\end{equation}
\item No such states exist in the neighborhood of $\CZ(\Gamma_1)$.
\end{enumerate}

\begin{figure}[htp]
\centering
\includegraphics[scale=0.5,angle=90,trim=0 0 0 0]{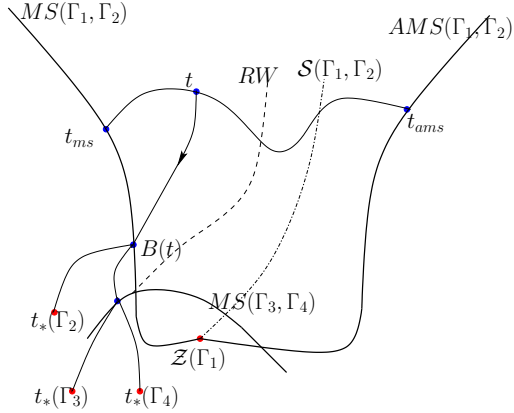}
\caption{Charge $\Gamma_1$ is realized as a bound state of
$\Gamma_3+ \Gamma_4$. The dashed line is the recombination wall $RW$ between $(\Gamma_2,(\Gamma_3,\Gamma_4))$ and $(\Gamma_4,(\Gamma_2,\Gamma_3))$+$(\Gamma_3,(\Gamma_2,\Gamma_4))$. } \label{fig:TSWleft}
\end{figure}

There are known examples of both possibilities. The first possibility, as discussed in the Introduction, gives rise the conjugation phenomenon across $\CS(\Gamma_1, \Gamma_2)$. We will  define the wall $\CS(\Gamma_1,\Gamma_2)$, subject to the constraint (\ref{eq:BSTwallHilbSpace}), to be the conjugation wall.

In the second case the charge $\Gamma_1$ is not populated around $\CZ(\Gamma_1)$, but by assumption, it is populated in the neighborhood of $t_{ms}$ and thus should be realized as a multicentered solution \cite{Denef:2000nb}. An example of this situation, when $\Gamma_1$ is a bound state of $\Gamma_3 + \Gamma_4$, is given in figure \ref{fig:TSWleft} \footnote{ We depict charges $\Gamma_2$, $\Gamma_3$ and $\Gamma_4$ as single-centered attractor flows, but the discussion is applicable to the most general case with all charges being some multicentered configurations. The notation $t_*(\Gamma)$ in the figure is used for the regular attractor point of charge $\Gamma$.}. The bound state of $\Gamma_3 + \Gamma_4$ has to decay as one approaches $\CZ(\Gamma_1)$, so that $MS(\Gamma_3, \Gamma_4)$ has to separate some region around  $\CZ(\Gamma_1)$ from the path $\CP$, as in figure \ref{fig:TSWleft}. Introducing the notation of nested lists to denote different attractor tree topologies the bound state just described is denoted as $(\Gamma_2, (\Gamma_3,\Gamma_4))$. Let's also denote
\begin{eqnarray}\label{eq:gammaijk}
&& \Gamma_{ij}: =  \langle \Gamma_i, \Gamma_j \rangle,  \nonumber\\
&& \Gamma_{ij,k}: =  \Gamma_{ik}+\Gamma_{jk},\quad \Gamma_{k,ij}:=-\Gamma_{ij,k}
\end{eqnarray}
for $i,j,k \in (2,3,4)$ and all different. The set of points in moduli space, such that attractor flow for
charge $\Gamma_{total} = \Gamma_2+\Gamma_3+\Gamma_4$ with at least one $\Gamma_{ij,k}$ non-zero, starting at
those points, passes through the locus where all three central
charges are aligned defines a {\it recombination wall} (RW)
\cite{Denef:2007vg}, which we will denote by

\begin{equation}\label{eq:TCdef}
 RW(\Gamma_2,\Gamma_3,\Gamma_4)=\{ t\vert (\Gamma_2+\Gamma_3+\Gamma_4) \mbox{-flow from t
crashes on} \: MS(\Gamma_3 + \Gamma_4,\Gamma_2) \cap MS(\Gamma_3,\Gamma_4) \}.
\end{equation}
The definition is in fact symmetric as will be explained in Section \ref{sec:TCwalls}. It is clear from the picture that as we move $t$ along $\CP$ from $t_{ms}$ and before crossing the $\CS(\Gamma_1,\Gamma_2)$ wall, we will hit $RW(\Gamma_2,\Gamma_3, \Gamma_4 )$. As discussed in the Introduction the structure of attractor trees describing the state changes across the recombination wall. Section \ref{sec:TCwalls} gives a detailed account of how the bound states transform across $RW(\Gamma_2,\Gamma_3, \Gamma_4 )$ wall and once again the puzzle from the Introduction gets resolved.

\section{Conjugation Walls and Fermi Flips}\label{sec:BST-walls}

In this section we describe what happens to the bound state when the background moduli cross a conjugation wall. Changing slightly the notation from the previous section, we will be interested in   bound states of a single particle of charge $\Gamma$ with  one or more particles with
 charges proportional to a primitive charge $\gamma$ where $\langle \Gamma, \gamma \rangle \not=0$.
 Our considerations will force us  to consider, more generally, a particle of charge $\Gamma + m \gamma$
 bound to one or more particles whose charges are proportional to $\gamma$.
 %
%

\subsection{Rules of the game}\label{subsec:GroundRules}

  We assume  that  $\CZ(\gamma)$ is located on the boundary of the moduli space at a finite distance.
   This locus is complex codimension one in moduli space $\CM$ and can be thought of as lying on a
   real codimension one boundary of the covering space $\widetilde{\CM}$. We will reduce arguments
   to one-complex-dimensional slices so one should keep in mind the analogy of $\widetilde{\CM}$
   as the upper half-plane with coordinate $\tau$ and $\CM$ as the unit disk with parameter
   $q=e^{2\pi i \tau}$. Then $\widetilde{\CZ}(\gamma)$ is $\IQ \cup \{ i \infty \}$ and
   $\CZ(\gamma)$ is $q=0$.  All our arguments should be understood as pertaining
 to some sufficiently small and generic neighborhood $\CU$ of $\CZ(\gamma)$ in $\CM$.

 The lattice $\Lambda$ of electromagnetic charges forms a local system over $\CM$. That is,
 there is a flat connection on $\Lambda$.  Moreover,
 the Hilbert space $\CH^{\rm one-particle}$ of \emph{all} one-particle states has a flat connection on $\CM$, and
 furthermore $\CH^{\rm one-particle}$ has a compatible grading by $\Lambda$. Typically,
 the local system $\Lambda$ will have nontrivial monodromy around $\CZ(\gamma)$.
 We will assume that $\gamma$ is monodromy invariant.
 \footnote{We  might need to
 pass to a finite cover of $\CU$ if $Z(\gamma;t)$ has a multiple
  zero on $\CZ(\gamma)$.  }

We will make some assumptions about the nature of certain BPS spaces in $\CU$. First, we assume
that
    $\CH( \ell_i \gamma;t)|_{t \in \CZ(\gamma)} \ne \emptyset$ for some collection of integers $\ell_i$.\footnote{ Here we deviate slightly
from the notation of \cite{Denef:2007vg} by using $\CH$ for the
reduced statespace of single-particle BPS states where the half-hypermultiplet
degrees of freedom from the center of mass have been factored out.
Thus $\CH$ was denoted by $\CH'$ in \cite{Denef:2007vg} and hence
the BPS index -- the second helicity supertrace -- is given by
$\Omega(\Gamma;t)= {\rm Tr}_{\CH(\Gamma;t)}(-1)^{2J_3}$ in this paper. }
Second, we assume that these spaces are ``constant'' or $t$-independent in $\CU$.
By ``constant'' we mean   there is
a flat connection  on the vector bundle of BPS states of charge $\ell \gamma$  whose
fiber at $t$ is  $\CH(\ell\gamma;t)$. Using the flat connection we trivialize the bundle and
just speak of $\CH(\ell\gamma)$. Third, it can very well happen that there is a linearly
independent charge $\gamma'$ with the same vanishing locus $\CZ(\gamma) = \CZ(\gamma')$.
However, we assume that if such charges arise they are not populated, that is,
$\CH(n \gamma + m \gamma';t)=0$ in $\CU$ whenever $m\not=0$. As we will see in
Section \ref{sec:Massless-VM}, this is a crucial assumption; one which is not always satisfied in
physically interesting situations.

Finally, returning to our charge $\Gamma$
such that $\langle \Gamma,\gamma\rangle\not=0$ we make some assumptions about $\CH(\Gamma;t)$.
Again, by taking $\CU$ sufficiently small we know that the only relevant walls of
(anti)marginal stability are in $W(\gamma,\Gamma)$.
%
%
 As we have explained, the locus $\CZ(\gamma)$ divides $W(\gamma,\Gamma)$ into two connected
 components, $\CU \cap MS(\gamma,\Gamma)$ and $\CU \cap AMS(\gamma, \Gamma)$. We assume that our neighborhood
 of $\CU$ is sufficiently small that, for all $n\in \IZ$,
  $\CH( \Gamma + n \gamma ;t)$ is ``constant'' on these
 two components in the sense explained above.
 %
 %
 Therefore we can speak of
  well-defined spaces $\CH^{ms}(\Gamma+n\gamma)$ and $\CH^{ams}(\Gamma+n\gamma)$. We assume that  $\CH^{ms}(\Gamma)$  is nonzero, but we do \emph{not} assume that $\CH^{ms}(\Gamma+n\gamma) \cong \CH^{ams}(\Gamma+ n\gamma )$.
  Indeed, such a statement is meaningless if $\Gamma$ is not invariant under the monodromy action
  around $\CZ(\gamma)$.

\subsection{Review of halo states}

\subsubsection{Multicentered Halo Solutions}

Four dimensional N=2 supergravity has stationary multicentered BPS black hole solutions \cite{Behrndt:1997ny,Denef:2000nb,LopesCardoso:2000qm}, which are typically true bound states with constrained center positions whenever the centers have mutually nonlocal charges \cite{Denef:2000nb,LopesCardoso:2000qm}.
It was shown in \cite{Denef:2002ru} that there is a distinguished class of multicentered
solutions of supergravity known as \emph{halo solutions}.  In these solutions there is one center,
known as the \emph{core} with a charge $\Gamma$ while all the other centers carry charges
proportional to a primitive charge $\gamma$. The name derives from the fact that
when the solution exists all the halo centers must lie on a sphere of fixed radius.
For total charge of the form $\Gamma + n \gamma$ the halo radius is
\begin{equation}\label{eq:haloRadius}
R_n(t) = \half \langle \Gamma, \gamma \rangle \frac{ \vert Z(\Gamma
+ n \gamma; t)\vert }{ {\rm Im} Z(\Gamma;t) \overline{
Z(\gamma;t)}}
\end{equation}
 The total halo charge $n \gamma$ might be divided up between
different halo centers in different ways corresponding to several
centers of charges $\ell_i \gamma$, with   $\sum \ell_i = n$.

Multi-centered halo configurations might or might not constitute acceptable
solutions to supergravity.
The existence criterion for acceptable multi-centered solutions of supergravity are
rather complex and difficult to check in general. However, for halo solutions
there are two simple necessary and sufficient criteria for existence:

\begin{enumerate}

\item  The  halo centers all must have parallel charges. That is the charges
must be of the form  $\ell_i \gamma$ where the integers
$\ell_i$   all have the same sign.

\item The single-centered attractor flow from $t$
 with total charge $\Gamma + n \gamma$ must split on a wall of \emph{marginal stability}
 $MS(\gamma,\Gamma)$ if $n>0$ and it must split on a wall of \emph{anti-marginal stability}
 if $n<0$.

\end{enumerate}

The first criterion is easy to understand. As we cross a wall of marginal stability the
halo radius $R_n(t)$ goes to infinity. If some particles had $\ell_i$ of opposite sign
then energy could not be conserved. Alternatively, if there were particles of opposite
sign we could bring them together adiabatically and annihilate them. Thus, the original configuration
could not have been BPS.

\subsubsection{The shell approximation}

The multi-centered halo solutions  are rather intricate, and lead to rather complicated variations of
moduli $t(\vec x)$ in space.    Some useful intuition can be gleaned by
examining a much simplified ``shell approximation'' to the multicentered solutions.
This shell approximation is closely related to the split attractor flow
description.

\begin{figure}[htp]
\centering
\includegraphics[scale=1,angle=90,trim=0 0 0 0]{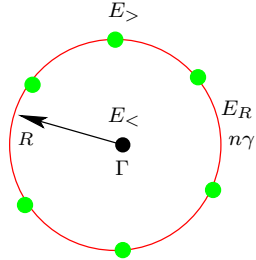}
\caption{Shell configuration.}
\label{fig:shell}
\end{figure}

In the ``shell approximation'' we replace the   multicentered supergravity solution by a
spherically symmetric shell solution \cite{Denef:2000nb} (see figure \ref{fig:shell}).
The supergravity field configuration is
radially symmetric. For   $r>R$ it is given by the attractor flow for
$\Gamma + n \gamma$. Following the lead of split attractor flow,
we choose the radius $R$ to be $R=R_n(t)$. Thus, the local
vectormultiplet moduli at $r=R$ are given by the point in $\CM$ where the
attractor flow of charge $\Gamma + n \gamma$ hits $W(\Gamma,
\gamma)$, denoted in what follows by $B_n(t)$. We next insert a
shell of uniform charge with total charge $n \gamma$ at the radius
$r=R$. Then, we continue the solution to $r<R$ using single
centered attractor flow for $\Gamma$. Let us compute the energy of
such a field configuration. The energy is a sum of three terms
$E_> + E_R + E_<$, the energy of the fields for $r>R$, $r=R$ and
$r<R$, respectively. These are given by

\begin{eqnarray}
&& E_> = \vert Z(\Gamma + n \gamma;t) \vert - \vert e^U Z(\Gamma + n
\gamma;B_n(t) \vert
\nonumber\\
&& E_R = \vert e^U Z(  n \gamma;B_n(t)) \vert \nonumber\\
&& E_< = \vert e^U Z( \Gamma;B_n(t)) \vert,
\end{eqnarray}
with the total energy given by the sum
\begin{equation}\label{eq:shell-energy}
E = \vert Z(\Gamma + n \gamma;t) \vert + 2\theta \vert e^U Z(
n \gamma;B_n(t)) \vert,
\end{equation}
where $\theta = 0$ if $B_n(t)$ lies on
$MS(\gamma,\Gamma)$ and $\theta = 1$ if $B_n(t)$ lies on $AMS(\gamma,\Gamma)$.
When $\theta=1$ the field configuration is certainly not BPS and
might not even be a solution of the equations of motion.
Thus, we have recovered the second existence criterion for
halo boundstates mentioned above: $B_n(t)$ must lie on $MS(\gamma,\Gamma)$.

\subsubsection{Halo Fock Spaces}\label{subsubsec:Halo-Fock-Spaces}

Upon quantization multi-centered solutions of supergravity correspond to
states in the Hilbert space of BPS states. These states are to be thought of
as boundstates of BPS particles. For the halo solutions we have a
 \emph{core particle} of charge $\Gamma$  and \emph{halo particles} whose
 charges are proportional to $\gamma$. The corresponding quantum states are
known as \emph{halo states}.

Thus far we have been referring to a single core particle. In general
the core charge $\Gamma$ in a multi-centered solution might correspond
to several particles and  might
not even have a single-centered realization. Nevertheless, if the boundstate
radius $R_n(t)$ is large compared to any dimensions of the multi-centered
solution of charge $\Gamma$ then we can still meaningfully distinguish between
the core and the halo and we can speak meaningfully of the ``halo contribution
to the BPS space $\CH(\Gamma + n \gamma;t)$.''  We will restrict attention
to such regions and   think of the BPS space as a direct sum
\begin{equation}
\CH(\Gamma+n \gamma ;t) = \CH^{halo}(\Gamma+n \gamma ;t) \oplus \CH'(\Gamma+n \gamma ;t),
\end{equation}
where $\CH^{halo}(\Gamma+n \gamma ;t)$ is the halo contribution and $\CH'(\Gamma+n \gamma ;t)$ is the core contribution.
By taking a sufficiently small neighborhood $\CU$ of a generic point on $\CZ(\gamma)$
any possible mixing between $\CH^{halo}(\Gamma;t) $ and $\CH'(\Gamma;t)$ can be made small. By the correspondence principle we expect that we can focus on $\CH^{halo}(\Gamma;t) $.

Let us now recall the description of the halo states given in \cite{Denef:2007vg}.
It is useful to consider a ``generating Hilbert space''
\begin{equation}\label{eq:Halo-Subspace}
\oplus_{n\geq 0}\CH^{halo}(\Gamma + n \gamma;t).
\end{equation}
Since the halo particles are
mutually BPS and noninteracting,  the
Hilbert space of all halo-type boundstates with a core particle of charge $\Gamma$ form a
$\IZ_2$-graded Fock space:
\begin{equation}\label{eq:Fockspace-plus}
\CH^{halo}_{\Gamma}:=  \CH^{ms}(\Gamma)
\otimes_{\ell\geq 1} \CF\left[
\left(J_{\Gamma,\ell\gamma}\right)\otimes\CH(\ell\gamma)\right]
\end{equation}
Here $J_{\Gamma,\ell\gamma} = \half (\vert \langle \Gamma,
\ell\gamma \rangle \vert -1)$ is an  $SU(2)$ spin   and
$\left(J_{\Gamma,\ell\gamma}\right)$ is the corresponding
representation space of $SU(2)$ with generators $J_i^{(1)}$ while
 $\CH(\ell\gamma)$ is also an $SU(2)$ representation space with
 generators $J_i^{(2)}$. The
finite-dimensional vector space
$\left(J_{\Gamma,\ell\gamma}\right)\otimes\CH(\ell\gamma)$ is
$\IZ_2$-graded by $-(-1)^{2J_3^{(2)}}$ and the Fock space
construction is applied in the $\IZ_2$-graded sense. The physical
reason for this seemingly strange choice of $\IZ_2$-grading is
explained in detail in \cite{Denef:2002ru}. In particular, halo
particles which are {\it hypermultiplets} behave like free fermions
and halo particles which are {\it vectormultiplets} behave like free
bosons. Note that because  of our assumptions, the space (\ref{eq:Fockspace-plus}) does not
depend on $t$.

There are three important subtleties one must be mindful of when using (\ref{eq:Fockspace-plus}).

\begin{enumerate}

\item First, at a   given value of $t$ it is not true that all of the
Hilbert space (\ref{eq:Fockspace-plus}) contributes to (\ref{eq:Halo-Subspace}).
Only those states contribute   for which the corresponding
halo solutions exist. In particular, for contributions to
$\CH^{halo}(\Gamma+n\gamma;t)$ with $n>0$ the point $B_n(t)$ must lie on $MS(\gamma,\Gamma)$
and hence $t$ must lie on the appropriate side of the conjugation wall $\CS(n\gamma,\Gamma)$.
In Appendix \ref{app:Wall-arrangement} we discuss the  arrangement of the walls $\CS(n\gamma,\Gamma)$
as a function of $n$ (see  Figure \ref{fig:BSTwall2}) and the consequences for (\ref{eq:Halo-Subspace}).

\item The second subtlety is that
the   halo Fock space (\ref{eq:Fockspace-plus}) singles out a special ``core charge'' $\Gamma$.
When speaking of the ``halo contribution to the space $\CH(\Gamma + n \gamma; t)$''
we should bear in mind that there can also be core charges of
the form $\Gamma + m\gamma$ surrounded by halos of particles of total charge $(n-m)\gamma$,
where $m$ can be any integer, and thus when working out the contributions to
(\ref{eq:Halo-Subspace})
we   should really sum over such core charges:
\begin{equation}
\oplus_{m\in \IZ} \CH^{halo}_{\Gamma+ m\gamma}.
\end{equation}
Once again, only values of $m$ such that the corresponding supergravity solutions exist
will contribute to (\ref{eq:Halo-Subspace}).
Note that the   BST walls  $\CS((n-m)\gamma, \Gamma + m\gamma)$ are the same for
all $m \in \IZ$.

\item Third, it is possible that there is nontrivial ``mixing'' between different halo states. This
would result from tunneling amplitudes between halo particles and core states. It is clearly
exponentially suppressed for large halo radius, but might in principle be nonvanishing.
Such mixing would alter our description of the Hilbert space of BPS states. This would have an
important impact on our description of the  spin characters, but it would not impact
our   description of the BPS indices.

\end{enumerate}

In (\ref{eq:Halo-Subspace}) we have considered halo states with halo particles whose charge
is parallel to $\gamma$. There is an analogous story for halo states with halo particles
whose charge is anti-parallel to $\gamma$. For these we should sum over negative values of $n$
in the analog of (\ref{eq:Halo-Subspace}), the analogous Fock space
(\ref{eq:Fockspace-plus}) involves particles drawn from $\CH(-\ell\gamma)$ with $\ell\geq 1$,
etc.

\subsection{The puzzle }\label{sec:transformhalostates}

Let us now return to the situation described in the previous sections.
We have a path $\CP$ joining $t_{ms}$ to $t_{ams}$ as in
figure \ref{fig:BSTwall1}.
Now imagine moving $t$ along the path $\CP$ and consider boundstates of
total charge $\Gamma + n \gamma$ with $n>0$. For $t$ near $t_{ms}$
we know there are halo boundstates with halo particles of charge
parallel to $\gamma$.  For $t$ near $t_{ams}$ such boundstates cannot
exist. Indeed,  when  $t$ crosses the wall $\CS(n\gamma, \Gamma)$
the halo bound states will cease to exist. Again we ask: \emph{what happened
to these BPS states}?

%
%


The simplest thing that can happen is that BPS states smoothly pair up and become
non-BPS states as $t$ crosses $\CS(n\gamma, \Gamma)$. We called this the elevation
phenomenon.  An example of this will be given in Section \ref{subsec:KMP}, where $\CN=2$  vector- and hypermultiplets will pair up to form massive vectormultiplets. This mechanism is indeed
suggested by the shell model.   The field configuration we
have written is clearly a solution of the equations   of motion for
$t\in \CP_{ms}$, since it satisfies the BPS bound. On the other hand, it
 ceases to be a BPS configuration for $t\in \CP_{ams}$. Nevertheless, the energy as a function of $t$
 continuously increases from the BPS bound.
(Of course, this is not a proof that the states smoothly evolve into non-BPS states since the
field configuration for $t\in \CP_{ams}$ might no longer solve the equations of motion, but
we consider it suggestive.)
%
%

Nevertheless, it is clear that this standard mechanism cannot be the whole story. The reason is that one
can easily compute the contribution of the halo states to the BPS index $\Omega(\Gamma + n \gamma;t)$
from (\ref{eq:Fockspace-plus}), and this contribution is typically nonzero.  Indeed, this is what happens in the example of \cite{Jafferis:2008uf}. The lifting mechanism
can only apply to states whose total contribution to the index vanishes. When the index does not vanish, the wall being crossed is the conjugation wall and there must be
at least some other kind of phenomenon to account for what happened to the BPS states. In sections
\ref{subsec:Fermi} and \ref{subsec:Fadeout}
we describe two new phenomenon - the \emph{Fermi flip} and the \emph{ fadeout}  --  which can account
for the disappearance of BPS states which contribute to an index. In sections
\ref{sec:conifold} and \ref{subsec:KMP} we show how these mechanisms  nicely accounts  for the
fate of BPS boundstates in the neighborhood of common types of discriminant loci of Calabi-Yau
manifolds. In order to motivate the Fermi flip it is useful to try to write out quantitatively
the condition that all the indices   $\Omega(\Gamma + n \gamma;t)$, $n\in \IZ$
 are   continuous functions along the path $\CP$. We turn to this in the next subsection.
 %
%


\subsection{BPS Indices}\label{subsec:BPS-Indices}

BPS indices can only change across walls of marginal stability.
Let us see what this implies for our setup with a path $\CP$ connecting
$t_{ms}$ to $t_{ams}$. We will define the ``partition function'':
\begin{equation}\label{eq:partfunc}
    F(\Gamma; t) := \sum_{n=-\infty}^{\infty}  q^n \Omega(\Gamma+ n \gamma;t).
\end{equation}
%

%
This is a formal series in $q, q^{-1}$ and we will
demand its continuity along the  path $\CP$. As we have explained, we may focus on the contributions of the halo
states with core charges of the form $\Gamma + m \gamma$, $m\in \IZ$.

When $t$ is infinitesimally close to $t_{ms}$ all halo
states with halo particles of charge parallel to $\gamma$
will contribute to the partition function. No halo states with halo particles with charges
anti-parallel
$\gamma$ will contribute. Therefore, the limiting value as $t$ approaches the marginal
stability line factorizes:
\begin{equation}\label{eq:Ztms}
  F(\Gamma; t_{ms}^+)   = F^{\rm ms}_{\rm core} \cdot F^{\rm ms}_{\rm halo}
\end{equation}
\begin{equation}
  F^{\rm ms}_{\rm core}   :=
      \sum_{n=-\infty}^{\infty}  q^n \Omega^{ms}(\Gamma+ n \gamma)
\end{equation}
\begin{equation}
    F^{\rm ms}_{\rm halo}   :=   \prod_{k >0} (1- (-1)^{k |\langle \Gamma, \gamma\rangle |} q^k)^{k|\langle \Gamma, \gamma\rangle | \Omega(k \gamma)}.
\end{equation}
Here $t_{ms}^+$ means a point infinitesimally displaced from $MS(\gamma,\Gamma)$ into the stable region.
By the same token, near the point $t_{ams}$ all halo states with halo particles of
charge parallel to $-\gamma$ will contribute, while no such states with halo particles of charge
parallel to $\gamma$ will contribute.   Therefore we have the factorization:
\begin{equation}\label{eq:Zatms}
  F(\Gamma; t_{ams}^+)   = F^{\rm ams}_{\rm core} \cdot F^{\rm ams}_{\rm halo}
\end{equation}
\begin{equation}
  F^{\rm ams}_{\rm core}   :=
      \sum_{n=-\infty}^{\infty}  q^n \Omega^{ams}(\Gamma+ n \gamma)
\end{equation}
\begin{equation}
    F^{\rm ams}_{\rm halo}   :=  \prod_{k >0} (1- (-1)^{k |\langle \Gamma, \gamma\rangle |} q^{-k})^{k|\langle \Gamma, \gamma\rangle | \Omega(k \gamma)}.
\end{equation}

Since (for sufficiently small $\CU$)  our path does not cross any walls of marginal stability the two partition functions above must be equal:
\begin{equation}\label{eq:partfunceq}
    F(\Gamma; t_{ms}^+)=F(\Gamma; t_{ams}^+).
\end{equation}
Combining this continuity requirement with the above factorization statements leads to
some interesting constraints on BPS indices.

Note first that since there are no walls of marginal stability in the unstable region
$F(\Gamma;t)$ cannot jump in this region. This suggests that $F^{\rm ms}_{\rm core}$
and $F^{\rm ams}_{\rm core}$ must be identical, but that is not quite the case because
the charges live in a local system. In stating (\ref{eq:partfunceq}) we have implicitly
assumed that the local system has been trivialized throughout the closure of the
stable region in $\CU$. Therefore, we must choose a ``cut'' in the unstable region.
(See Figure \ref{fig:MonodromySign} and   Section    \ref{subsec:monodromy} below.)
Taking into account the monodromy of the local system    we see that instead
\begin{equation}\label{eq:continuity3}
F^{\rm ms}_{\rm core} = q^{-I} F^{\rm ams}_{\rm core}
\end{equation}
for some integer $I$.
Equating the coefficient of $q^n$ on both sides gives:
\begin{equation}\label{eq:indexrel}
    \Omega^{ms}(\Gamma+ n \gamma) = \Omega^{ams}(\Gamma+ (n+I) \gamma).
\end{equation}

Now, equations (\ref{eq:partfunceq}) and (\ref{eq:continuity3}) together would seem to imply
\begin{equation}\label{eq:trick}
F^{\rm ams}_{\rm halo} = q^{-I} F^{\rm ms}_{\rm halo}
 \end{equation}
and indeed formal manipulation of the product formulae above lead to such an identity with
 \begin{equation}
   I :=\vert \langle \Gamma, \gamma\rangle \vert  \sum_{k=1}^\infty k^2 \Omega(k\gamma). \label{eq:I}
\end{equation}
However, we must  stress that (\ref{eq:trick}) is only a formal identity!   The left-hand side is a series in negative powers of $q$. On the other hand, the   right-hand side is a series in positive powers of $q$ times $q^{-I}$, and thus the power series is bounded below.  This necessarily implies   that the power series is
bounded both above and below and moreover  that $F^{\rm ms}_{\rm halo}$, and  $F^{\rm ams}_{\rm halo}$
must be \emph{polynomials} and finally that $I \ge 0$ with $I=0$ only when
$F^{\rm ms}_{\rm halo}= F^{\rm ams}_{\rm halo}$ is a constant in $q$.
 Thus, if we are in a situation where (\ref{eq:trick})
holds then we can conclude:

\begin{enumerate}

\item
At a generic point of a discriminant locus, if $\Omega(k\gamma)\not=0$ for some $k$ then
the quantity $I$ defined in (\ref{eq:I}) must be \emph{positive}. In particular,
it is impossible to have $\Omega(k\gamma)\leq 0$ for all $k\in\IZ$.

\item The spectrum must be such that the product
\begin{equation}\label{eq:Math-Pred2}
P(q) := \prod_{k >0} (1-   q^k)^{k  \Omega(k \gamma)}
\end{equation}
is a \emph{polynomial} in $q$.

\end{enumerate}

The quantity $\sum_{k>0} k^2 \Omega(k\gamma)$ has a nice physical interpretation,
associated with the key insights of \cite{Seiberg:1994rs} and \cite{Strominger:1995cz}.  It is
the coefficient of the $\beta$-function for the $U(1)$ coupling defined by the direction
$\gamma$ in the charge lattice, that is
\begin{equation}
4\pi i \mu \frac{\p}{\p \mu}   \tau = I
\end{equation}
where $\mu$ is the low energy scale at which the coupling is measured.
Thus, our conclusion would seem to be that, given the hypotheses of
Section \ref{subsec:GroundRules} the low energy field theory should be IR free.

Before stating this conclusion we must hasten to add that there is a logical gap  in the
above derivation. The difficulty is that   one must be careful because
manipulation with formal power series in $q,q^{-1}$ can be tricky.
As a simple example note that
$$(\sum_{n\in \IZ} q^n ) (1-q) =0, $$
 so
formal power series can have zero-divisors. From the mathematical standpoint
we must consider three cases: In the first case,  $F^{\rm ms}_{\rm halo}$ is
a polynomial in $q$ and the above reasoning holds. In the second case
$F^{\rm ms}_{\rm halo}$ is a rational function. In the third case
$F^{\rm ms}_{\rm halo}$ is an infinite product.

In this paper we will have nothing to say about the third case, other than to
note that it can happen. (For example the $D0$ halo factor around a $D6$ brane
is a copy of the McMahon function \cite{Denef:2007vg}.)
The second case, where
$F^{\rm ms}_{\rm halo}$ is a rational function would seem to be very physical
since while (half)hypermultiplets have $\Omega = +1$, vectormultiplets have $\Omega = -2$,
which can lead to nontrivial denominators in the product formula for $F^{\rm ms}_{\rm halo}$.
Somewhat surprisingly, as we discuss in Section \ref{sec:Massless-VM}  below, in all
examples we have analyzed, points in moduli space leading to massless vectormultiplets
violate the hypotheses stated in Section \ref{subsec:GroundRules}. Indeed, we show
in   Section \ref{subsec:Rational-case} below that the second case leads to some rather
peculiar physical predictions, and we suspect there are no examples.

Thus we conclude that \emph{ if we assume: 1.) the hypotheses of Section \ref{subsec:GroundRules}
2.) the low energy effective field theory is a conventional field theory, and 3.) the halo
factor is not an infinite product,  then $I>0$
and (\ref{eq:Math-Pred2}) is a polynomial in $q$. }

\bigskip
\textbf{Remark}: The formula (\ref{eq:I}) for $I$ can also be derived
from the Kontsevich-Soibelman wall-crossing formula using the relation between
that formula and monodromy pointed out in \cite{Gaiotto:2009hg} and elaborated in
\cite{Andriyash:2010qv}. In particular, the requirement that the product of KS
transformations denoted $U_{k\gamma}$ in \cite{Andriyash:2010qv} in fact has a
well-defined action on $F(\Gamma;t)$ leads to an alternative argument in favor of
(\ref{eq:Math-Pred2}). A version of this argument is given in Section \ref{sec:FHSV}
below.


%
%

\subsection{The Fermi Flip  }\label{subsec:Fermi}

In this section we describe one way in which the Hilbert spaces of halo states
can change upon crossing conjugation walls.

Restricting attention to the halo subsector of Hilbert space the discussion of
subsection \ref{subsubsec:Halo-Fock-Spaces} shows that
for $t = t_{ms}^+$ the Hilbert space of halo states of total charge
$\Gamma + n \gamma$ is
\begin{equation}\label{eq:GammanLeft}
    \CH(\Gamma+n \gamma;t)|_{t \in \CP_{ms}} \simeq \oplus_{m \ge 0} \Big( \CH^{ms}(\Gamma+(n-m) \gamma) \otimes
    \biggl\{ \otimes_{\ell \ge 1} \CF[ (J_{\Gamma, \ell \gamma}) \otimes \CH(\ell \gamma)] \biggr\}_m \Big).
\end{equation}
The Fermi Fock spaces are graded by a $U(1)$ charge corresponding to $\gamma$ and the subscript $m$
means the subspace of the Fermi Fock space of total $U(1)$ charge $m \gamma$. For $t\in \CP_{ms}$
only halo particles with charges parallel to $\gamma$ can contribute, and in particular the only
 nonzero contributions come from $m\geq 0$. The sum on $m$ comes about
because it is possible to have different core charges.

After crossing the  wall $\CS(n \gamma, \Gamma)$ the Hilbert space becomes:
\begin{equation}\label{eq:GammanRight}
    \CH(\Gamma+n \gamma;t)|_{t \in \CP_{ams}} \simeq \oplus_{m \ge 0} \Big( \CH^{ams}(\Gamma+(n+m) \gamma   ) \otimes \biggl\{ \otimes_{\ell \ge 1} \CF[ (J_{\Gamma, \ell \gamma}) \otimes \CH(-\ell \gamma)] \biggr\}_{-m} \Big).
\end{equation}

Now, it is possible for   (\ref{eq:GammanLeft}) and (\ref{eq:GammanRight}) to be isomorphic
through the following mechanism.
For simplicity of exposition suppose that
   $\CH^{ms}(\Gamma) \ne \emptyset$ but  $\CH^{ms}(\Gamma+n \gamma) = \emptyset$ for $n \ne 0$. It follows from (\ref{eq:indexrel}) that $\Omega^{ams}(\Gamma + m \gamma)$ can only be nonzero for $m=I$, and this
suggests that $\CH^{ams}(\Gamma + m \gamma)=0$ unless $m=I$. We will make that assumption. Then
(\ref{eq:GammanLeft}) and (\ref{eq:GammanRight}) simplify to
\begin{equation}\label{eq:GammanSimpleA}
 \CH(\Gamma+n \gamma;t)|_{t \in \CP_{ms}} \cong  \CH^{ms}(\Gamma) \otimes \left[ \otimes_{\ell \ge 1} \CF[ (J_{\Gamma, \ell \gamma}) \otimes \CH(\ell \gamma)] \right]_n
 \end{equation}
 \begin{equation}\label{eq:GammanSimpleB}
  \CH(\Gamma+n \gamma;t)|_{t \in \CP_{ams}} \cong  \CH^{ams}(\Gamma + I \gamma) \otimes \left[ \otimes_{\ell \ge 1} \CF[ (J_{\Gamma, \ell \gamma}) \otimes \CH(-\ell \gamma)] \right]_{n-I}
\end{equation}
In  (\ref{eq:GammanSimpleA}) we must have $n\geq 0$ while in (\ref{eq:GammanSimpleB})
  we must have $n-I \leq 0$. Thus,
there can only be non-empty spaces for $0\leq n \leq I$. This means that the  Fock space
(\ref{eq:Fockspace-plus})
must be  finite dimensional, i.e. the halo particles of charge proportional to $  \gamma$ must be fermionic.

If we  put $n=I$ and equate (\ref{eq:GammanSimpleA}) with (\ref{eq:GammanSimpleB}) then we find that
\begin{equation}\label{eq:fermisea}
\CH^{ams}(\Gamma + I \gamma) \cong  \CH^{ms}(\Gamma) \otimes \CL
\end{equation}
where $\CL$ is the complex line:
\begin{equation}\label{eq:Filled}
\CL = \Lambda^{max} \biggl (\oplus_{\ell\geq 1}
   (J_{\Gamma, \ell \gamma}) \otimes \CH(\ell \gamma) \biggr).
   \end{equation}
  One can view  (\ref{eq:Filled}) as the entirely filled Fermi Fock space,
  which of course furnishes a ``flipped Fermi sea.''
  More generally, equating the Hilbert spaces at $t=t_{bst}$ we find
\begin{equation}\label{eq:Fermi-Flip}
\CL \otimes \biggl\{\otimes_{\ell\geq 1} \CF[ (J_{\Gamma, \ell \gamma}) \otimes \CH(-\ell \gamma)] \biggr\}_{n-I}
  \cong \biggl\{ \otimes_{\ell \geq 1}  \CF[(J_{\Gamma, \ell \gamma}) \otimes \CH(\ell \gamma)] \biggr\}_n
  \end{equation}

The equation (\ref{eq:Fermi-Flip}) suggests the following interpretation. We should associate to the
halo particles a Clifford algebra.\footnote{Presumably this is simply the algebra of BPS states
 \cite{Harvey:1996gc,Moore:1997ar,KS-CohoHall}. }
On the LHS of (\ref{eq:Fermi-Flip})  we have a subspace of a
 Fock space with creation operators associated to particles of charge $-\ell\gamma$, $\ell>0$.
On the RHS we have a subspace of a Fock space with creation operators associated to particles of
charge $\ell \gamma$, $\ell>0$.  The isomorphism corresponds to a Bogolyubov transformation
that exchanges creation and annihilation operators. The transformation of Fock vacua may be
referred to as ``flipping the Fermi sea.''  An example of this situation is
the conifold point, which we consider in greater detail in Section
\ref{sec:conifold}. In that case the only available halo particles have  charge $\pm \gamma$ and
form
a hypermultiplet,   so all the assumptions are met.

The Fermi flip  nicely accounts for how the BPS spaces change across
conjugation walls when the halo particles are all fermionic.
In cases when $\CH(\ell\gamma)$ contains bosonic degrees of freedom for some $\ell$, we do
not expect an isomorphism between
$\CH(\Gamma+n \gamma;t)|_{t \in \CP_{ms}}$
and $\CH(\Gamma+n \gamma;t)|_{t \in \CP_{ams}}$
and in fact in  section \ref{subsec:spinChar} we will see that in general it cannot be the
case.   (Nevertheless, the index is continuous.)

\subsection{The Fadeout}\label{subsec:Fadeout}

The Fermi flip described in section \ref{subsec:Fermi} implies that as $t$
crosses  the conjugation wall $\CS(n\gamma, \Gamma)$ the supergravity
description of the boundstate changes in an interesting way. Consider first the case $n=I$.
For $t\in \CP_{ms}$ there is a
core charge at a single point in $\IR^3$ and it is surrounded by a halo of particles with charges parallel
to $\gamma$ of total charge $I$.  After crossing the BST wall we have a single core charge of total
charge $\Gamma + I \gamma$ and \emph{no} halo particles! More generally, for $0 \leq n \leq I$,  a state with core charge $\Gamma$
and halo particles with charge parallel to $\gamma$  of total halo charge $n\gamma$ evolves into a state with core charge $\Gamma + I \gamma$ and halo particles with charge anti-parallel to $\gamma$ with total
halo charge $- (I-n)\gamma$.  This sounds like a very discontinuous
process, but, remarkably, the process is in fact physically smooth. Nothing violent happens to our
boundstates. In particular we stress that the boundstate
radius $R_n(t)$  is finite and smooth  in the neighborhood of $t_{bst}$.

First, let us address how a state with halo particles of charge parallel to $\gamma$ can
smoothly evolve into a state with halo particles of charge anti-parallel to $\gamma$.
Recall that as $t$ crosses the conjugation wall the central charge $Z(\gamma;t(\vec x))$
vanishes at the halo radius (in the shell approximation).  Now, let us consider a probe BPS halo particle of charge $\gamma$ in an
attractor background of charge $\Gamma + n \gamma$. It has Lagrangian \cite{Denef:2002ru}
\begin{equation}
L = - 2 e^U \vert Z(\gamma; t(\vec x))\vert (1 - \cos (\alpha_\gamma - \alpha) )
\end{equation}
where $\alpha_\gamma$ is the phase of $Z(\gamma; t(\vec x))$ and $\alpha$ is the
phase of $Z(\Gamma + n \gamma; t(\vec x))$. The term proportional to the cosine
comes from the interaction with the electromagnetic field, and the other term comes
 from the rest mass of the BPS particle. Since $Z(\gamma;t(\vec x)) \to 0$
on the halo radius $R_n(t)$ as   $t$  crosses $\CS(n\gamma, \Gamma)$ we see that the halo particles
have both a mass and coupling to the background gauge fields which approaches
zero. Thus, as $t$ crosses the BST wall $\CS(n\gamma, \Gamma)$ the halo particles
decouple from any possible \emph{local} physical measurement! We call this process the \emph{  fadeout}.

The fadeout takes care of the halo particles, but the reader might still be
disturbed because our description of the core charge has changed discontinuously as $t$ crosses the
BST wall, namely  a core of charge $\Gamma$ appears to have jumped suddenly to a core of charge $\Gamma+I \gamma$.
 The process is in fact physically smooth, but this aspect is best
explained after we have discussed monodromy in Section \ref{subsec:monodromy}

\subsection{Spin character}\label{subsec:spinChar}

Hilbert spaces of BPS states are representations of the spatial rotation group $SU(2)$.
As such they are completely classified by their character. In this section write out
what the description of halo states of section \ref{subsubsec:Halo-Fock-Spaces}  above implies for the
spin character.

To begin, let us define the spin character of halo particles to be:
\begin{equation}
{\rm Tr}_{\CH(\ell\gamma;t)}(-z)^{2J_3} := \Omega(\ell\gamma;-z;t)=
\sum_{-M_{\ell}}^{M_{\ell}} a_{m,\ell} z^m
\end{equation}
By assumption the integers $a_{m,\ell}$ are independent of $t\in
\CU$.  Because of our $\IZ_2$ grading the particles contributing to
$m$ even (for which $a_{m,\ell}>0$) correspond to \emph{fermionic}
particles in the Fock space while those contributing to $m$ odd (for
which $a_{m,\ell}<0$) correspond to \emph{bosonic} particles in the
Fock space. In particular, if $M_{\ell}\leq 1$ then  $a_0$ counts
\emph{hypermultiplets} in four dimensions and $a_{\pm 1}$ counts
\emph{vectormultiplets} in four dimensions. By rotational invariance
$a_{m,\ell} = a_{-m,\ell}$. By CPT invariance $a_{m,\ell} =
a_{m,-\ell}$. In particular,
$\Omega(\ell \gamma;y;t) = \Omega(-\ell\gamma;y;t)=\Omega(\ell\gamma;y^{-1};t)$.

Now we introduce a generating function for the spin characters of
the positive halo Fock spaces:

\begin{equation}
F(q,y;t):= \sum_{n \in \IZ} q^{n} {\rm
Tr}_{\CH^{halo}(\Gamma+n \gamma; t)}  y^{2J_3}
\end{equation}
This is a formal power series in   $q$ whose
coefficients are finite Laurent polynomials in $y$.  The
contribution to this generating function from boundstate halo
particles of charge $\pm \ell \gamma$ is

\begin{equation}
F^{(\pm\ell)}(q,y) =  \prod_{m=-M_{\ell}}^{M_{\ell}} \prod_{j=-
J_{\Gamma,\ell\gamma}}^{J_{\Gamma,\ell\gamma}} (1 + (-1)^m y^{2j+m}
q^{\pm \ell} )^{a_{m,\ell}}
\end{equation}
and in the spirit of our discussion around (\ref{eq:Ztms})-(\ref{eq:Zatms}), we can write down the full generating
function at points $t_{ms}^+$ and $t_{ams}^+$:
\begin{equation}\label{eq:Zms}
F(q,y;t_{ms}^+)  = \Omega^{ms}(\Gamma;y) \prod_{\ell=1}^\infty
F^{(\ell)}(q,y)
\end{equation}
\begin{equation}\label{eq:Zams}
F(q,y;t_{ams}^+)  = \Omega^{ams}(\Gamma;y) \prod_{\ell=1}^\infty
F^{(-\ell)}(q,y)
\end{equation}
where $\Omega^{ms}(\gamma;y)$ is the spin character of $\CH^{ms}(\Gamma)$, etc.

It is easy to see that when $\gamma$ has bosonic internal degrees of freedom, $F(q,y;t_{ms}^+)$ is
an infinite series in positive powers of $q$. The cancellation mechanism mentioned
below (\ref{eq:Math-Pred2}) is no longer operative. Similarly, in this case $F(q,y;t_{ams}^+)$
will be an infinite series in negative powers of $q$. Thus,   (\ref{eq:Zms}) and (\ref{eq:Zams}) can never be equal.
 We conclude that when halo particles include bosonic degrees of freedom the spin character must change
 across conjugation walls.

On the other hand when the only BPS particles  with charge parallel to  $\gamma$  are fermionic, we can apply
the analog of (\ref{eq:trick}), which states that:
\begin{equation}
    \prod_{\ell=1}^\infty F^{(\ell)}(q,y) = q^I \prod_{\ell=1}^\infty F^{(-\ell)}(q,y),
\end{equation}
In this case, the spin character will be smooth across conjugation walls provided the analog of (\ref{eq:continuity3}), is satisfied:
\begin{equation}
    \sum_{m} \Omega^{ms}(\Gamma+m \gamma;y) q^m = \sum_{m} \Omega^{ams}(\Gamma+m \gamma;y) q^{m-I}.
\end{equation}
which would follow, for example, if $\CH^{ms}(\Gamma + n \gamma) \cong \CH^{ams}(\Gamma + (n+I)\gamma)$
upon parallel transport with the flat connection.
When all halo particles are fermionic this is quite a reasonable condition, as we explain in
section \ref{subsec:monodromy}.

\subsection{Monodromy}\label{subsec:monodromy}

It is now time to understand the meaning of the identity (\ref{eq:indexrel}).
First let us note that the choice of $\Gamma$ is rather general.  After all, there
  will be many charges $\Gamma$ which are not local with $\gamma$ and furthermore
support regular attractor points and hence support single-centered black hole solutions in the supergravity approximation. Indeed in the local system of charges   $\Lambda$  (trivialized on $\widetilde{\CM}$),
 there should be an open set of such charges  in $\Lambda\otimes \IR$.

\begin{figure}[htp]
\centering
\includegraphics[scale=0.5,angle=90,trim=0 0 0 0]{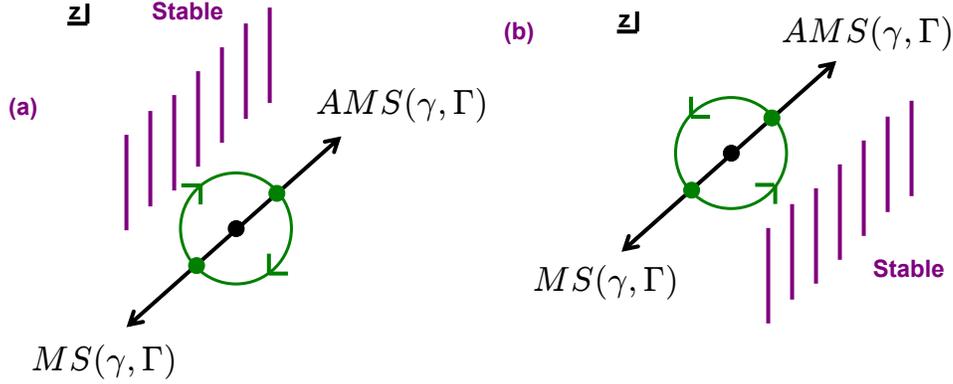}
\caption{A figure of the $z$-plane where $Z(\gamma;t) = z$.
When $\langle \Gamma,\gamma\rangle <0$ the stable region is the shaded region in
(a), and when $\langle \Gamma,\gamma\rangle >0$ it is the shaded region in (b).
The transformation of charge $\Gamma \to \Gamma + I \gamma$ corresponds to a monodromy
transformation around the closed paths indicated in green.}
\label{fig:MonodromySign}
\end{figure}
Since the charges are sections of a
local system $\Lambda$  when comparing indices in an equation such as
(\ref{eq:indexrel}) we must specify a path along which charges have been
 parallel transported.  We   have implicitly assumed that the charges are related by parallel
transport through the stable region. On the other hand, the BPS Hilbert space should
remain ``constant'' in the unstable region, and hence equation (\ref{eq:indexrel})
is naturally explained if $\Lambda$ undergoes a monodromy transformation
\begin{equation}\label{eq:monodromy}
    \Gamma \to M\cdot \Gamma = \Gamma+ I \gamma.
\end{equation}
along a closed path winding once around $\CZ(\gamma)$. The direction of the path is determined
by noting that in the argument used in section \ref{subsec:BPS-Indices}
 we parallel transport the charge $\Gamma$ from $MS(\gamma, \Gamma)$ to
$AMS(\gamma,\Gamma)$ through the stable region. Since the spaces are ``constant'' in the
unstable region a path beginning on   $AMS(\gamma,\Gamma)$ and passing through the unstable
region does not change $\Gamma$, and hence we should consider a closed path that begins on
$AMS(\gamma,\Gamma)$, first passes through the unstable region to $MS(\gamma,\Gamma)$ and
then returns through the stable region back to $AMS(\gamma,\gamma)$, as
 shown in figure \ref{fig:MonodromySign}. Thus, the
  sign of the winding is correlated with the sign of $\langle \Gamma, \gamma \rangle$
and hence we can say that the monodromy transformation for a \emph{clockwise} oriented
curve of winding number one is
\begin{equation}\label{eq:monodromyp}
    \Gamma \to M\cdot \Gamma = \Gamma+ \bar I \gamma.
\end{equation}
where
\begin{equation}\label{eq:Def-Ibar}
\bar I = \langle \Gamma, \gamma\rangle \sum_{k=1}^\infty k^2 \Omega(k\gamma)
\end{equation}

\begin{figure}[htp]
\centering
\includegraphics[scale=0.6,angle=90,trim=0 0 0 0]{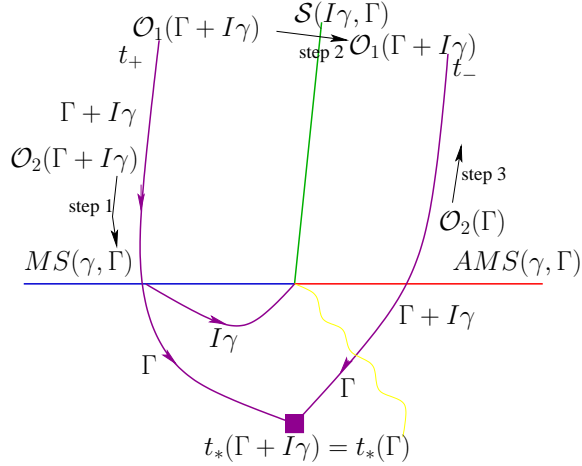}
\caption{The steps in the Gedankenexperiment shown in terms of paths on moduli space.
The conjugation wall is shown in green. There is a cut for the local system $\Lambda$ shown in
yellow, and the attractor flows are illustrated in purple.
The entire experiment involves parallel transport of the charge lattice once around
the locus $\CZ(\gamma)$ in $\CM$.}
\label{fig:thoughtExp}
\end{figure}

\subsubsection{Monodromy and the Fermi Flip: A Gedankenexperiment}

A monodromy transformation of the form (\ref{eq:monodromyp}) around $\CZ(\gamma)$ also
nicely explains how  the Fermi flip transformation of Section
\ref{subsec:Fermi} above is in fact a continuous
physical process. Again, let us work locally on moduli space in the neighborhood $\CU$
of $\CZ(\gamma)$. Let us imagine there are two observers $\CO_1$ and $\CO_2$ in a laboratory
 located very far from  the halo core, effectively at infinite radius. The vectormultiplet moduli
 at this radius are denoted $t$.
  Let us suppose the background modulus $t$
is initially at $t_{bst}^+$, on the $\CP_{ms}$  side of $\CS(I\gamma, \Gamma)$.
Both $\CO_1$ and $\CO_2$ can measure the total charge within a fixed radius $r$.
(For example, they can measure fluxes with local test particles and integrate the
flux.) They both measure the charge of the boundstate to be $\Gamma + I \gamma$.
We now consider a four step experiment.  In step one, one observer, say $\CO_2$,
 travels radially inward toward the core
(potentially observing attractor flow of the vectormultiplet moduli along the way).
As $\CO_2$ passes through the radius $R_n(t_{bst}^+)$ there will be some mild disturbance,
but, because of the fadeout phenomenon, this disturbance will be arbitrarily mild.
For $r< R_n(t_{bst}^+)$, as $r$ decreases to the horizon at $r=0$
$\CO_2$ measures total charge $\Gamma$, and concludes that
the core has charge $\Gamma$.
In step two, observer $\CO_1$ changes the vectormultiplet moduli $t$, crossing the conjugation wall from
$t_{bst}^+$ to $t_{bst}^-$ proceeding from $\CP_{ms}$ to $\CP_{ams}$. Nothing discontinuous
has happened either to $\CO_1$ or to $\CO_2$. In particular
$\CO_1$ continues to measure charge $\Gamma + I \gamma$ of the boundstate. At the same
time, $\CO_2$ also sees nothing discontinuous happening and continues to measure
the charge $\Gamma$.  In step three the observer $\CO_2$
travels radially outward from $r=0$ back to the laboratory of $\CO_1$. In this third step
$\CO_2$ notes that no halo is encountered. Now, for the final step  four of our experiment
$\CO_1$ and $\CO_2$
compare their results for the electromagnetic charge of the core. $\CO_2$ agrees that there is
a single-centered boundstate, and declares its charge to be $\Gamma$, while $\CO_1$ insists
that the charge is $\Gamma + I \gamma$. They are both right, because the Gedankenexperiment
we have just described involves a closed loop in
moduli space around $\CZ(\gamma)$ as shown in figure \ref{fig:thoughtExp}. In our description
of the Fermi flip in section \ref{subsec:Fermi} we used the viewpoint of $\CO_1$. However, in order to investigate
if something discontinuous has happened to the core while crossing $\CS(I \gamma, \Gamma)$
we must send out the observer $\CO_2$ to report core activity from the scene of the crime.
As we have explained, $\CO_2$ saw nothing dramatic happening.

\subsubsection{Area code walls and basins of attraction}

There is an interesting interpretation of conjugation walls in terms of
walls between basins of attraction for attractor flow, $i.e.$
``area code walls'' \cite{Moore:1998zu}. This relation will be used in  the
 covering space description of our Gedankenexperiment in Section \ref{subsubsec:Ged-again}
 below.

Let us assume that $\Gamma + I \gamma$ has at least one regular attractor point.
Consider the attractor flows for $\Gamma + I \gamma$ on the covering space $\widetilde{\CM}$.
The   wall $\widetilde \CS(I\gamma, \Gamma)$ is a set of attractor flows. If this
wall separates $\widetilde{\CM}$ into more
than one component then, since attractor flows cannot intersect except at a
regular attractor point or a singular point the attractor flows cannot
cross the wall $\widetilde \CS(I\gamma, \Gamma)$.

\begin{figure}[htp]
\centering
\includegraphics[scale=0.4,angle=90,trim=0 0 0 0]{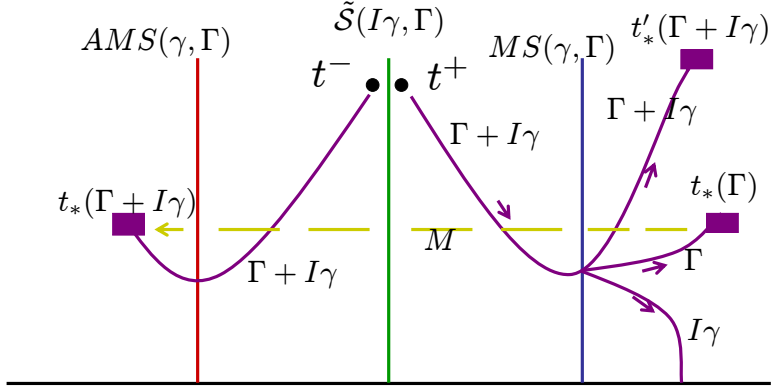}
\caption{Illustrating why the  conjugation wall $\widetilde \CS(I\gamma, \Gamma)$ is an area code wall.
The attractor flow for $\Gamma + I \gamma$ (shown in purple)
from $t^+$ has a split attractor flow, splitting into flows for $\Gamma$ and $I\gamma$. We assume
the attractor
 flow for $\Gamma$ has a regular attractor point at $t_*(\Gamma)$. In the Fermi flip scenario
the action of the monodromy group $M$ (shown in gold) on $t_*(\Gamma)$ produces the regular attractor point
$t_*(\Gamma + I \gamma)$ for flows which begin from $t^-$. Note that $t^-$ is  on the other side of the
conjugation wall $\widetilde  \CS(I\gamma, \Gamma)$. On the other hand,   the $\Gamma+ I \gamma$ flow
from $t^+$ can  be continued from its intersection with $MS(\gamma,\Gamma)$
 and, unless $t_*(\Gamma)$ is a regular attractor point
of rank two, the flow will continue and end on a point other than $t_*(\Gamma)$. We
have denoted this distinct point by $t_*'(\Gamma + I \gamma)$. We claim that $t_*'(\Gamma + I \gamma)$
is also distinct from $t_*(\Gamma + I \gamma)$, and therefore  $\widetilde \CS(I\gamma, \Gamma)$
separates basins of attraction for $\Gamma + I \gamma$ flow.    In our  local
model, where the inverse attractor flow on $\widetilde \CS(I\gamma, \Gamma)$ extends infinitely far upwards,
 the flow cannot cross back to the point $t_*(\Gamma + I \gamma)$ on the left because in order
to do so it would have to cross $\widetilde \CS(I\gamma, \Gamma)$, which is impossible. (In addition it would also have to cross
two  walls $MS(\gamma, \Gamma)$ and $AMS(\gamma, \Gamma)$, which is also impossible by Property 1
of Appendix A.) Thus - in our local model -  if the attractor flow for $\Gamma + I \gamma$ from $t^+$
terminates on a regular attractor point it must be a distinct point  from
$t_*(\Gamma + I \gamma)$.  }
\label{fig:AreaCode}
\end{figure}
It is difficult to give a completely general argument that $\widetilde \CS(I\gamma, \Gamma)$
is an area code wall because one must take into account global properties of covering
space. However, a very  natural scenario is illustrated and explained in figure \ref{fig:AreaCode}. In this case,
$\widetilde  \CS(I\gamma, \Gamma)$ is a wall between basins of attraction for the flow
$\Gamma + I \gamma$.

\subsubsection{Gedanken Again}\label{subsubsec:Ged-again}

\begin{figure}[htp]
\centering
\includegraphics[scale=0.7,angle=90,trim=0 0 0 0]{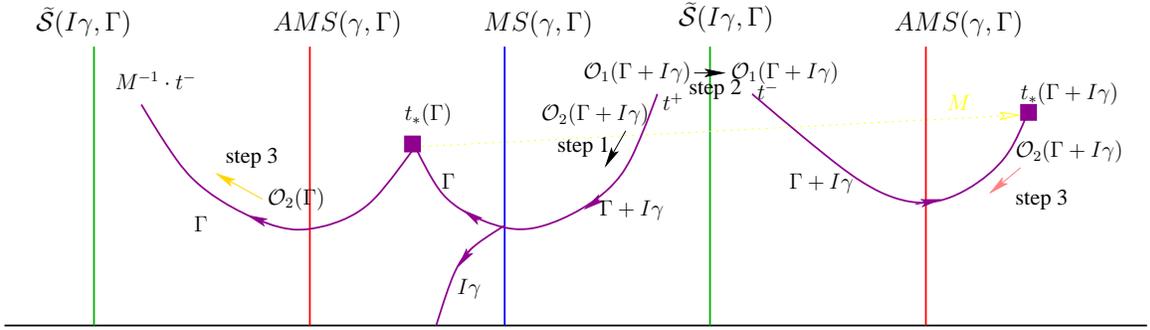}
\caption{Steps in the Gedankenexperiment as described in the covering space. The blue vertical
lines are the marginal stability walls, the red vertical lines are the anti-marginal stability walls and
the green vertical lines are the conjugation walls. Attractor flows are shown in purple. The group of
Deck transformations is generated by $M$ and corresponds to a shift by $+1$. In step 3 $\CO_2$ can
choose to reverse the attractor flow from $t_*(\Gamma)$ as indicated in gold. Alternatively,
$\CO_2$ can   make a gauge transformation while at $r=0$ and reverse the attractor
flow from $t_*(\Gamma + I \gamma)$ as indicated in pink. Note the
discontinuous nature of the attractor flow as $t$ crosses the conjugation wall.}
\label{fig:Cover-Expt}
\end{figure}

Since discussions of this nature are apt to cause confusion it is worthwhile to describe
the same experiment using the language of the covering space $\widetilde{\CU}$ of
$\CU$. Now we must take into account the action of a
 gauge transformation  by a generator $M$ of the
covering (i.e. modular) group. Under this transformation all physical quantities are
invariant, so for example
\begin{equation}
Z(M\cdot \Gamma; M\cdot t ) = Z(\Gamma;t)
\end{equation}
and,
\begin{equation}
\CH(M\cdot \Gamma; M\cdot t ) \cong \CH(\Gamma;t).
\end{equation}

We illustrate the Gedankenexperiment expressed in the language of the
 covering space in  figure \ref{fig:Cover-Expt}.  A  crucial new point
comes at step 3 where $\CO_2$ makes the return trip in $\IR^3$ radially back outward to infinity.
Now,  if $\CO_2$ traverses the inverse attractor flow for charge $\Gamma + I\gamma$ from $t_*(\Gamma)$,
as shown in step 3 of figure \ref{fig:Cover-Expt} she measures the charge $\Gamma$ of the core, but
ends up expressing her measurements
in terms of the point in $ M^{-1}\cdot t^-_{bst} \in  \widetilde{\CU}$  which differs from the point $t^-_{bst}$
used by $\CO_1$. In order to compare results with $\CO_1$ the experimenters must
be on the same page - which in our case literally corresponds to being on
 the same sheet of the covering space.
One of the two experimenters must transform the data by the action of $M^{\pm 1}$. Once this is done they
will   \emph{agree} on the total charge of the single-centered state, as they must,
since the local system $\Lambda$ has been trivialized on $\widetilde{\CU}$.
Alternatively, as indicated by the pink arrow in step three of figure \ref{fig:Cover-Expt}
$\CO_2$ might make a gauge transform while at $r=0$ in order for the  return trip to $r=\infty$
to take her data to $t^-_{bst}$. Thus she starts   not from  $t_*(\Gamma)$  but rather
from its modular image
\begin{equation}
t_*(\Gamma + I \gamma) = t_*(M\cdot \Gamma) = M \cdot t_*(\Gamma)
\end{equation}
It is at this point that $\CO_2$ decides the core has charge $\Gamma + I \gamma$.
 She does not think this is a discontinuous change from step 2, because
she has merely applied a gauge transformation. In step 4, $\CO_1$ and $\CO_2$ are now
on the same page and once again, they agree on the value of the core charge.

\subsubsection{Remark:  Relation to the monodromy of the derived category}

A formula for the autoequivalences in the derived category induced by monodromy around
loci where branes become massless was conjectured in \cite{Aspinwall:2002nw}. (See
Conjecture 1 on page 6.) If an object $\textbf{A}$ in the derived category becomes massless
on a component of the discriminant locus $\CZ$ (and we assume $\textbf{A}$ corresponds to a
single D-brane which is stable in some sufficiently small neighborhood $\CU$ of $\CZ$)
then when one considers a loop
in $\CU$ around $\CZ$ which winds once around $\CZ$
the monodromy action on the   objects in the derived category is claimed to be:
\begin{equation}\label{eq:AJK}
\textbf{B} \to {\rm Cone}\left({\rm Hom}(\textbf{A},\textbf{B})\otimes \textbf{A} \to \textbf{B} \right).
\end{equation}
If one takes the Chern character of this equation, setting $\ch(\textbf{A}) = k \gamma$
and $\ch(\textbf{B})= \Gamma$,
   then (\ref{eq:AJK}) becomes the standard
Lefshetz formula
\begin{equation}
\Gamma \to \Gamma  +  \langle \Gamma,k \gamma \rangle k\gamma.
\end{equation}
There is therefore some tension between (\ref{eq:AJK}) and our
  formula (\ref{eq:monodromy}), (\ref{eq:Def-Ibar}). The latter involves a \emph{sum} over all charges
  parallel to $\gamma$ and is moreover weighted by the BPS index $\Omega(k \gamma)$.
      Since the group of  autoequivalences of the derived
category does not depend on a stability condition our expression involving $I$ might seem
somewhat strange. However, since we have explicitly assumed that the $\Omega(\gamma;t)$
are constant in the neighborhood of $\CZ(\gamma)$ there is not a strict contradiction here.
Clearly this point needs to be understood better.\footnote{We thank Emanuel Diaconescu
for useful discussions about this point.}

\subsection{The case when $F_{\rm halo}$   is  a rational function }\label{subsec:Rational-case}

Let us now return to the logical possibility, mentioned in Section \ref{subsec:BPS-Indices},
that $F_{\rm halo}^{\rm ms}(q)$ is a rational function of $q$. We will show that this
leads to some     physical predictions which are so peculiar that we suspect that there
are no examples.

To begin we prove a small Lemma from High School Mathematics:

\bigskip

\textbf{Lemma}:  Let  $R(q)$  be a rational function of $q$ with a convergent power series
around $q=0$,  $R(q) = 1 + \CO(q)$. Suppose $R$  has poles, and all the poles
lie on the unit circle. Then there exists a positive integer $L$ such that
$R(q)$ has a power series expansion of the form

\begin{equation}
R(q) = \sum_{r=0}^{L-1} \sum_{n\geq 0}  a_{r,n} q^{r + nL}
\end{equation}

where, for some $r, n_0 $ we have  $a_{r,n} \not=0$  for all $n\geq n_0$.

\bigskip
\emph{
Proof:} First note that $R(q)$  has a continued fraction expansion of the form:
\begin{equation}\label{eq:cont-frac}
R(q) = r(q) + \sum_{\rho, s}   \frac{a_{\rho,s} }{(1 - \rho q)^s }
\end{equation}
where $r(q)$ is a polynomial and the sum over
$\rho$ runs over a finite set of roots of unity and $s $ runs
over a finite set of positive integers. The coefficients $a_{\rho,s}$
are complex numbers.

Note that if $\rho$ is an $L^{th}$  root of unity then
\begin{equation}\label{eq:qs-exp}
\frac{1 }{(1 - \rho q)^s }=    \sum_{r=0}^{L-1} \sum_{n\geq 0}  p_r(n)  q^{r + nL}
\end{equation}
where, for each $r$,  $p_r(n)$ is a polynomial in $n$.  ($p_r(n)$ might depend on $r$ and is of order $(s-1) $).
Now let $L$ be any integer such that all $\rho$ which occur in (\ref{eq:cont-frac})  are $L^{th}$ roots of unity
and observe that a non vanishing polynomial can have at most a finite number of roots $\blacksquare$

Now we claim that if $F^{\rm ms}_{\rm halo} $ is a rational function of $q$  with
poles on the complex $q$-plane,  (as can happen if there are bosons in the halo)
then $F(\Gamma;t)$   is a finite Laurent polynomial
in $q$. In particular,  $F_{\rm core}^{\rm ms}(q)$ is a finite Laurent polynomial in $q$ whose set of
zeroes includes the set of poles of  $F^{\rm ms}_{\rm halo} (q)$  and $F^{\rm ms}_{\rm halo} (q^{-1} )$, counted
with multiplicity.

To prove this we need   need three ingredients:
First, as we have seen,  it  follows from the monodromy   that we have (\ref{eq:continuity3}),
an identity of   formal series in $q, q^{-1}$. Second,
continuity of the index away from walls of marginal stability implies (\ref{eq:partfunceq}).
Third, finiteness of the number of attractor flow trees implies that if we write
$F_{\rm core}^{\rm ms}(q) = \sum b_m q^m$  and $F^{\rm ms}_{\rm halo}(q) = \sum c_n q^n$, then, for all $N$,
  \begin{equation}
  d_N = \sum_{n+m=N}  b_m c_n
  \end{equation}
is a \emph{finite} sum.

Now if  $F^{\rm ms}_{\rm halo}(q)$ has a pole then by the Lemma above  we know that
for some $L, r, n_0$  we have  $a_{r,n} \not=0$  for all $ n\geq n_0 $ in the expansion:
\begin{equation}
F^{\rm ms}_{\rm halo}(q) = \sum_{r=0}^{L-1} \sum_{n\geq 0}  a_{r,n} q^{r + nL}
\end{equation}
Next, suppose the exponents of $q$ in the expansion of $F_{\rm core}^{\rm ms}(q)$  is
unbounded below. That is, $\{ m\vert b_m \not=0 \}$  is unbounded below.
Then choosing $L$ as above, it must be that for some residue $r'$  and  some  $n_0'$
we have
 $b_{-r - n L } \not=0$   for   $n> n_0'$ .  Then the coefficient of $q^{r-r'}$ contains the infinitely many terms
\begin{equation}
\sum_{n > {{\rm max}[n_0,n_0'] } }   b_{-r' -n L} a_{r,n}
\end{equation}
But this violates the ingredient 3 above, based on the finiteness of the number of attractor
flow trees, and hence we  conclude that  $F_{\rm core}^{\rm ms}(q)$  has coefficients
bounded blow.

On the other hand,    $F^{\rm ams}_{\rm halo}(q) = F^{\rm ms}_{\rm halo}(q^{-1})$   so exactly the same reasoning
implies that  $F_{\rm core}^{\rm ams}(q)$  has coefficients bounded above. It now follows from
equation (\ref{eq:continuity3})   that $F_{\rm core}^{\rm ms}(q)$  is a finite Laurent polynomial.
Applying the same reasoning and now using the equality of indexes,  (\ref{eq:partfunceq})  shows that
$F_{\rm core}^{\rm ms}(q) F^{\rm ms}_{\rm halo}(q)$  must also be a finite Laurent polynomial.
Therefore, the poles of $F^{\rm ms}_{\rm halo}(q)$ must also be zeroes of $F_{\rm core}^{\rm ms}(q)$, including
multiplicity.  But the same reasoning  applied to  $q^I  F_{\rm core}^{\rm ms}(q) F^{\rm ams}_{\rm halo}(q)$
shows that the poles of $F^{\rm ams}_{\rm halo}(q)$  must also be zeroes of $F_{\rm core}^{\rm ms}(q)$.
This concludes the proof $\blacksquare$

Now, the above structure of $F_{\rm halo}$ and $F_{\rm core}$ is rather odd.
The poles of $F_{\rm halo}$ only depend on $\gamma$. On the other hand,
  there is a wide
choice of $\Gamma$'s for which the rules of the game in Section \ref{subsec:GroundRules}
apply. Indeed, we expect that there is an open set of such elements in the space of
charges. For all these charges $F_{\rm core}$ must be a Laurent polynomial with
zeroes at the poles of $F_{\rm halo}$. This means that for some collection of roots
of unity $\rho$ (depending only on $\gamma$) we must have
\begin{equation}
\sum_{n\in \IZ}  \Omega^{ms}(\Gamma + n \gamma) \rho^n =0
\end{equation}
for all charges $\Gamma$ mutually nonlocal with respect to $\gamma$ and supporting, say,
single-centered attractor flows. Such constraints seem to us physically unreasonable.
We certainly know of no examples, and we find the above a compelling argument that
$F^{\rm ms}_{\rm halo}$ is either a finite Laurent expansion or an infinite product.

\section{An Example: Conifold-Like Singularities}\label{sec:conifold}

In this Section we discuss in detail a simple example of the setup of Section \ref{sec:BST-walls},
in which the spectrum of massless states of charge
$\gamma$ consists of a single hypermultiplet. Such massless BPS states are present when a Calabi-Yau manifold develops a conifold
singularity. In the IIA picture, the halos are made of light D2/D0
branes wrapping a rational curve, bound to a heavy core D6 brane
filling the entire Calabi-Yau manifold. The local geometry of the
Calabi-Yau near the rational curve is ${\cal O}(-1) \oplus {\cal
O}(-1) \rightarrow \mathbb{P}^1$, and a hypermultiplet becomes
massless at the singular point in the K\"{a}hler moduli space
where the class of the rational curve vanishes. The massless state is a pure fermion with $\Omega(\gamma)=1$ and from the discussion in Section \ref{subsec:Fermi}  we know the structure of the halo bound states of core D6 with D2/D0 particles near the singularity, their BPS indices and Hilbert spaces. In particular, the Hilbert spaces of halo states transform smoothly across the conjugation walls through the Fermi flip of Section \ref{subsec:Fermi}. The generating function of the index of such BPS states as a function of the K\"{a}hler moduli was first computed in \cite{NagaoNakajima, Nagao} from the quiver category point of view, generalizing the results in special chambers found in \cite{qfoam} and \cite{Szendroi}.

\subsection{The resolved conifold}\label{subsec:conifold}

 Since the motivation for the present paper was the puzzle raised in \cite{Jafferis:2008uf}, we now briefly review the setting of that    puzzle   and describe it's relation to the general discussion above. We will describe how the local geometry
of the covering space near the conifold point fits together
with the patch of the moduli space in which the volume of the
entire Calabi-Yau is taken very large. (This   was used
in \cite{Jafferis:2008uf} to derive the partition function of
D6/D2/D0 bound states using the semi-primitive wall crossing formula.)
Consider the local limit of a Calabi Yau 3-fold X, with only one homology class,
a rigid rational curve, dual to $\beta \in H^4(X, \IZ)$, remaining
small. The K\"ahler parameter is $t = z \CP + L e^{i \phi}
\CP'$, where $L \to \infty$ in the local limit, $\CP \beta =
1$, for a positive integral class $\CP$, and $\CP' \beta = 0$ for
a semi-positive class $\CP'$. This parameterization is only valid
for $\phi \in (0, \pi)$ and $Im(z)>0$, which corresponds to a
patch in the full covering space. In this patch the large volume expression for the periods can be used, and for instance (here $Z_h$ is the holomorphic central charge):

\begin{equation}
    Z_h(1) = \frac{1}{3} L^3 e^{3 i\phi} \quad , \quad Z_h(\beta) = z \quad {\rm and } \quad Z_h(dV) = -1.
\end{equation}

\begin{figure}[htp]
\centering
\includegraphics[scale=1,angle=0,trim=0 0 0 0]{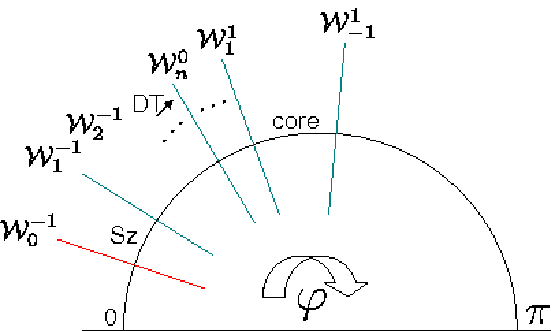}
\caption{.}
\label{fig:Conifoldwalls}
\end{figure}

BPS states with charges of the form $\Gamma_{m,n} = 1-m \beta + n
dV$, where $dV$ is a generator of $H^6(X;\IZ)$,  are realized as multicentered solutions in supergravity
description. As argued in \cite{Jafferis:2008uf}, in the neighborhood of the wall $\phi = \frac{1}{3} \arg z +\frac{\pi}{3}$ only the pure D6 brane with charge $\Gamma_0=1$ exists as a single centered object (out of the set of
objects with charges of the form $\Gamma_{m,n}$).
 We are interested in configurations that consist of this core $D6$ charge
surrounded by halos of $D2$-$D0$ particles $\gamma_{m,n} = -m
\beta + n dV$.  \cite{Jafferis:2008uf} computed the D6/D2/D0
partition function defined by

\begin{equation}\label{eq:Zd6d2d0}
 F(u,v; t_\infty)
:= \sum_{N \in \IZ, \beta \in H^4(X,\IZ)} u^N v^\beta \Omega(1-\beta
+ N dV; t_\infty),
\end{equation}
in all chambers of the K\"ahler cone, parametrized in the local
limit $L \to \infty$ by $(z, \phi)$. Fixing some value of
$z$ the $\phi$ interval can be divided into chambers, bounded by
(anti)marginal stability walls $\CW^m_n$ of D6 with
$\gamma_{m,n}$ as in figure \ref{fig:Conifoldwalls}, found to be

\begin{eqnarray}\label{eq:Wwalls}
&& \CW^m_n = \{ (z,\phi): \phi = \frac{1}{3} arg (z+ n/m) + \frac{\pi}{3} \} \nonumber\\
&& \CW^{-m}_n = \{ (z,\phi): \phi = \frac{1}{3} arg (z- n/m)  \} \nonumber\\
&&  \CW^{-m}_{-n} = \{ (z,\phi): \phi = \frac{1}{3} arg (z+ n/m) \},     \nonumber\\
\end{eqnarray}
with both $n \ge 0$, $m \ge 0$. One can see that there is a path from $\CW^m_n$ to $\CW^{-m}_{-n}$
along the $\phi$ direction and we can form ``positive''
(``negative'') halos around the D6 with particles $\gamma_{m,n}$
($-\gamma_{m,n}$) near $\CW^m_n$ ($\CW^{-m}_{-n}$), but these
halos cannot exist near $\CW^{-m}_{-n}$ ($\CW^m_n$ ). As all
particles $\gamma_{m,n}$ are mutually local, we can focus on one
of them to see the resolution. Also the index of particles with
$|m|>1$ is zero, so we can take some charge $\gamma_{1,n}$ as an
example. The moduli space has conifold singularities at the loci
$z = n$, where  $\gamma_{\pm 1, \mp n}$ becomes massless. Locally
around each point $z = n$ the moduli space will look like an
infinite sheeted cover of the $z$-plane, described by a coordinate
$u_n =\log (z-n)$, and the charge lattice will have the monodromy
$\Gamma \mapsto \Gamma - \langle \Gamma, \ \gamma_{1,n} \rangle
\gamma_{1,n}$.  However, it is clear that in the large $L$
limit for fixed $u_n, \phi$, and to leading order in $1/L$,
the periods will be given exactly as above. Therefore walls of
marginal stability will approach being periodic in the $u_n$-plane
for large $L$. We now describe the conjugation behavior in a single
patch $Im(z) > 0, \ \phi \in (0,\pi)$, which will be repeated
periodically in the $u_n$-plane.

The location of the conjugation walls for $\Gamma$ with each of the
$\gamma_{1,n}$ can be determined precisely in the limit of large
$L$. In terms of the period vector $\Omega(t)$ the attractor
flow of the total charge is determined by

\begin{equation}\label{eq:attreq}
2 e^{-U} \Im(e^{-i\alpha} \Omega(t) ) = -\tau \Gamma +
H_{\infty}(t_{\infty}),
\end{equation}
where $\tau$ is the parameter along the flow and $\Gamma =
\Gamma_0 + \gamma_{1,n}$. We are interested in finding the locus
$t_{\infty}$, such that $t(\tau)$ will intersect the discriminant
locus of vanishing $Z(\gamma_{1,n})$. Note that such a $t(\tau)$
necessarily satisfies
\begin{equation}\label{eq:BSTeq}
\im \left[ \bar Z(\gamma_{1,n};t) Z(\Gamma_0;t) \right] = 0,
\end{equation}
which immediately determines the value of $\tau$ at the
intersection, due to the linearity of the attractor equation for
the periods. Now we impose that this $t$ is on the discriminant
locus. Of course, to solve (\ref{eq:BSTeq}) and find the location
of the conjugation wall we need to know the periods of D6 and D2/D0 close
to the singularity, where they do get corrections from the D2/D0
becoming massless. It turns out that in the limit of large
$L$ the point $t(\tau)$ will also have an order $L$
component in the direction of $\CP'$. Thus in the neighborhood of
the conjugation wall the corrections to the periods at $t(\tau)$ will
still be of order $1/L$, which allows us to solve
(\ref{eq:BSTeq}) in this limit (see Appendix \ref{app:walls}). The
conjugation walls $\CS(\gamma_{m,n}, \Gamma_0)$ turn out to be
located at
\begin{equation}\label{eq:BSTlocation}
\phi =
\frac{1}{2} \arg \left(-\frac{m}{n}z -1\right) + \frac{\pi}{2},
\end{equation}
in the patch of the Teichm\"uller space that we are
discussing. Comparing (\ref{eq:Wwalls}) and (\ref{eq:BSTlocation}) we see that the conjugation wall is indeed located in the region of bound state
stability between the corresponding marginal and anti-marginal stability walls. Notice that the conjugation walls for all charges $k \gamma_{m,n}$ coincide, unlike the picture in $z$-plane near the singularity that we had before in Figure \ref{fig:BSTwall2}. This is of course an artifact of using the large volume expression for the periods, or in other words as $L \to \infty$  all the conjugation walls for $k \gamma_{m,n}$ asymptote to (\ref{eq:BSTlocation}) .

The particles $\gamma_{m,n}$ that inhabit the halos in this
example have indices given by
\begin{align}
    & \Omega(\pm \beta + n dV) = 1, \textrm{for all } n, \nonumber\\
    & \Omega(n dV) = -2, \textrm{for } n \neq 0,
\end{align}
where the D2/D0 states with charges $\pm \beta + n dV$ are
the free hypermultiplet at the conifold point and its images under
the large gauge transformations $B \mapsto B - n {\cal P}$. The
attractor flows of objects with only D0 charge flow to large
volume, where $n$ D0 branes can form precisely one bound state,
giving a massless vector multiplet. The expression
(\ref{eq:BSTlocation}) shows that there are no conjugation walls for the
D6 brane with the D0 particles, confirming the general argument
that $\Omega(\gamma)$ must be positive. This can also be seen
directly: the attractor flow of the D6 with D0's can never reach
the large volume point since then the period of the D6 brane would
be increasingly well approximated by the large volume form, which
grows without bound at large volume, contradicting the gradient
flow.

On the other hand, $\gamma_{1,n}$ particles will have conjugation walls with the D6. Upon crossing those, a filled Fermi sea of $n$ $\gamma_{1,n}$ particles appears at the halo radius. From Section \ref{subsec:BPS-Indices} we know that the partition function for each individual halo particle $\gamma_{1,n}$ stays constant across the conjugation walls, so we can write the partition function for halos of all $\gamma_{1,n}$'s in all chambers of (\ref{eq:Wwalls}) as\footnote{The first two lines were already presented in \cite{Jafferis:2008uf}, but the third line is a new result.}

\begin{eqnarray}\label{eq:partitionFunctions}
&& F(u,v; [{\cal W}^1_n {\cal W}^1_{n+1}]) = \prod_{j=1}^n \left(1 - (-u)^j v\right)^j \nonumber\\
&& F(u,v; [{\cal W}^{-1}_{n+1} {\cal W}^{-1}_n]) = \prod_{j > 0}
\left( 1 - (-u)^j \right)^{-2j} \left(1 - (-u)^j v \right)^j
\prod_{k > n} \left(1 - (-u)^k v^{-1} \right)^k \nonumber\\
&& F(u,v; [{\cal W}^{-1}_{-n-1} {\cal W}^{-1}_{-n}]) = \left( \prod_{\ell = 1}^n (u^\ell v)^\ell \right)
 \prod_{j > 0}
\left( 1 - (-u)^j \right)^{-2j} \prod_{j > 0} \left(1 - (-u)^j
v^{-1} \right)^j \prod_{k>n}\left(1 - (-u)^k v \right)^k, \nonumber\\
\end{eqnarray}
with $n \ge 0$. The last expression is evaluated in the chamber $[{\cal W}^{-1}_{-n-1} {\cal W}^{-1}_{-n}]$, where all halos of particles $- \gamma_{1,k}$, $k=1..n$ have decayed after crossing $MS(\Gamma, -\gamma_{1,k})$ from stable to unstable side, and the factor $\prod_{\ell = 1}^n (u^\ell v)^\ell$ accounts for the change of the core charge $\Gamma_0$ due to monodromy. It comes about through the identity

\begin{equation}\label{eq:rearrangement}
    (1-(-u)^\ell v)^\ell =( u^\ell v)^\ell \: (1-(-u)^{-\ell} v^{-1})^\ell,
\end{equation}
where the second factor is the contribution of all halos of $-\gamma_{1,n}$ particles. After crossing the conjugation walls $\CS(\gamma_{1,\ell}, \Gamma_0)$, $\ell = 1..n$ the core charge changes according to the monodromy

\begin{equation}\label{eq:coreChargeMonodromy}
    \Gamma_0 \to \Gamma_0 - \sum_{\ell=1}^n \langle \Gamma, \ \gamma_{1,\ell} \rangle \gamma_{1,\ell} = \Gamma_0 + \sum_{\ell=1}^n \ell \gamma_{1,\ell}  .
\end{equation}
In particular, the ``pure D6'' brane with core charge $\Gamma_0$   {\it does  not  exist} after crossing the first conjugation wall. It   gets replaced according to (\ref{eq:coreChargeMonodromy}). Also note that to be more precise one might want to rewrite (\ref{eq:partitionFunctions})  using the left-hand side of (\ref{eq:rearrangement}) before crossing the wall $\CS(\gamma_{1,\ell}, \Gamma_0)$, and right-hand side after crossing it. This is straightforward using (\ref{eq:BSTlocation}) and (\ref{eq:Wwalls}), but the result would look rather messy which is why we don't do it here.

Note that for fixed $z$, taking $\phi$ to zero involves crossing
infinitely many conjugation walls and marginal stability walls. In fact,
the limits $\phi \rightarrow 0$ and $L \rightarrow \infty$
do not commute, since when $\phi$ is too close to 0, the imaginary
part of the K\"ahler form becomes small, and the large volume
approximation to the periods breaks down. Moreover, $$\lim_{\phi \rightarrow 0}\left(
\lim_{L \rightarrow \infty} Z(u,v; t_\infty = z {\cal P} +
L e^{i\phi} {\cal P}^\prime) \right)$$ does not exist. To proceed further would
require specifying a particular compact Calabi-Yau with a conifold
degeneration.

\section{Recombinaton walls}\label{sec:TCwalls}

\subsection{BPS index}

In this Section we describe what happens to the BPS state when it crosses the recombination wall. Going back to Figure \ref{fig:TSWleft}, let us first describe the $RW(\Gamma_2,\Gamma_3,\Gamma_4)$ in more detail. As mentioned above (\ref{eq:TCdef}) at least one of $\Gamma_{ij, k}$ is nonzero. Using an identity $\Gamma_{ij, k}+\Gamma_{jk, i}+\Gamma_{ki, j}=0$ we see that two of them have to be non-zero. Without loss of generality we have  $\Gamma_{34,2} \ne 0$, $\Gamma_{24,3} \ne 0$. In what follows we also take $\Gamma_{23,4} \ne 0$ and comment on the case $\Gamma_{23,4} = 0$ in the end of this section. Although the definition appears to depend on the entire attractor flow, the position of this wall can be expressed
entirely in terms of the periods evaluated at $t_\infty$. By definition, the recombination wall consists of points
$\{t|\tau_{ms}(\Gamma_{total}, \Gamma_2; t) = \tau_{ms}(\Gamma_{total}, \Gamma_3; t)\}$, as the central charge
of $\Gamma_4$ will then also be aligned at the point
$t(\tau_{ms})$. Thus by equation (\ref{eq:tau_ms}), it is given by

\begin{equation}\label{eq:thresholdstab}
    \Im\left[
\frac{Z(\Gamma_3+\Gamma_4; t) \bar{Z(\Gamma_2; t )}}{\langle
\Gamma_3+\Gamma_4, \Gamma_2\rangle}\right] =
\Im\left[\frac{Z(\Gamma_2+\Gamma_4;t) \bar{Z(\Gamma_3;t)}}{\langle
\Gamma_2+\Gamma_4, \Gamma_3 \rangle} \right].
\end{equation}
This equation can be expressed in the form

\begin{eqnarray}\label{eq:TCMS}
&& \Im \: Z(\Gamma_a;t) \bar Z(\Gamma_b;t)=0, \: \: \mbox{where} \nonumber\\
&& \Gamma_a = \Gamma_2 \langle
\Gamma_{total}, \Gamma_3 \rangle - \Gamma_3 \langle
\Gamma_{total}, \Gamma_2 \rangle\nonumber\\
&& \Gamma_b =\Gamma_{total}.
\end{eqnarray}
(\ref{eq:TCMS}) can be interpreted as the MS wall for charges $\Gamma_a$ and $\Gamma_b$, satisfying $\langle \Gamma_a, \Gamma_b\rangle =0$.

\begin{figure}[htp]
\centering
\includegraphics[scale=0.5,angle=90,trim=0 0 0 0]{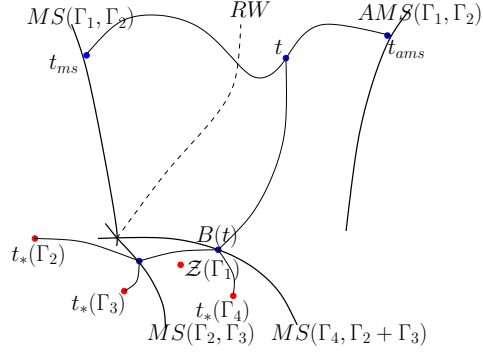}
\caption{Bound state of charges $\Gamma_4 + (\Gamma_3 + \Gamma_2)$ on the right of the recombination wall.}
\label{fig:TSWright}
\end{figure}

Returning to Figure \ref{fig:TSWleft}, we see that the path $\CP$ necessarily
crosses the recombination wall, and we claim that the bound state $\Gamma_2 + (\Gamma_3 +
\Gamma_4)$ on the left  is transformed into two
bound states $\Gamma_3 + (\Gamma_2 + \Gamma_4)$ and $\Gamma_4 +
(\Gamma_3 + \Gamma_2)$  on the right. Figure \ref{fig:TSWright} shows one
of the two bound states on the right of the recombination wall. In \cite{Denef:2007vg, DdBvdB} this situation was illustrated in particular examples, and it was found that on the level of
the index the transition is smooth, so that the two bound states
on the right have exactly the same index as the one on the left.
Note that our puzzle from the Introduction gets resolved since the bound state of $\Gamma_1=\Gamma_3+\Gamma_4$ and
$\Gamma_2$ does not exist when $t$ reaches $t_{ams}$ and the total charge $\Gamma_2+\Gamma_3+\Gamma_4$ has
a different realization.

Now we will explain in complete generality why the above claim is true:
that is, why there are precisely two attractor trees on one side of the wall and one tree on the other side, and why the net BPS index is the same and also why the spin character does not change.
\footnote{In some independent
work Jan Manschot found  another version of this proof  \cite{Manschot:2010xp}.}
 The locus where all three central charges align is complex codimension 1 in the moduli space, and we can parametrize the transverse plane with complex coordinate $z$, $z=0$ being the alignment point. To understand the phenomenon it is enough to work in a small neighbourhood of $z=0$. There will be six walls of marginal stability for pairs of charges $(\Gamma_2, \Gamma_3+ \Gamma_4)$, $(\Gamma_3, \Gamma_4+ \Gamma_2)$, $(\Gamma_4, \Gamma_2+ \Gamma_3)$, $(\Gamma_2, \Gamma_3)$, $(\Gamma_3, \Gamma_4)$, $(\Gamma_4, \Gamma_2)$, all intersecting at $z=0$. To understand which attractor trees exist on each side of the recombination wall, we will find the intersection points of the attractor flow with $(\Gamma_2, \Gamma_3+ \Gamma_4)$, $(\Gamma_3, \Gamma_4+ \Gamma_2)$, $(\Gamma_4, \Gamma_2+ \Gamma_3)$, and then see whether these points lie on the stable side of $MS(\Gamma_3, \Gamma_4)$ , $MS(\Gamma_4, \Gamma_2)$ and $MS(\Gamma_2, \Gamma_3)$  respectively.

In a small neighbourhood of $z=0$ we can expand all central charges in powers of $z$ to first order

\begin{eqnarray}
Z(\Gamma_i;z) \approx   Z(\Gamma_i;0) + \partial_z Z(\Gamma_i;z)|_{z=0} \: z,
\end{eqnarray}
for $i=2,3,4$. We will assume that the two terms in this expansion are nonzero.
In this approximation the marginal stability walls can be written as

\begin{eqnarray}\label{eq:rhodef}
&&  MS(\Gamma_i, \Gamma_j) = \{z: \Im \: \rho_{ij} \:\bar z =0 \}, \nonumber\\
&&  MS(\Gamma_i+\Gamma_j,\Gamma_k) = \{z: \Im \: (\rho_{ik} +\rho_{jk}) \:\bar z =0 \}, \nonumber\\
&& \rho_{ij}:= Z(\Gamma_i;0) \: \partial_{\bar z} \bar Z(\Gamma_j;\bar z)|_{\bar z=0}- Z(\Gamma_j;0) \: \partial_{\bar z} \bar Z(\Gamma_i;\bar z)|_{\bar z=0}.
\end{eqnarray}
The recombination wall of the total charge
$\Gamma_2+\Gamma_3+\Gamma_4$
is determined from (\ref{eq:thresholdstab}) and
%
%
can be
written in parametric form as

\begin{equation}\label{eq:TCWflow}
    z(s) = s \frac{\left( \rho_{23} \Gamma_{23,4} + \rho_{24} \Gamma_{3,42} + \rho_{43} \Gamma_{2,34} \right)}{\left| \rho_{23} \Gamma_{23,4} + \rho_{24} \Gamma_{3,42} + \rho_{43} \Gamma_{2,43} \right|},
\end{equation}
where $s \in \IR$ is a parameter. The attractor flows on the two
sides of the recombination wall can be analogously written as

\begin{equation}\label{eq:TCWflows}
    z_{\pm}(s) = (\pm i \varepsilon + s)\frac{\left( \rho_{23} \Gamma_{23,4} + \rho_{24} \Gamma_{3,42} + \rho_{43} \Gamma_{2,34} \right)}{\left| \rho_{23} \Gamma_{23,4} + \rho_{24} \Gamma_{3,42} + \rho_{43} \Gamma_{2,43} \right|},
\end{equation}
where $\varepsilon >0$ is a very small shift off the recombinationW. Let us
choose for definiteness the plus sign in (\ref{eq:TCWflows}), find
the value of $s$ where the flow intersects  the wall $MS(\Gamma_2,
\Gamma_3+ \Gamma_4)$, and check if this point is on the
stable side of the wall $MS(\Gamma_3, \Gamma_4)$. Some simple
algebra yields the condition:

\begin{equation}\label{eq:123condition}
    -\frac{\Gamma_{34}}{\Gamma_{34,2}} \frac{\Im \rho_{23} \bar \rho_{34} + \Im \rho_{34} \bar \rho_{42}}{\Im \rho_{42} \bar \rho_{23} + \Im \rho_{23} \bar \rho_{34} + \Im \rho_{34} \bar \rho_{42}}>0
\end{equation}
If this condition is true, then the attractor flow $z_+(s)$ will cross $MS(\Gamma_2, \Gamma_3+ \Gamma_4)$ on the stable side of $MS(\Gamma_3, \Gamma_4)$ and the attractor tree  $(\Gamma_2, (\Gamma_3, \Gamma_4))$ will exist. Using the definition of $\rho_{ij}$ from (\ref{eq:rhodef}) together with the fact that all the $Z(\Gamma_i,0)$ have
 the same phase, we can, after a bit more algebra, rewrite (\ref{eq:123condition}) as

\begin{equation}\label{eq:123condition1}
    -\frac{\Gamma_{34}}{\Gamma_{34,2}} \frac{|Z(\Gamma_3;0)| + |Z(\Gamma_4;0)|}{|Z(\Gamma_2;0)| +|Z(\Gamma_3;0)| + |Z(\Gamma_4;0)|}>0.
\end{equation}
Repeating the calculation for the two remaining trees $(\Gamma_3, (\Gamma_4, \Gamma_2))$ and $(\Gamma_4, (\Gamma_2, \Gamma_3))$ and getting rid of positive factors we get

\begin{eqnarray}\label{eq:existenceoftrees}
(\Gamma_2, (\Gamma_3, \Gamma_4)): & -\Gamma_{34} \: \Gamma_{34,2} >0 \nonumber\\
(\Gamma_3, (\Gamma_4, \Gamma_2)): & -\Gamma_{42} \: \Gamma_{42,3} >0 \nonumber\\
(\Gamma_4, (\Gamma_2, \Gamma_3)): & -\Gamma_{23} \: \Gamma_{23,4} >0. \nonumber\\
\end{eqnarray}
This is the main result of the above calculation. Taking into account that

\begin{equation}\label{eq:identity}
    \Gamma_{34} \: \Gamma_{34,2} +\Gamma_{42} \: \Gamma_{42,3} + \Gamma_{23} \: \Gamma_{23,4} = 0,
\end{equation}
we see that it is impossible to have all three trees to be present on one side of the recombination wall. Furthermore,
 since choosing the attractor flow on the other side of the recombination wall $z_-(s)$ will give the same existence conditions with $>$ exchanged with $<$, precisely the trees that exist on one side of the recombination wall will cease to exist on the other and vise versa. It is also clear now how the index gets preserved across the recombination wall: the index of each tree $(\Gamma_i, (\Gamma_j, \Gamma_k))$  is given by\footnote{We do not write explicitly the dependence of the BPS index on the moduli, meaning that all indices are evaluated at $z=0$.}

$$ \Omega_{(\Gamma_i, (\Gamma_j, \Gamma_k))} = (-1)^{\Gamma_{ij}+\Gamma_{jk}+\Gamma_{ki}} \left|(\Gamma_{ij} - \Gamma_{ki}) \Gamma_{jk} \right|\Omega(\Gamma_i) \Omega(\Gamma_j) \Omega(\Gamma_k),$$
and if the trees $(\Gamma_i, (\Gamma_j, \Gamma_k))$ and $(\Gamma_{j}, (\Gamma_{k}, \Gamma_{i}))$ exist on the $"+"$ side of the recombination wall and $(\Gamma_{k}, (\Gamma_{i}, \Gamma_{j}))$ exists on the $"-"$ side, then from (\ref{eq:existenceoftrees}) we have

\begin{eqnarray}\label{eq:exampleconfig}
&& \Gamma_{jk}(\Gamma_{ij}-\Gamma_{ki}) >0 \nonumber\\
&& \Gamma_{ki}(\Gamma_{jk}-\Gamma_{ij}) >0 \nonumber\\
&& \Gamma_{ij}(\Gamma_{ki}-\Gamma_{jk}) <0, \nonumber\\
\end{eqnarray}
which leads to $$ \Omega_{(\Gamma_i, (\Gamma_j, \Gamma_k))}+ \Omega_{(\Gamma_j, (\Gamma_k, \Gamma_i))} =  \Omega_{(\Gamma_k, (\Gamma_i, \Gamma_j))}.$$

One thing to mention here is the fact that by looking at the marginal stability walls and recombination wall on the $z$-plane for given charges $\Gamma_1$, $\Gamma_2$, $\Gamma_3$ it is not possible to determine on  which
side of the recombination wall two trees exist and on which side only one tree exists.
Resolving this ambiguity   requires the knowledge of the actual periods.

Now let's comment on the case when not all $\Gamma_{ij,k} \ne 0$, but rather $\Gamma_{34,2} \ne 0$, $\Gamma_{24,3} \ne 0$ and $\Gamma_{23,4} =0$.
According to (\ref{eq:TCMS}), the recombination wall coincides with $MS(\Gamma_4, \Gamma_2+\Gamma_3)$. It is clear that there will be a single configuration on each side of $RW(\Gamma_2, \Gamma_3, \Gamma_4)$, $(\Gamma_2, (\Gamma_3, \Gamma_4))$ on one side and $(\Gamma_3, (\Gamma_4, \Gamma_2))$  on the other side. BPS index will again be preserved through (\ref{eq:identity}). The recombination wall in this particular case is called the Threshold Stability(TS) wall, introduced first in \cite{DdBvdB}. Physically, in $(\Gamma_2, (\Gamma_3, \Gamma_4))$ configuration charge $\Gamma_4$ is bound to $\Gamma_3$, and after crossing the recombination wall it leaves $\Gamma_3$ and binds to $\Gamma_2$. Exactly on the recombination wall we have a bound state $\Gamma_2+\Gamma_3+\Gamma_4$ at threshold, as described in \cite{DdBvdB}.

\subsection{Spin character}

Finally let us show that the spin character is invariant across the recombination wall. The spin character of the bound state of two charges $\Gamma_j$ and $\Gamma_k$ can be written as

\begin{equation}
    \Omega_{(\Gamma_j, \Gamma_k)}(y)= \sum_{m = -\frac{|\Gamma_{jk}|-1}{2}}^{\frac{|\Gamma_{jk}|-1}{2}} y^{2 m} \: \Omega(\Gamma_j, y) \Omega(\Gamma_k, y) = \frac{y^{|\Gamma_{jk}|}-y^{-|\Gamma_{jk}|}}{y-y^{-1}} \: \Omega(\Gamma_j, y) \Omega(\Gamma_k, y)  .
\end{equation}
Generalizing this to a three-centered configuration $(\Gamma_i, (\Gamma_j, \Gamma_k))$ one gets

\begin{equation}\label{eq:spinchar3center}
    \Omega_{(\Gamma_i, (\Gamma_j, \Gamma_k))}(y)= \frac{1}{(y-y^{-1})^2} (y^{|\Gamma_{jk}|}-y^{-|\Gamma_{jk}|} ) (y^{|\Gamma_{ij}-\Gamma_{ki}|}-y^{-|\Gamma_{ij}-\Gamma_{ki}|} )\: \Omega(\Gamma_i, y) \Omega(\Gamma_j, y) \Omega(\Gamma_k, y)  .
\end{equation}
The conservation of the spin character across the recombination wall is the consequence of a simple identity:

\begin{equation}
    (y^{a}-y^{-a} ) (y^{c-b}-y^{-c+b} ) +
(y^{b}-y^{-b} ) (y^{a-c}-y^{-a+c} ) +
(y^{c}-y^{-c} ) (y^{b-a}-y^{-b+a} ) \equiv 0,
\end{equation}
true for any $a,b,c$ and $y$. Consider, for example, the case (\ref{eq:exampleconfig}) again. In this case $\Gamma_{jk}$ and $\Gamma_{ij}-\Gamma_{ki}$ have the same sign and thus their absolute values in (\ref{eq:spinchar3center}) can be replaced simultaneously by their actual values. The same is true for $\Gamma_{ki}$ and $\Gamma_{jk}-\Gamma_{ij}$ entering the expression for $\Omega_{(\Gamma_j, (\Gamma_k, \Gamma_i))}(y)$, but $\Gamma_{ij}$ and $\Gamma_{ki}-\Gamma_{jk}$ have different signs and so droping the absolute value signs gives additional minus sign in the expression for $\Omega_{(\Gamma_k, (\Gamma_i, \Gamma_j))}(y)$, leading to the desired result:

\begin{equation}
    \Omega_{(\Gamma_i, (\Gamma_j, \Gamma_k))}(y)+\Omega_{(\Gamma_j, (\Gamma_k, \Gamma_i))}(y)=\Omega_{(\Gamma_j, (\Gamma_k, \Gamma_i))}(y).
\end{equation}

\subsection{Attractor Flow Conjecture revisited} \label{sec:sfconj}

It is interesting to see what happens to the moduli space of the supergravity solutions as one crosses the recombination wall. Recall that the Split Attractor Flow Conjecture (SACF) \cite{Denef:2000nb} states that the components of the moduli spaces of the multicentered BPS solutions
with constituent charges $\Gamma_i$ and background $t_{\infty}$ , are in 1-1 correspondence with the
attractor flow trees beginning at $t_{\infty}$ and terminating on attractor points for $\Gamma_i$. Thus we expect to observe that the number of components of the supergravity solution changes discontinously as we cross the recombination wall.

Let us write explicitly the stability conditions for the supergravity solution with three charges $\Gamma_2$, $\Gamma_3$, $\Gamma_4$:

\begin{eqnarray}\label{eq:integrability}
&& -1 + \frac{-\Gamma_{42}}{-\Gamma_{23}+ \theta_2 x_{23} } + \frac{\Gamma_{34}}{\Gamma_{23}+ \theta_3 x_{23} } \ge 0\nonumber\\
&& 1 - \frac{-\Gamma_{42}}{-\Gamma_{23}+ \theta_2 x_{23} } + \frac{\Gamma_{34}}{\Gamma_{23}+ \theta_3 x_{23} } \ge 0\nonumber\\
&& 1 + \frac{-\Gamma_{42}}{-\Gamma_{23}+ \theta_2 x_{23} } - \frac{\Gamma_{34}}{\Gamma_{23}+ \theta_3 x_{23} } \ge 0 .
\end{eqnarray}

Here the moduli space is 1-dimensional and we chose to parametrize it with $x_{23}$ - the distance between charges $\Gamma_2$ and $\Gamma_3$. The $\theta's$ are defined as follows:

\begin{equation}
    \theta_i = 2 \Im \left[ Z(\Gamma_i, t_{\infty}) \bar Z(\Gamma_{tot}, t_{\infty})\right].
\end{equation}

In this case the moduli space will be represented by one or two intervals in $x_{23}$. Let us suppose that the configuration $(\Gamma_2,(\Gamma_3, \Gamma_4))$ exists on  the left of the recombination wall, at some point $t_L$, and $(\Gamma_3,(\Gamma_4, \Gamma_2))$, $(\Gamma_4,(\Gamma_2, \Gamma_3))$ on the right at a point $t_R$. It is easy to see that for $t_R$ and $t_L$ sufficiently close to the recombination wall there will be only one  component of the moduli space in some open region, containing these two points.\footnote{Finding the roots of numerators and denominators of the rational functions entering (\ref{eq:integrability}) one can check than none of them become equal on the recombination wall, which means that the number of components cannot change as one crosses this wall.} This means that the SAFC as it was originally formulated {\it does not hold}!

Nevertheless it is clear that there is a relation between attractor trees and the components of the moduli space. To understand this relation let us first look at the moduli space parametrized by the absolute value squared of the angular momentum of the configuration. As discussed in \cite{deBoer:2008zn} each component of the moduli space will be an interval of the form $[J^2_d, J^2_u]$ with $J^2_{d,u}$ determined by the intersection numbers of the charges. For the example at hand there will an interval for each topology of the attractor tree:

\begin{eqnarray}
&&(\Gamma_2,(\Gamma_3, \Gamma_4)): \: I_{2,34} = [\Gamma_{42}- \Gamma_{23}-\Gamma_{34},\Gamma_{42}- \Gamma_{23}+\Gamma_{34} ] \nonumber\\
&&(\Gamma_3,(\Gamma_4, \Gamma_2)): \: I_{3,42} =[\Gamma_{42}- \Gamma_{23}+\Gamma_{34},\Gamma_{42}+ \Gamma_{23}-\Gamma_{34} ] \nonumber\\
&&(\Gamma_4,(\Gamma_2, \Gamma_3)): \: I_{4,23} =[\Gamma_{42}+ \Gamma_{23}-\Gamma_{34},\Gamma_{42}- \Gamma_{23}-\Gamma_{34} ] .
\end{eqnarray}
We see that $I_{2,34} = I_{3,42} \amalg I_{4,23}$ and the moduli space always consist of only one interval $I_{2,34}$, which becomes partitioned into two on the right of the recombination wall. This leads us to a modified version of the SACF as follows: {\it The classical BPS configuration space and the quantum BPS Hilbert space are partitioned by attractor flow trees.} The partitioning in the classical case is defined as follows: start with some value of the background moduli, then adiabatically deform it by dialing the moduli at infinity along the attractor flow for the total charge. If there are several configurations of attractor flow trees then upon crossing MS walls they will decay and the corresponding components of the moduli space will disappear. As we saw above different components do not have to be disjoint, but the point is that the change of the moduli space will be discontinuous which allows to identify the part of the moduli space with the attractor flow tree.  The quantum case is analogous, although we now have to allow for evolution into linear superpositions of different decay outcomes if there are multiple trees. The Hilbert space of BPS states will be partitioned in states which have only nonzero amplitudes to decay  adiabatically into the constituents of each corresponding attractor tree.

\section{Massless Vectormultiplets }\label{sec:Massless-VM}

In this section we discuss a   class of examples  where  the spectrum at the singularity contains massless vector multiplets.  The BPS index of a vectormultiplet has $\Omega  = -2$ and therefore (\ref{eq:Math-Pred2}) might very well be an infinite series and not a finite polynomial.   In various important developments in string theory, such as geometrical engineering of gauge theories and heterotic/type II duality, these kinds of singularities played a key role. Some of  the models with massless vectormultiplets which have appeared in this
  literature appear to conform
to our basic assumptions in Section \ref{subsec:GroundRules} and thus threaten to pose counterexamples
to our prediction \eqref{eq:Math-Pred2}. In this Section and the next we examine these examples
and demonstrate that in fact there are no counterexamples to our prediction.

We divide the zoo of examples into three groups:
%
%
%
%

\begin{itemize}
    \item singularities with the spectrum of an asymptotically free gauge theory,
    \item conformal fixed points of gauge theories with vanishing $\beta$-function,
    \item Theories with electric spectrum (with respect to some
    duality frame) being that of IR free gauge theory.
\end{itemize}


\subsection{Asymptotically free gauge theories}

Consider the simplest example of $SU(2)$ $N_f=0$. This can emerge  at a singularity in type II  string compactification,
 and a simple example was discussed in \cite{Aspinwall:1995mh}. The CY in this case is a K3 fibration over $\IP^1$, which develops an $A_1$ singularity. Classically at the singularity there is a massless vector multiplet with index $\Omega(\gamma) = -2$ and this clearly contradicts our conclusions from Section \ref{subsec:BPS-Indices}. In particular, the quantity $I$ which was associated with the beta-function there is negative
 and the product $P(q)$ of equation \eqref{eq:Math-Pred2} is
 \begin{equation}
 P(q) = \frac{1}{(1-q)^2 }
 \end{equation}
 and is certainly not a polynomial!

 However the full quantum moduli space does not have a singularity with massless vector multiplets at finite distance in the moduli space (\cite{Kachru:1995wm}, \cite{Kachru:1995fv}, \cite{Katz:1996fh}). Indeed one recovers the full moduli space of the $SU(2)$ gauge theory, including the strong coupling region, in a certain double scaling limit on IIA side. Part of the IIA string moduli space can be parameterized by the K\"ahler moduli $t_b$ and $t_f$ of the base of K3 fibration and it's fiber respectively. There will be two discriminant loci, corresponding to the monopole and the dyon.  These   intersect on the boundary of the moduli space $t_b=0$. It is on this codimension 2 intersection where one expects to have massless vector multiplets. This boundary is a nongeneric point on the discriminant
 locus and so this example does not meet the requirements of Section \ref{subsec:GroundRules}. In fact if we try to consider bound states of some massive black hole $\Gamma$ with the $W$-boson $\gamma$ near the codimension 2 locus where the $W$-boson is massless, we find that the attractor flow of $\Gamma+\gamma$ will never pass through this locus, because the attractor flow will always have the direction away from the boundary divisor $t_b=0$. Thus for any $t_{\infty}$ the flow of $\Gamma+\gamma$ will intersect $MS(\Gamma, \gamma)$ at some point with $t_b \ne 0$ and the $W$-boson will be realized as a bound state of well-separated monopole and dyon. As we move $t_\infty$ in the $t_f$ plane around the origin our paradox is resolved through the recombination process of the 3-centered bound state $\Gamma+(\gamma_{monopole}+\gamma_{dyon})$.

It is natural to assume that this  conclusion extends to all cases where one engineers an asymptotically free gauge theory: there will be no places in the moduli space at finite distance where on a codimension 1 locus a vector multiplet becomes massless. The singularity will always "split" and there will be a number of conifold-like singularities, around each of which the picture is as described in Section \ref{subsec:BPS-Indices}.

%
%

\subsection{Conformal fixed points}

These theories are conformal fixed points of gauge theories with vanishing beta-function $I=0$. It is known that in such theories the spectrum necessarily contains mutually non-local populated charges that becomes massless at the conformal point. This violates our assumptions that there is only one charge $\gamma$ (and possibly some other parallel charges), that is massless and populated at the singularity. Although there is no notion of particles in such theories, away from the superconformal point the theory does contain particles. One can form halo bound states of these light particles with some massive black hole near the superconformal point and talk about wall-crossing phenomenon. We examine an example of this situation in Section \ref{sec:FHSV}.

\subsection{Electrically IR-free gauge theories}\label{subsec:IRfree}

This class of theories at first sight seems to conform to our assumptions from Section \ref{subsec:GroundRules}. The light spectrum near the singularity is that of an IR free gauge theory. Examples include the  model \cite{Katz:1996ht} and  the model based on a chain of heterotic/IIA duals first discussed in  \cite{Aldazabal:1995yw} and further analyzed in \cite{Berglund:1996uy}. In Section \ref{subsec:ExtrTran} we consider an example where the electrically charged spectrum gives a non-polynomial expression for (\ref{eq:Math-Pred2}).

We will now argue that although the spectrum near the singularity is that of an IR free gauge theory, at the singularity itself there is a violation of the central assumption of \ref{subsec:GroundRules}
that the only massless BPS particles have charge parallel to $\gamma$. If we parametrize the plane transverse to the singularity by the expectation value of the adjoint scalar $v$ from the light vector multiplet then the $W$-boson will have mass $\sim |v|$. On the other hand, the theory will also have BPS monopoles that can be reliably constructed as large, smooth, classical solutions to the YM field equations. The mass of the monopole is, as usual, proportional to the vacuum expectation value of scalars from the vector multiplet $\frac{|v|}{g^2(v)}$, where $g(v)$ is the coupling at the scale $v$. At energies smaller than the monopole mass the dependence of the running coupling on the scale $v$ is given by $\frac{1}{g(v)^2} = \beta \log(\mu/v)$. As the
energy scale set by $v$ goes to zero   the relation between the mass of the monopole and $W$-boson does not get spoiled by the quantum corrections, and in the IR limit these masses are still proportional. Taking $v \to 0$ and keeping $\mu$ to be some fixed string scale, both masses of $W$-boson and monopole go to zero. Thus we have mutually non-local massless states at $v=0$, violating a key assumption of \ref{subsec:GroundRules}.

We should remark that some care is required when interpreting the above massless monopole. The ratio of
monopole mass to W-boson mass goes to to infinity as $v \to 0$ so with an appropriate cutoff the IR
free theory is indeed a free theory. On the other hand,  one generally expects when one approaches a
 locus with mutually nonlocal massless particles the theory should become
a nontrivially interacting conformal field theory. We believe that there are some important order of
limits questions here. In particular, the monopole also becomes larger and more diffuse, since
  its typical length scale is set by $v^{-1}$. For purposes of BPS statecounting and the computation
  of $\Omega$ one should include this particle. For purposes of the computation of loop diagrams
  one should exclude it.

%

We will examine two examples of this type. In   Section \ref{subsec:KMP} we consider a model whose electric spectrum is that of an IR free gauge theory which has
 massless vectors provided the hypermultiplet moduli are tuned appropriately. In this case, there are cancelations between vectors and adjoint hypers so that (\ref{eq:Math-Pred2}) is still polynomial. Moreover, the massless monopoles have vanishing index, and thus do not affect the partition function of the index of BPS states.
In Section \ref{subsec:ExtrTran} we consider another famous example of this kind. As a check of our conclusion about the spectrum we show that there are two dual CY periods, vanishing at the singularity. The above
 reasoning implies they are both populated. Note that our picture of the massless spectrum at the singularity is very different from the one advocated in \cite{Berglund:1996uy}.

\section{Examples with massless vectors}\label{sec:MasslessVM-examples}

\subsection{The FHSV Model}\label{sec:FHSV}

In this section we describe a particular example of a model where
one has   massless vector multiplets at certain places on the
moduli space and the theory at the singularity is superconformal. The model is
referred to as the FHSV model \cite{Ferrara:1995yx}. Due to the high amount of symmetry the moduli space of this theory is known exactly. From the $S$-duality symmetry we also can make a good guess
 about the   massless spectrum at the singularity. This spectrum contains both electrically and magnetically charged states. The purpose of this subsection is to illustrate the  very nontrivial wall-crossing phenomenon around
 such a singularity.

\subsubsection{Basic Setup for the model}

We recall that the FHSV model is  an example of a type II compactification with  a heterotic dual
which is in fact simply an asymmetric orbifold of the heterotic string on $T^6$. Both the
vectormultiplet and hypermultiplet moduli spaces are known exactly. The vectormutiplet moduli
space has universal cover
\begin{equation}\label{eq:FHSV-vm}
   \widetilde{ \CM_V }= \frac{SU(1,1)}{U(1)} \times\frac{SO(10,2)}{SO(10) \times SO(2)}.
\end{equation}

It is convenient to choose coordinates on \ref{eq:FHSV-vm}
(we follow conventions of \cite{Harvey:1995fq}). Let $\IC^{1,r}$
 denote the  $r+1$ dimensional complex vector space equipped with a Lorentzian bilinear form
 of signature $(+,-^{r})$.
 Let  $\CH^{1,r}$ be  the subspace of with
positive definite imaginary part. Then our coordinates are
 $(\tau, \vec y)$, where $\tau \in
\CH^{1,1}$ and $\vec y$ is a ``tube domain'' coordinate of $\CH^{1,9}$.  We also introduce $u:=(\vec y, 1,
-\frac{\vec y^2}{2}) \in \IC^{2,10}$. The lattice of electric charges
is $II^{2,10} = II^{1,9}\oplus II^{1,1}$ with a quadratic form:

\begin{equation}
    ( v,v ) = \vec v^2+ 2 v_+ v_- \quad {\rm where} \: \: v:=(\vec v,v_+, v_-) \in II^{2,10}.
\end{equation}
Elements of the full electromagnetic lattice $\Lambda = II^{2,10} \oplus II^{2,10}$ will have the form $(q,p)$,
with $q,p \in II^{2,10}$. Using the quadratic form on $II^{2,10}$ we construct the symplectic form on
$\Lambda$
\begin{equation}
\langle \Gamma, \Gamma' \rangle = (q,p') - (p,q')
\end{equation}
where $\Gamma = (q,p)$ and $\Gamma' = (q',p')$
The holomorphic central charge  can be written as:

\begin{equation}
    Z_h(\Gamma) =  (\vec q \cdot \vec y)- q_+\frac{\vec y^2}{2} + q_- + \tau \left( (\vec p \cdot \vec y)- p_+\frac{\vec y^2}{2} + p_- \right).
\end{equation}

The model has two kinds of singularities on the moduli space with enhanced gauge symmetry. The first kind is given by

\begin{equation}
    \vec \alpha \cdot \vec y  = 0, \quad {\rm with} \: \: \vec \alpha^2 = -2,
\end{equation}
and corresponds to $\CN=4$ $SU(2)$ SYM gauge theory in the infrared. The second kind is the locus
\begin{equation}
    \left( (\vec 0, -1,1), (\vec y, 1, -\frac{\vec y^2}{2}) \right) = 1+ \frac{\vec y^2}{2} = 0,
\end{equation}
and corresponds to $\CN=2$ $N_f=4$ $SU(2)$ gauge theory.
We will describe them on the same footing as loci where
\begin{equation}
    ( \alpha, u ) = 0, \quad {\rm where} \: \: ( \alpha,\alpha ) = -2.
\end{equation}
for $\alpha \in II^{2,10}$.

The superconformal theory at the two singularities has $S$-duality symmetry but the
 spectrum  of  massless states at the singularity and the structure of halo states will be
 different in the two cases. In both cases there  will be  two BPS states with mutually non-local charges $\gamma = (\alpha,0)$ and $\gamma_D = (0,\alpha)$, $\langle \gamma, \gamma_D\rangle = -2$, that become massless. The spectrum in the case of the $SU(2)$, $\CN=4$ singularity is given by
\begin{align}\label{eq:spectrN4}
    \CH(m \gamma +n \gamma_D) \ne \emptyset,  \: \Omega(m \gamma +n \gamma_D)= 0 \: &&{\rm if } \: \quad gcd(m,n)=1,\nonumber\\
    \CH(m \gamma +n \gamma_D) = \emptyset, \: &&{\rm otherwise}.
\end{align}
In case of the $SU(2)$, $N_f=4$ singularity the spectrum is
\begin{align}\label{eq:spectrN2}
    &\CH(m \gamma +n \gamma_D) \ne \emptyset, \quad \Omega(m \gamma +n \gamma_D)= 8, && gcd(m,n)=1,  \nonumber\\
    &\CH(m \gamma +n \gamma_D) \ne \emptyset, \quad \Omega(m \gamma +n \gamma_D)= -2, && gcd(m,n)=2, \nonumber\\
    &\CH(m \gamma +n \gamma_D) = \emptyset, &&{\rm otherwise}.
\end{align}

We are interested in the behavior of BPS indices and the structure of Hilbert spaces of charges
\begin{equation}
\Gamma_{m,n}:= \Gamma+ m \gamma+ n \gamma_D
 \end{equation}
 around the superconformal point. Here $\Gamma= (q,p)$ is some charge, mutually non-local to $\gamma$ and $\gamma_D$, such that   $\Omega(\Gamma) \ne 0$ and constant in the neighborhood of $\CZ(\gamma)$. (Let us say it supports a
 heavy single-centered black hole with a regular attractor point near $\CZ(\gamma)$.)  It is convenient to choose basis so that $\langle \Gamma, \gamma \rangle =0$ and  $a:=\frac{1}{2}\langle \Gamma, \gamma_D \rangle <0$. In particular we are taking:
 \begin{equation}\label{eq:def-of-a}
 (p,\alpha) =0  \qquad\qquad (q, \alpha) = 2a < 0 .
 \end{equation}

\begin{figure}[htp]
\centering
\includegraphics[scale=0.4,angle=90,trim=0 0 0 0]{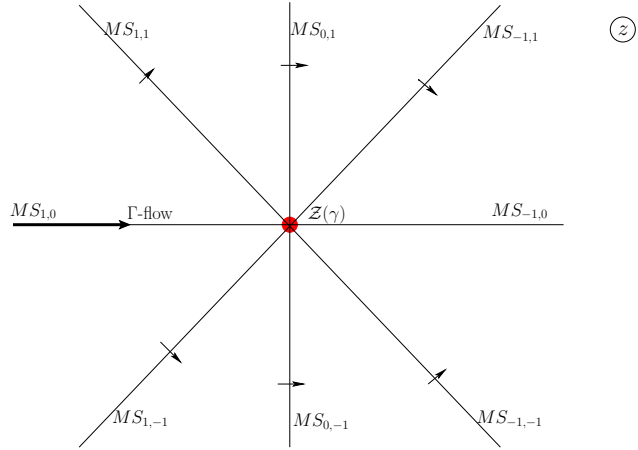}
\caption{Marginal stability walls $MS_{m,n}$ for charge $\Gamma$ with charges $\gamma_{m,n} = m \gamma+ n \gamma_D$. The walls form a dense set, labeled by the rational numbers in lowest terms. The arrows point in the
direction of stable to unstable region appropriate to that wall of marginal stability.}
\label{fig:FHSVmswalls}
\end{figure}

\subsubsection{Attractor flows and walls of marginal stability}

As another preliminary we describe relevant walls of marginal stability
and the attractor flows.
It is convenient to parameterize the plane transverse
to $\CZ(\gamma)$ by $z:=( \alpha, u )$, the period of $\gamma$, and
to project the walls of marginal stability and the attractor flows into this plane.

First, let us plot the walls of marginal stability. Let $\gamma_{m,n}:= m \gamma + n \gamma_D$.
There will be a dense set of walls of marginal stability $MS_{m,n}:=MS(\Gamma, \gamma_{m,n})$
 for all $(m,n) \in \IZ^2$, $gcd(m,n)=1$. For the    $N_f=4$ case there will also be
 such walls for $gcd(m,n)=2$. As usual these walls will sit in the locus:
\begin{equation}
    \Im \: Z(\Gamma; \tau,y) (m \bar z + n \bar \tau \bar z) =0.
\end{equation}
The marginal stability
 walls will end at $z=0$. We work in an arbitrarily small neighborhood of $z=0$ and hence can treat
$Z(\Gamma;t)$ as a constant. We will  normalize $Z(\Gamma,t)$ to be  $Z(\Gamma,t) =-1$ at $z=0$
 and moreover we will take
$\tau=i$ for simplicity. In the linear approximation the walls $MS_{m,n}$ will be
\begin{equation}
MS_{m,n} = \{ \rho  (-m + i n) : \rho >0 \}.
\end{equation}
%
%
The marginal stability walls are in the
 upper half-plane for $n>0$ and for $m \to + \infty$ they asymptote to the negative $x$ axis,
 while for $m \to - \infty$ they asymptote to the positive $x$ axis. Some of the walls are illustrated in
Figure \ref{fig:FHSVmswalls}. Again, because we are working at small $z$ we can in fact identify
\begin{equation}
MS(\Gamma_{m,n}, \gamma_{m',n'}) \cong MS(\Gamma, \gamma_{m',n'}),
\end{equation}
an approximation which will be used throughout.

The stable side of the walls is readily computed from
\begin{equation}
\langle \Gamma, \gamma_{m,n} \rangle  \Im \: Z_\Gamma \bar Z_{\gamma_{m,n}} = 2na(nx+my)
\end{equation}
where $z = x+ i y$. Note that for $x \to +\infty$  the dominant
term in the expression, $2a n^2 x$,  is always negative,
hence the unstable side is always on the right in Figure
 \ref{fig:FHSVmswalls}, as indicated by the arrows.

Next, let us turn to the attractor flows. In the small $z$ approximation
 the attractor flows for $\Gamma_{m,n}$ can be written for both cases in a uniform way:
\begin{equation}\label{eq:bstwallzdot}
    \dot z \sim   - (\alpha, q) + 2 m + \bar \tau  \left(  - (\alpha, p) + 2 n \right),
\end{equation}
which, for our choice of parameters $\tau=i$ and \eqref{eq:def-of-a}
 is simply
\begin{equation}\label{eq:easy-attractor-G}
\Gamma_{m,n}-{\rm flow}: \qquad  \dot z = (m-a) - i n ,
\end{equation}
and similarly we have
\begin{equation}\label{eq:easy-attractor-g}
\gamma_{m,n}-{\rm flow}: \qquad  \dot z = m - i n .
\end{equation}
Here we have neglected the variation in $\tau$, again using the small $z$ approximation.
(We also rescaled time by a factor of $2$.)

In order to prove \eqref{eq:easy-attractor-G} and  \eqref{eq:easy-attractor-g}
let us start with the  $\CN=4$ singularity with
$z = \vec \alpha\cdot \vec y$.  Writing the attractor equation as in
(\ref{eq:attrgradient}), we get
\begin{equation}
    \dot z = \alpha_a \dot y^a \sim - \alpha_a g^{a \bar b} \partial_{\bar y^{\bar b}} \left(  |Z_h(\Gamma_{m,n};\tau,y)|e^{K/2}\right)|_{z=0}.
\end{equation}
Taking into account that the K\"ahler potential is given by $K = - \log \left( 4 (\Im \vec y)^2\right) - \log \left( \Im \tau \right)$ gives
\begin{equation}
    \dot z \sim   - \vec \alpha\cdot \vec q + 2 m + \bar \tau  \left(  - \vec \alpha\cdot \vec p  + 2 n \right).
\end{equation}
Repeating the same calculation for the $\CN=2$ singularity we get:
\begin{eqnarray}
&&  \dot z = \left( 1+ \frac{\vec y^2}{2}\right)^{.} = y_a \dot y^a \sim - y_a g^{a \bar b} \partial_{\bar y^{\bar b}} \left(  |Z_h(\Gamma_{n,m};\tau,y)|e^{K/2}\right)|_{z=0}, \nonumber\\
&& \dot z \sim   - (q_+ - q_-) + 2 m + \bar \tau  \left(  - (p_+ - p_-) + 2 n \right).
\end{eqnarray}
thus establishing \eqref{eq:easy-attractor-G}, from which one  can also deduce \eqref{eq:easy-attractor-g}.

We remark that

\begin{enumerate}

\item The attractor flows for $\Gamma_{m,n}$ are parallel to the marginal stability
walls $MS_{(m-a), n}$. In particular, the flows for $\Gamma$ itself are parallel to
the $x$-axis in the direction of increasing $x$.

\item In particular, the bound state transformation
 wall $\CS(\gamma_{m,n}, \Gamma)$ is precisely the marginal stability
wall $MS_{m-a, n}$. We will see that it is really a hybrid of
conjugation and recombination walls in this example.

\item The attractor flow for $\gamma_{m,n}$ is parallel to
the walls   of marginal stability $MS_{m,n}$

\item Attractor flows always proceed from stable to unstable regions,
in accord with Property 3 of \ref{app:properties}.

\end{enumerate}

\subsubsection{Monodromy}\label{subsec:FHSV-monodromy}

It will be important in our story below to take into account the $\IZ_2$-monodromy
of the local system of charges around $z=0$.  The $z$-plane is simply a double-cover
of the moduli space under $z \to - z$. The action on a general charge $\lambda \in \Lambda= II^{2,10} \oplus II^{2,10}$ is
\begin{equation}\label{eq:Z2mon}
    M\cdot \lambda = \lambda - \langle  \lambda , \gamma\rangle \gamma_D + \langle \lambda , \gamma_D\rangle \gamma.
\end{equation}
This takes $\gamma \to - \gamma$, $\gamma_D \to - \gamma_D$, and is the identity on charges
orthogonal to both $\gamma,\gamma_D$. Thus, if we write
 $ \Gamma = \Gamma_0 - a \gamma$ where $\Gamma_0$ is orthogonal to $\gamma $ and $\gamma_D$
 then the  monodromy image is $\Gamma_M := M\cdot\Gamma =  \Gamma + 2 a \gamma=\Gamma_0 + a \gamma $.

Since there are no basins of attraction in the FHSV  model, both charges $\Gamma$ and $\Gamma_M$  will be populated charges in the neighborhood of the singularity and will have isomorphic Hilbert spaces.

\subsubsection{Attractor flow trees}

Now let us turn our attention to the attractor flow trees for $\Gamma_{m,n}$.

We are interested in attractor flow trees relevant to considering $\Gamma$ as a core charge,
that is, trees of the form:
\begin{equation}\label{eq:Flow-Trees}
    \Gamma_{m,n} \to (\gamma_{m_{1},n_{1}}+(\gamma_{m_{2},n_{2}}+...(\gamma_{m_{L-1},n_{L-1}}+(\Gamma+ \gamma_{m_L,n_L}))...)).
\end{equation}
Of course, charge conservation requires
\begin{equation}
   \sum_i m_i=m \qquad \qquad  \sum_i n_i=n.
\end{equation}

The attractor flow trees are systematically constructed from two principles:
First the flow must split from the stable to the unstable side on the wall of
marginal stability and second the charges must be conserved at each vertex of
the tree.

In Appendix \ref{appendix:FHSV} we give an algorithm for enumerating the trees and
show, in particular, that the number of such trees is finite. If $n>0$ then the
initial point of the tree can only be in the upper half plane. Moreover,
the initial point must be to the right of the BST wall $\CS(\gamma_{m,n},\Gamma)$,
otherwise there are no acceptable trees. The reason for this is that simple
geometry forces the  trees that
begin on the left of the BST wall  $\CS(\gamma_{m,n},\Gamma)$ to intersect
marginal stability walls $MS_{m',n'}$ in a direction from unstable to stable side.
But this is a
forbidden vertex. This would appear to pose a serious problem for continuity of the
index. We discuss that point in the next subsection.

\begin{figure}[htp]
\centering
\includegraphics[scale=0.7,angle=90,trim=0 0 0 0]{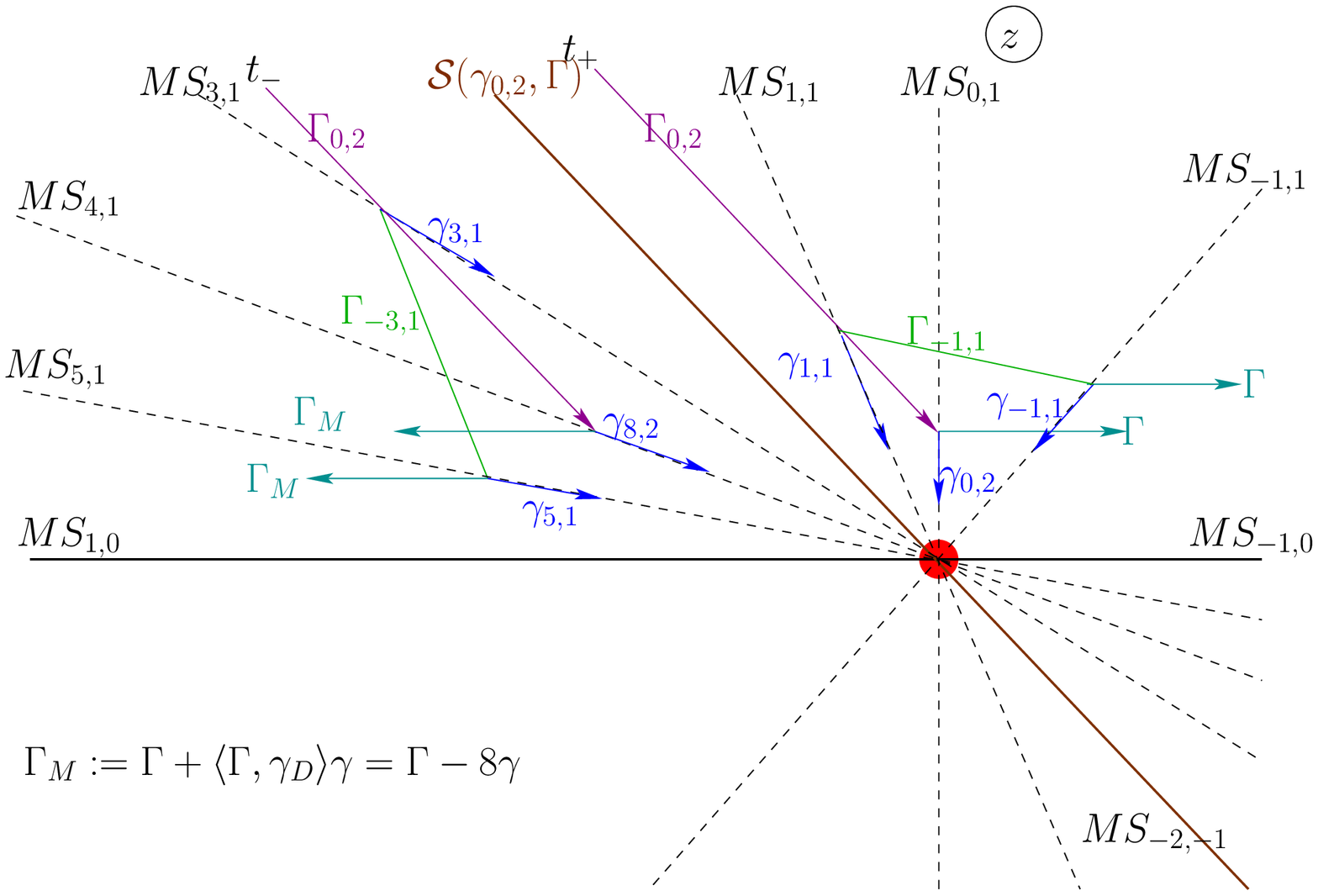}
\caption{Attractor trees, contributing to the realization of charge $\Gamma_{0,2}$ on two sides of $\CS(\gamma_{0,2}, \Gamma)$. We have chosen a=-4. Attractor flows of core with halos are purple, green and cyan colored line, attractor flows of halo particles are blue lines.}
\label{fig:FHSVhalos}
\end{figure}

As an example, consider the attractor trees contributing to the realization of charge $\Gamma_{0,2}$.
These are shown in  Figure \ref{fig:FHSVhalos}.
Starting at $t_+$ there will be only two valid trees. One is a two centered solution $\Gamma+ \gamma_{0,2}$
and the other is   a 3-centered solution $(\Gamma+\gamma_{1,1})+\gamma_{-1,1}$. If, on the other
hand, we move the inital point $t_\infty$ from $t_+$ to $t_-$ across the BST wall then these two attractor flow
trees cease to exist!

\subsubsection{Continuity of the index}

In the previous subsection we remarked that attractor flows of type \eqref{eq:Flow-Trees}
with final core charge $\Gamma$ do not exist for initial point $t_\infty$ to the left
of the BST wall $\CS(\gamma_{m,n},\Gamma)$. As we noted, this would appear to pose a
serious problem for the index.
The only way the index can be continuous across $\CS(\gamma_{m,n},\Gamma)$ is if there exists another core charge that can form halos with $\gamma_{m',n'}$-particles of the same total charge $\Gamma_{m,n}$.
Because of the $\IZ_2$-monodromy there is indeed another natural core charge, namely
$\Gamma_M = \Gamma +2a\gamma = \Gamma_{2a,0}$. The attractor flows for $\Gamma_M$ are parallel
to the $x$-axis but the flow is to the left. Similarly, the stable and unstable sides of
all the walls of marginal stability are flipped. (All this becomes more obvious
if we write $\Gamma= \Gamma_0 - a \gamma$ and $\Gamma_M = \Gamma_0 + a \gamma$
as in Section \ref{subsec:FHSV-monodromy}.)
One can write out conditions similar to those in Appendix \ref{appendix:FHSV}
for enumerating the attractor flow trees corresponding to core charge $\Gamma_M$. Again there
will be finitely many such trees. In particular, the initial point for a flow tree
with terminating with core charge $\Gamma_M$ must lie to the \emph{left} of the
BST wall
\begin{equation}
\CS(\gamma_{m,n},\Gamma)=
\CS(\gamma_{m-2a,n},\Gamma_M)
\end{equation}

Continuity of the index leads us to expect, and hence we conjecture, the following:
\emph{The sum of contributions to  the index from flow trees
terminating on core $\Gamma$ with initial point $t_+$ infinitesimally to the right of $\CS(\gamma_{m,n},\Gamma)$
is equal to the sum of contributions to the index from flow trees terminating on the
core charge $\Gamma_M$ with inital point $t_-$ infinitesimally to the left of $\CS(\gamma_{m,n},\Gamma)$ .}

As a simple check on this idea  consider charges of type $\Gamma_{m,1}$. These only support a single
branch. At a point $t_+$ just to the right of the wall $\CS(\gamma_{m,1},\Gamma)$ there is
only a single tree $\Gamma_{m,1} \to  \Gamma + \gamma_{m,1}$. This contributes to the index
\begin{equation}
\Delta \Omega(\Gamma_{m,1} \to  \Gamma + \gamma_{m,1}) = (-1)^{\langle \Gamma, \gamma_{m,1}\rangle-1}
\vert \langle \Gamma, \gamma_{m,1}\rangle\vert \Omega(\gamma_{m,1}) \Omega(\Gamma) = - 16 a \Omega(\Gamma).
\end{equation}
At a point $t_-$ just to the left of the wall this tree does not exist, but the tree
$\Gamma_{m,1} \to \Gamma_{M} + \gamma_{m-2a,1}$ does exist. The latter contributes
\begin{equation}
\Delta \Omega(\Gamma_{m,1} \to  \Gamma_{M} + \gamma_{m-2a,1}) = (-1)^{\langle \Gamma_M, \gamma_{m-2a,1}\rangle-1}
\vert \langle \Gamma_M, \gamma_{m-2a,1}\rangle\vert \Omega(\gamma_{m-2a,1}) \Omega(\Gamma_M) = - 16 a \Omega(\Gamma_M).
\end{equation}
Now, thanks to the $\IZ_2$ monodromy $\Omega(\Gamma) =  \Omega(\Gamma_M)$ and so indeed the contributions
to the index are continuous.

As a second check  we return to the   example of the previous section. For $t_+$ on
the right of $\CS(\Gamma_{0,2})$ with $a=-4$ the  tree with one branching contributes
\begin{equation}
\Delta \Omega = - 16 \Omega(\gamma_{0,2}) \Omega(\Gamma) =  32 \Omega(\Gamma)
\end{equation}
and the second tree contributes
\begin{equation}
\Delta \Omega = 32 \Omega(\gamma_{1,1}) \Omega(\gamma_{-1,1}) \Omega(\Gamma) =  2^{11} \Omega(\Gamma)
\end{equation}
On the other hand, going  back to Figure \ref{fig:FHSVhalos}, starting from $t_-$ there will be two solutions with the core charge $\Gamma_M$ of the form $\Gamma_M+ \gamma_{8,2}$ and $(\Gamma_M+ \gamma_{5,1})+\gamma_{3,2}$.
The first tree contributes $32 \Omega(\Gamma_M)$ and the second $2^{11} \Omega(\Gamma_M)$.

For charges of type $\Gamma_{m,2}$ we must take into account trees with one and two branches and
the computation becomes more elaborate. We have performed this check and the index is continuous.
The computation is very similar to that in given in the next subsection.
 In general, upon crossing the BST wall we have conjugation - since the core charge is
 replaced by a monodromy image - at the same time as  recombination - so the walls in this
 example exhibit a  hybrid  of the  conjugation and recombination mechanisms.

\subsubsection{Wall-Crossing near a superconformal point}

\begin{figure}[htp]
\centering
\includegraphics[scale=0.4,angle=90,trim=0 0 0 0]{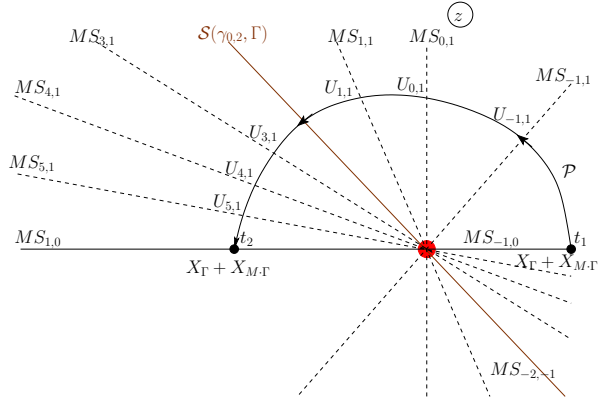}
\caption{Transformation of the partition function through operators $U_{m,n}$ along a path $\CP$ from $t_1$ to $t_2$. }
\label{fig:FHSVpartfunc}
\end{figure}

In the companion paper \cite{Andriyash:2010qv} we give a simple proof of the Kontsevich-Soibelman
wall-crossing formula based on supergravity halos. Moreover, as explained in Section 4 of
\cite{Andriyash:2010qv}, the line of reasoning adopted there suggests a generalization of
the KSWCF. The FHSV model provides a nice example in which to illustrate the ideas.

Following  \cite{Andriyash:2010qv} we consider the partition function

\begin{equation}
	F(q, p;t) = \sum_{m,n} q^{m-a} p^n \: \fro_{\Gamma}(\gamma_{m,n};t),
\end{equation}
where $\fro_{\Gamma}(\gamma_{m,n};t)$ are the ``framed" BPS indices described in \cite{Andriyash:2010qv}.
The sum here runs over $n \geq 0$ and $m\in \IZ$. For our purposes these framed BPS indices can be
identified with the halo contributions to the total index $\Omega(\Gamma_{m,n};t)$ with $\Omega(\Gamma)$
factored out. (This is also what we were considering in Section \ref{subsec:BPS-Indices}.)

Let us  consider the path $\CP$ shown in
Figure  \ref{fig:FHSVpartfunc} going from $t_1$ to $t_2$. Using the reasoning of \cite{Andriyash:2010qv} we see that the partition function at a point $z=x+ i y$ on $\CP$ is given in terms of that at $z=t_1$ by
\begin{equation}\label{eq:partfuncT}
	F(q, p;z) = \prod_{my + nx <0 \, \& \, n > 0 } U_{m,n} F(q, p;t_1).
\end{equation}
Here the operators $U_{m,n}$ are defined in terms of basic
KS-transformations:
\begin{equation}
T_{m,n}   := (1- q^m p^n)^{D_{m,n} }
\end{equation}
where $D_{m,n} $ is a differential operator
defined by $D_{m,n} q^\alpha p^\beta : = 2(n\alpha - m \beta) q^\alpha p^\beta  $
and
\begin{equation}
U_{m,n} = T_{m,n}^8 T_{2m,2n}^{-2}.
\end{equation}
The $U_{m,n}$ are only defined for $gcd(m,n)=1$ and this restriction
is understood on the product \eqref{eq:partfuncT} and similar products below.
 The restriction $my+nx<0$ on the terms in
the product
applies because only the walls $MS_{m,n}$  (defined by $my+nx=0$) that have been crossed
while moving along $\CP$ should be included. Finally, the factors in the product are ordered
so that terms with  increasing argument $\arg(-m+in)$ are placed to the left.

If we identify the framed BPS index with the index of
states which can be described as halo states around a core of
charge $\Gamma$ then at $z=t_1$ there are no halo states and hence the core
 simply contributes a factor of $q^{-a}$. We have already seen in the halo description
 that such states do not give a continuous index across the BST walls and we should
 expect trouble here too if we only include $q^{-a}$ in $F(q,p;t_1)$. Indeed, the examples
 below will bear that out. Thus we should include the monodromy image $\Gamma_M$
 and its halo states. Again, at $z=t_1$ the only halos around $\Gamma_M$ are the single
 core state itself. Recalling that $\Gamma_M = \Gamma + 2a \gamma$ we see that
 including these two cores gives
\begin{equation}\label{eq:F-at-t1}
F(q,p;t_1)=q^a + q^{-a}.
\end{equation}
Substituting this into \eqref{eq:partfuncT} and expanding as a series in $p, q^{\pm 1}$ we
observe that the number of terms in the expansion of the product
contributing to a given monomial $q^{m-a} p^n$ is finite.
Indeed, we can observe that we need to choose a partition of $n$ to account for the
power of $p$. The power of $q$ is more complicated. The walls crossed in the
first quadrant all have $m<0$, but for $x<0$ there will be a finite number of
walls with $0\leq m< -nx/y$. Once these nonnegative values of $m$ have been chosen, the
remaining negative values of $m$ constitute a partition, and therefore there are
only finitely many choices. Thus the infinite product will be well-defined.

As an example of the issues involved let us examine the product
\begin{equation}
\prod_{my + nx <0 \, \& \, n > 0 }  U_{m,n}  (q^a + q^{-a})
\end{equation}
and let us extract the coefficients of $q^{m-a} p $ and $q^{m-a}p^2$.

We first consider the case of $q^{m-a} p $.
There are only two terms which can contribute to $q^{m-a} p$. First, there is
the term coming from the expansion of
\begin{equation}
(1-q^{m}p)^{8 D_{m,1} } q^{-a} = (1-q^m p)^{-16a} q^{-a}
\end{equation}
which enters the product when  $my+x <0$, i.e. when $m < -x/y$  and contributes   $16a$
to the coefficient. The other term which can contribute comes from the expansion of
\begin{equation}
(1-q^{m-2a}p)^{8 D_{m-2a,1} } q^{a}= (1-q^{m-2a} p)^{16a} q^{-a}
\end{equation}
This contributes $-16a$ to the coefficient and enters the product when $m-2a < -x/y$.
Thus the coefficient of $q^{m-a} p$  (that is, the framed BPS degeneracy) is given by
\begin{center}
\renewcommand{\arraystretch}{1.3}
\begin{tabular}{|c|c|c|c|c|c|c|}
\hline
  $    -\frac{x}{y} < m $ & $0$
\\
\hline    $  m  <  -\frac{x}{y}< m-2a $ & $16a$
\\
\hline    $ m-2a <  -\frac{x}{y}  $ & $0$
\\
\hline
\end{tabular}
\end{center}
Now, a short computation shows
 that the BST wall $\CS(\gamma_{m,1}, \Gamma)$ is given by $-x/y = m-a$, and hence
 the index is continues across it.

A slightly more elaborate computation is required to compute the coefficient
of $q^{m-a}p^2$. The power of $p^2$ can come from a single factor, or from
two distinct factors. We have listed the cases in the table below
 together with the contribution of that factor and the range in
which it applies: \footnote{We have taken $m$ to be even for simplicity. A slightly
different computation applies when $m$ is odd.}

\begin{center}
\renewcommand{\arraystretch}{1.3}
\begin{tabular}{|c||c|c|c|c|c|c|}
\hline $I$     & $ (1-q^m p^2)^{-2 D_{m,2}} q^{-a} $ & $ -8 a q^{m-a} p^2    $ & $ \frac{m}{2} <  -\frac{x}{y} $
\\
\hline  $II$ & $ (1-q^{m-2a} p^2)^{-2 D_{m-2a,2}} q^{a} $ & $ 8 a q^{m-a} p^2   $ & $ \frac{m}{2}-a <  -\frac{x}{y} $
\\
\hline $III$ & $ (1- q^{\frac{m}{2}} p)^{8 D_{\frac{m}{2},1}} q^{-a} $ & $8a (16a+1) q^{m-a} p^2    $ & $ \frac{m}{2} <  -\frac{x}{y} $
\\
\hline  $IV$ & $ (1-q^{\frac{m}{2} -a}  p)^{8 D_{\frac{m}{2} -a,1}} q^{a} $ & $ 8a(16a-1) q^{m-a} p^2     $ & $ \frac{m}{2}-a <  -\frac{x}{y} $
\\
\hline $V_\mu$ & $ (1-q^{\mu} p)^{8D_{\mu,1}} (1-q^{\mu_2} p)^{8D_{\mu_2,1}} q^{-a} $ & $ -2^8 a(m-a-2\mu)   $ & $ \frac{m}{2} <\mu <-\frac{x}{y}, \,\, \mu+\mu_2 = m$
\\
\hline $VI_\mu$ & $ (1-q^{\mu} p)^{8D_{\mu,1}} (1-q^{\mu_2} p)^{8D_{\mu_2,1}} q^{a} $ & $ 2^8 a(m-a-2\mu)   $ & $ \frac{m}{2}-a <\mu <-\frac{x}{y}, \,\, \mu+\mu_2 = m-2a$
\\
\hline
\end{tabular}
\end{center}\label{tab:prod-exp-expl}

The range of $\mu$ in the last two rows is derived as follows. The
expression
\begin{equation}
 (1-q^{\mu_1} p)^{8D_{\mu_1,1}} (1-q^{\mu_2} p)^{8D_{\mu_2,1}} q^{-a}
\end{equation}
can contribute to $q^{m-a}p^2$ when $\mu_1 + \mu_2 = m$. The factors are properly
ordered for
$$0 < {\rm Im} (-\mu_1 + i ) \overline{( - \mu_2 +i)} = \mu_1 - \mu_2$$
and hence $m/2 < \mu_1$. On the other hand, for the second factor to contribute we must have
$\mu_1 y + x <0$ and hence $\mu_1 < -x/y$. The range for $VI_\mu$ is derived similarly.

When $-x/y> m/2$ we can evaluate the contribution of $V_\mu$
using the identity
\begin{equation}\label{eq:sum-ident}
\sum_{\frac{m}{2}<\mu \leq N } (m-a - 2\mu)  = \left(\frac{a+1}{2}\right)^2 - \left(N - \frac{m-a}{2} + \frac{1}{2}\right)^2.
\end{equation}

Now we can add up the contributions. For $-\frac{x}{y} < \frac{m}{2} $ there are no contributions, and
the coefficient is $0$. In the range  $ \frac{m}{2}  <  -\frac{x}{y}<\frac{m}{2} -a  $ terms of
types $I$, $III$ and $V_\mu$ all contribute. The  sum of the contributions of type $I$ and $III$
is $2^7 a^2$. Using \eqref{eq:sum-ident} we can evaluate the contribution of terms of type $V_\mu$
and thus derive the index
\begin{equation}\label{eq:m2-index}
2^7 a^2 - 2^6 a(a+1)^2 + 2^8 a \left(N_{x,y} - \frac{m-a}{2} + \frac{1}{2}\right)^2
\end{equation}
where $N_{x,y} = \lfloor -x/y \rfloor$. This is to be evaluated for $-x/y$ nonintegral.

When $ \frac{m}{2} -a <  -\frac{x}{y} $ the index is zero.
The way this comes about is interesting: The terms $I$ and $II$ cancel and $III$ and $IV$ add up
to give $2^8 a^2 $. Moreover $V_\mu$ and $VI_\mu$ together have canceling terms for
$ \frac{m}{2}-a < \mu$ and the sum of these two terms becomes the constant (as a function
of $x,y$) given by
\begin{equation}\label{eq:five-sum}
-2^8 a \sum_{ \frac{m}{2} < \mu \leq \frac{m}{2}-a } (m-a-2\mu).
\end{equation}
The range of this sum can be written as $a \leq m-a -2\mu < -a$ and hence
\eqref{eq:five-sum} is trivially
equal to $-2^8 a^2$, thus leading to total index $0$.

In particular for our example $\Gamma_{0,2}$ with $a=-4$ discussed in Figure \ref{fig:FHSVhalos}
above
we have

\begin{center}
\renewcommand{\arraystretch}{1.3}
\begin{tabular}{|c|c|c|c|c|c|c|}
\hline
  $    -\frac{x}{y} < 0 $ & $0$
\\
\hline    $ 0 <  -\frac{x}{y}<1 $ & $2^{11}$
\\
\hline    $ 1 <  -\frac{x}{y}< 3 $ & $2^{12}$
\\
\hline    $ 3 <  -\frac{x}{y}< 4 $ & $2^{11}$
\\
\hline $ 4 <  -\frac{x}{y} $ & $0 $
\\
\hline
\end{tabular}
\end{center}

A number of interesting lessons can be drawn from these examples:

\begin{enumerate}

\item The BST wall $\CS(\gamma_{m,2}, \Gamma)$ is at $-\frac{x}{y} = \frac{m-a}{2}$.
Note that at $t_+$ on the right of this wall $N_{x,y} =  \frac{m-a}{2}-1$ and at $t_-$
on the left $N_{x,y} =  \frac{m-a}{2}$. Thus, from equation \eqref{eq:m2-index}
we see that the index is indeed constant across the BST wall.

\item Different terms from the product can be identified with
different kinds of attractor trees. Terms of type $I$ correspond
to single-branched flows with core $\Gamma$ while terms
of type $II$ correspond to single-branched flows with core $\Gamma_M$.
 Terms of type $III$ and $IV$ correspond to 2-particle haloes with core $\Gamma$ and $\Gamma_M$,
respectively.
  Some of the terms of type  $V_\mu$ and $VI_\mu$ can be associated with
  two-branched trees with cores $\Gamma$ and $\Gamma_M$, respectively.

\item If we considered only one core charge $\Gamma$, corresponding to $F(q,p,t_1)=q^{-a}$
then  terms of type $V_\mu$ would not give a well-defined index as the point $z$ on
$\CP$ approaches $t_2$ since $N_{x,y}$ goes to infinity and the sum of terms
of type $V_\mu$ grows without bound. On the other hand, when we
include the monodromy image $q^{a}$ there is a term-by-term cancelation between
$V_\mu$ and $VI_\mu$ so that the index is in fact well-defined.

\end{enumerate}

%
%


We expect the features of the above example to hold for general coefficients of $q^{m-a}p^n$ with $n>2$.
In particular, following the path $\CP$   all the way from $t_1$ to $t_2$  should yield an
identity of the form
\begin{equation}
	F(q, p;t_2) = \prod_{\frac{m}{n} \in (-\infty, \infty), \: n>0} U_{m,n} \:F(q, p;t_1).
\end{equation}
Just like $t_1$, at  $t_2$ all halo states are unstable and the partition functions have only two contributions from core charges $\Gamma$ and $\Gamma_M$ so:
\begin{equation}
F(q, p;t_2) = q^{-a} + q^a
\end{equation}
and hence we arrive at the following (somewhat strange) identity for formal power series in $q$ and $p$:
\begin{equation}
	q^{-a} + q^a= \prod_{\frac{m}{n} \in (-\infty, \infty), \: n>0} U_{m,n} \left(q^{-a} + q^a \right).
\end{equation}

It is worth stressing that the operator $\prod_{\frac{m}{n} \in (-\infty, \infty), \: n>0} U_{m,n}$
does not act well on general power series or even on general rational functions of $q,p$. As we have
seen above, it does have a well-defined action on $q^a + q^{-a}$ (and hence on finite
sums of such terms). Thus, requiring that such
infinite products have well defined actions on partition functions puts a nontrivial constraint
on the spectrum of BPS states. This is the sense in which the approach of \cite{Andriyash:2010qv}
constitutes a generalization of the Kontsevich-Soibelman wall-crossing formula.

\subsection{Massless vectors with adjoint hypermultiplets}\label{subsec:KMP}

In this section we consider an example with massless vector multiplets at the singularity, introduced by Katz, Morrison and Plesser (KMP) in \cite{Katz:1996ht}. The massless spectrum at the singularity is that of an $\CN=2$, $SU(2)$ gauge theory with some number of adjoint hypermultiplets.  One distinctive feature of this example is that there are massless vector bosons at the singularity in the K\"ahler moduli space only at some special value of hypermultiplet (complex structure) moduli. If one moves off this special value of hypermultiplet moduli, the $\CN=2$ vector multiplet will become massive and the massless spectrum will consist of some number of hypermultiplets. At the special
locus where there are massless vectormultiplets we can invoke the discussion of
 Section \ref{subsec:IRfree}  above  to conclude that the spectrum will also include massless monopoles/dyons.
Note that when moving away from the special locus the monopoles will become confined due to the dual Meissner effect and will not be present in the massless spectrum either. This means that the monopoles must have zero BPS index.

As we discussed in more detail in Section \ref{sec:FHSV} at the special locus the   spectrum leads to a very complicated wall-crossing phenomena of halos states and there must be some hybrid BST walls, across which both conjugation and recombination take place. On the other hand the BPS index is preserved exactly as in the conifold like case, i.e. the index jumps across marginal stability walls of the core charge with hypermultiplet halo particles. The rest of the massless spectrum has index zero and thus across marginal stability walls with halo particles containing $W$-boson and monopole charges, as well as across BST walls for such halos, the index is trivially constant.
Thus the index of halo states changes exactly the same way as for generic complex structure with a conifold-like singularity. In the latter case the conditions of Section \ref{subsec:GroundRules} are met, and our main puzzle is resolved through the presence of the conjugation walls.

For completeness, in the remainder of this Section we recall a few details of the KMP model
and justify the above statements a bit more.
KMP considered the compactification of type II string theory
on a Calabi-Yau  which is a $K3$ fibration over a genus $g$ curve
$\CC$. In particular, the fibration has a curve $\CC$ of $A_{N-1}$
singularities, which corresponds to $SU(N)$ enhanced gauge
symmetry. Close to the singularity there is a field theory
description in terms of $\CN=2$ $SU(N)$ gauge theory with $g$
adjoint hypermultiplets, with the Lagrangian

\begin{eqnarray}
&& 2 \pi \CL = \Im \left[  {\rm Tr} \int d^4 \theta \left( M^\dagger_i e^V M^i + \tilde M^\dagger_i e^V \tilde M^i \right) +\Phi^\dagger e^V \Phi + \frac{\tau}{2} \int d^2 \theta  {\rm Tr} W^2 + i \int d^2 \theta   \CW \right]   \nonumber\\
&& \CW = {\rm Tr} \tilde M^i [\Phi,M_i ].
\end{eqnarray}
Here, in   $( M^i ,\tilde M^i )$ are $g$ adjoint hypermultiplets, representing complex structure moduli, $V$ is an
$\CN=1$ vector multiplet superfield with field strength $W$, $\Phi$ is an $\CN=1$ adjoint chiral multiplet, representing K\"ahler moduli, and the scalar potential is given by

\begin{eqnarray}\label{eq:scpotential}
&& \CV = {\rm Tr} \left[ [m_i, m^{\dagger i}]^2 + [\tilde m_i, \tilde m^{\dagger i}]^2 + [\phi, \phi^{\dagger}]^2 + 2 [ m^{\dagger i},\phi] [ \phi^\dagger ,m_{i}] + 2 [ \tilde m^{\dagger i},\phi] [ \phi^\dagger ,\tilde m_{i}]+ \right. \nonumber\\
&& \left. + 2 [ \tilde m^\dagger _j,m^{\dagger j}] [ m_i, \tilde m^i]\right],
\end{eqnarray}
where $m_i$, $\tilde m_i$, and $\phi$ are the  scalar components
of the corresponding superfields. The moduli space of this theory
can be parametrized by diagonal traceless matrices $\phi$ and $(m_i, \tilde m_i)$ and has
the form $\IC^{N-1} \times \IC^{2 g (N-1)}$, where the first
factor comes from vector multiplet moduli, and the second - from
hypermultiplets. The point of enhanced $SU(N)$ gauge symmetry
occurs at codimension $N-1$ in vector moduli space and in
codimension $2 g (N-1)$ in hypermltiplet moduli space, which
translates to codimension $N-1$ in K\"ahler  and  $2 g (N-1)$ in
complex structure moduli spaces in the full IIA string theory. As
we are interested in singularities of codimension 1 in K\"ahler
moduli space, we put $N=2$.

The special locus $m_i=0, \tilde m_i=0$ is a complex plane, parametrized by
$a$, $\phi = a \sigma_3$. The spectrum is that of a $U(1)$
$N=2$ SYM with $g$ hypermultiplets, that is enhanced to $SU(2)$ at
the origin. The multiplet with charge 1 under this $U(1)$ that
becomes massless corresponds to $W^+$ boson in field theory and to
a state with charge $\gamma$ in string theory. It has second
helicity supertrace $2g-2$ and thus

$$ \Omega(\gamma) = 2g-2,$$
where $2 g$ comes from $g$ hypermultiplets  and $-2$ comes from the vectormultiplet. The modulus $a$ corresponds to the period of the charge $\gamma$ in string theory. Denoting by $a_D$ the period of the dual charge
$\gamma_D$, $\langle \gamma, \gamma_D\rangle=-2$,
\cite{Katz:1996ht} give the monodromy of these periods around the
singularity in the $a$-plane to be

\begin{equation}
    \left\{ \begin{array}{l}
 a \to a \\
a_D \to a_D - 4( g -1)a,
\end{array} \right.
\end{equation}
which translates to the monodromy of the charge lattice as follows
\begin{equation}\label{eq:monodromyKMP}
    \Gamma \to \Gamma - (2g-2) \langle \Gamma, \gamma\rangle \gamma = \Gamma - \Omega(\gamma) \langle \Gamma, \gamma\rangle \gamma,
\end{equation}
for any charge $\Gamma$. As we mentioned above, going around the singularity the Hilbert spaces of halo states $\Gamma+m\gamma+n\gamma_D$ will change in a complicated way due to the presence of hybrid conjugation/recombination walls.

Now we change the complex structure away from the special point, for
example take non-zero $m_1 \ne 0$. Careful examination of the
scalar potential (\ref{eq:scpotential}) shows  that in this case
the point $a=0$ does not have enhanced gauge symmetry, but there
is  a massive $N=4$  vector multiplet\footnote{Here for instance $V^+ := V^1 + i V^2$.}  $(V^+,
\Phi^+)$with
charge $+1$ under the unbroken $U(1)$ as well as $(g-1)$ massless
$N=2$ hypermultiplets $(M_i^+, \tilde M_i^+)$, $i = 2,..,g$ of the
same charge. This massless spectrum has the same BPS index but
different spin character than the one at the special complex structure value.
In particular the Hilbert space of $\gamma$ became purely
fermionic. From IIA string theory point of view, there are
$2g -2$ spheres in the same homology class shrinking to zero (\cite{Katz:1996ht}) and the BPS index of massless state is $\Omega(\gamma) = 2 g -2$.  If we now go around the $a=0$ singularity, the picture will be exactly the same as in Sections \ref{subsec:BPS-Indices},\ref{subsec:Fermi}. In particular the monodromy of the charge lattice (\ref{eq:monodromyKMP}) is consistent with what we find in (\ref{eq:monodromy}),(\ref{eq:I}). The index will be preserved across the conjugation wall, the Hilbert spaces of halo states will undergo the Fermi flip and the spin character will also be preserved.

\subsection{Extremal Transitions}\label{subsec:ExtrTran}

In this Section we consider an example that was discussed in a number of papers on geometric engineering and heterotic/type II dualities.
It first appeared in the discussion of chains of heterotic/type II dual models in \cite{Aldazabal:1995yw}.
The spectrum at the singularity was analyzed from the type IIA side in \cite{Berglund:1996uy}.

We will be interested only in the last step of the chain of heterotic/II duals, connected by extremal transitions, as described in  \cite{Berglund:1996uy}. On the type II side there are two CY manifolds denoted by $X_4$ and $X_3$,
and given,  roughly, \footnote{We say roughly because the polytope, used to define $X_4$ in the language of toric geometry, has one additional vertex and correspond to a different toric variety than just $WP(1,1,2,6,10)$.} by  hypersurfaces  in
weighted projective spaces $WP(1,1,2,6,10)[20]^{4,190}$ and $WP(1,1,2,8,12)[24]^{3,243}$. Each of these models has a heterotic dual, given by   certain $\IZ_6$ orbifolds. The main hero is the manifold $X_4$ with four K\"ahler moduli $t_i$. It develops a singularity on the locus $t_4=0$ and transitions to the $X_3$ model.

The description of the singularity in the IIA language is the following: for $t_4 \ne 0$ $X_4$ contains a $\IP^1$ of blown-up $A_1$ singularities with 28 double points, where the blown-up $\IP^1$ splits into two. This family of $\IP^1$'s fibered over the base $\IP^1$ represents a divisor of $X_4$, which is nothing else than just a family of conics in $\IP^2$ over $\IP^1$ with 28 degenerate fibers. In the limit $t_4 \rightarrow 0$ the fiber $\IP^1$ shrinks and this gives rise to massless particles. One gets an enhanced gauge symmetry at this locus and the massless spectrum consists of a massless $SU(2)$ vectormultiplet and 28 hypermultiplets, fundamental under the new $SU(2)$. This has an interpretation on the heterotic side, where the IIA modulus $t_4$ corresponds to a Wilson line, of a point of perturbative enhanced gauge symmetry with hypermultiplets coming from the $E_8$ instanton degrees of freedom.

The transition to $X_3$ occurs at $t_4=0$ and on the IIA side consists of blowing-up the singularities and obtaining a new CY with  $N_v=3$  vector moduli and $N_h=243$ hypermultiplet moduli. The transition has a nice interpretation in field theory, where one goes from the Coulomb to Higgs  branch \cite{Greene:1995hu}.   On the Higgs branch the gauge group is completely broken and one is left with $2 \times 28 - 3 = 53$ additional hypermultiplet moduli.

One way to verify the spectrum at the singularity, that does not involve the heterotic dual of this model, is to compute the periods of $X_4$   and find their monodromy around the singularity. In this example it is possible to do this explicitly by using a Mellin-Barnes representation of the periods and analytically continuing the periods of $X_4$ from the large volume point to the neighborhood of the singularity. The advantage of the method of analytic continuation is that it automatically gives an integral symplectic basis for the periods, since we start with such a basis from the large volume point. Performing this computation one finds that there are two dual vanishing periods at the singularity, denoted $t_4$ and $t_4^D$, and given in terms of algebraic coordinates $z_i$ on the complex structure moduli space of the mirror $X_4$, by
\begin{align}\label{eq:periods}
t_4 &= \sqrt{z_4} (1 + O(z_i))  \nonumber\\
t_4^D &= -\frac{\sqrt{z_4}}{2 \pi i} \left( (24 \log z_4 + 2 \log z_2 + 4 \log z_3 + 8 \log z_1 )  (1 + O(z_i)) + O(z_i)\right)
\end{align}
These periods correspond to a pair of charges $\gamma, \gamma_D$ with $\langle \gamma, \gamma_D \rangle = 2$,
respectively.
From these periods we see that there is a monodromy around singularity in the $z_4$ plane of the form

\begin{align}
t_4 &\to - t_4 \nonumber\\
t_4^D &\to -t_4^D + 24 t_4,
\end{align}
which is the monodromy, expected in the $SU(2)$ gauge theory with 28 fundamental flavors. Nevertheless, as we argued in Section \ref{subsec:IRfree}, the spectrum will also contain massless monopoles, and the fact that the dual period $t_4^D$ vanishes at the singularity confirms this fact. The physical argument of Section \ref{subsec:IRfree}
implies that this charge is populated, but we can  supply further evidence for the existence of a monopole (and indeed an infinite tower of dyons) becoming massless together with the $W$ boson at the singularity. In IIA language the $W$ boson is a D2 brane wrapping the generic fiber of the shrinking divisor. The monopole is a D4 brane wrapping the divisor itself (perhaps with some flux). The intersection product of this D2 and D4 is 2 as it should be. Classically, when the D2 volume vanishes at $t_4=0$, the D4 volume vanishes also. On the other hand, the IR freeness of the theory implies that this fact is not spoiled by quantum corrections, as we have seen directly from the periods. So the $W$-boson and monopole become massless together.  The quantum corrections do not spoil this relation, although from the periods we see there is some logarithmic running of the proportionality factor $\frac{1}{g^2}$, again as expected from field theory with $\beta = 48$ and
$$
\frac{\mu}{v} = \frac{ z_1^{-1/6} z_2^{-1/24} z_3^{-1/12} }{\sqrt{z_4}}.
$$

In the mapping of IIA with heterotic variables one identifies $z_2 \sim e^{- 2 \pi S}$, where $S$ is the heterotic dilaton $S =  \frac{4 \pi}{g^2_0} - i \frac{\theta}{2 \pi}  $. Looking at the periods (\ref{eq:periods}) we see that in the limit of weak heterotic coupling $z_2 \to 0$ for fixed $t_4$  the dual period $t_4^D$ becomes infinitely massive and disappears  from the spectrum. This is as expected in the IR-free theory since the mass of the monopole is proportional to $\frac{1}{g^2}$. If we keep the coupling constant non-zero and take the limit $t_4 \to 0$ then we find that in fact both periods vanish at the singularity and the monopole becomes massless. Moreover, the monopole state cannot be seen in the perturbative spectrum in the IR limit, but it does become massless at the singularity. In this sense the perturbative heterotic string is  $misleading$ with regards to the massless spectrum. Of course there might be other dyonic massless states and the full massless spectrum depends on the UV completion of the field theory. To determine the full spectrum one has to do a detailed analyses of the singularity in the IIB picture and find all the special Lagrangian cycles, that shrink at the singularity. Regrettably, this computation appears to be out of reach at present. We want to stress that, as follows from the discussion in Section \ref{subsec:BPS-Indices}, the massless spectrum of the IR-free gauge theory with fundamental matter is \emph{inconsistent}  with the wall-crossing phenomenon. Indeed,
the massless electric spectrum gives for the product $P$ in \eqref{eq:Math-Pred2}
\begin{equation}
P(q) = \frac{(1-q)^{28}}{(1-q^2)^4}
\end{equation}
which is clearly not a polynomial.

If one goes through the extremal transition to the Higgs branch of the gauge theory, the monopoles get confined via the dual Meissner effect. Thus on that side of the extremal transition they will not be seen in the spectrum. If one wishes to go back and transition to the Coulomb branch
the flux tubes confining the monopoles shrink and become tensionless at the transition point, so one again gains free massless monopoles. For some related discussion see  \cite{Greene:1996dh}.

To summarize, this model has a spectrum at the singularity, that does not satisfy the requirements of Section \ref{subsec:GroundRules}. We still expect that the BPS indices are preserved as one crosses BST walls. To verify this we need to know the full massless spectrum at the singularity. We do not have this information, and even if we did, as discussed in Section \ref{sec:FHSV}, it is not clear how to sum all contributions to the given BPS state on both sides of the BST wall.

Similarly in the spirit of Section \ref{subsec:BPS-Indices} and the companion paper \cite{Andriyash:2010qv}, we can form a partition function of the BPS indices of states $\Gamma+ m\gamma+ n \gamma_D$. Here $\gamma$ is the charge of the hypermultiplet that becomes massless and $\gamma_D$ is the charge of magnetic monopole. The partition function changes as we go around the singularity and the net change is just the monodromy of the local system of charges. As we discuss in the companion paper this gives   a restriction on the spectrum at the singularity in the form of a generalized KS formula which relates the product of KS transformations around the singularity to the monodromy of the local system of charges. Unfortunately we cannot check this statement here since we do not know the full spectrum.

\begin{section}* {Acknowledgements}

We would like to thank Dionysis Anninos, Dieter van den Bleeken, Miranda Cheng, Wu-Yen Chuang, Clay Cordova, Emanuel Diaconescu, Tudor Dimofte, Davide Gaiotto, Sergei Gukov, Albrecht Klemm,
Andy Neitzke, Duco van Straten, David Morrison, Nathan Seiberg, and Edward Witten
for discussions and correspondence.   This work is supported by the DOE under grants
DE-FG02-96ER40959 and DE-FG02-91ER40654. DJ also wishes to acknowledge funding provided by the Association of Members of the Institute for Advanced Study, and would like to thank the Simons Center Workshop for hospitality during the completion of this work. GM would like to thank the Aspen Center for Physics and the Simons Center
Workshop for hospitality during the
completion of this work.

\end{section}

\appendix
\section{Properties of Attractor Flow trees}\label{app:properties}

In this appendix we review some properties of the attractor mechanims and attractor flow trees that are relevant to our discussion.

The attractor flow equations for the K\" ahler moduli $t^a$ for given charge $\Gamma$ have the form
\cite{FerraraSer95,Strominger:1996kf}:

\begin{eqnarray}\label{eq:attreqns}
&& \frac{d}{d\tau}t^a(\tau) = - 2 e^U g^{a \bar b}\partial_{\bar b} |Z(\Gamma;t(\tau)| \nonumber\\
&& \frac{d }{d \tau} {e^{-U}}  = |Z(\Gamma;t(\tau))|,
\end{eqnarray}
where $\tau$ is the parameter along the flow, $e^U$ is metric warp factor, $g_{a \bar b }$ is the metric on the moduli space. The attractor flow can be written in an integrated form, which is particularly useful for multicentered generalization \cite{Denef:2000nb}:

\begin{equation}\label{eq:localattr}
2 e^{-U} \Im(e^{-i\alpha} \Omega(t) ) = -\tau \Gamma + H_{\infty}(t_{\infty}),
\end{equation}
where $\Omega(t)$ is the period vector, $\alpha = \arg Z(\Gamma;t)$, $t_{\infty}$ is the starting point of the flow and

\begin{equation}\label{eq:Hinfty}
H_{\infty} = 2{\rm Im} \left[e^{-i \alpha_{\infty}} \Omega(t_{\infty})\right].
\end{equation}

Now we list and prove a number of properties of attractor flows that are useful in the main text.

{\em \noindent \textit{ {\bf Property 1:} For any two charges $\Gamma$ and $\gamma$ the attractor flow for charge $\Gamma + \gamma$ crosses the (anti)marginal stability locus $\Im \left[ \bar Z(\Gamma;t) Z(\gamma;t)\right]=0$ at most once.
}}

{\bf Proof:} (\ref{eq:localattr}) determines the value of the flow parameter $\tau$, for which the locus $\Im \left[ \bar Z(\Gamma;t) Z(\gamma;t)\right]=0$ is crossed to be

\begin{equation}\label{eq:tau_ms}
\tau_{ms}(t_{\infty}) = \frac{2 \Im \left[  Z(\Gamma;t_{\infty}) \overline{Z(\gamma;t_{\infty})}
\right]}{\langle \Gamma, \gamma \rangle |
Z(\Gamma+\gamma;t_{\infty})|} \end{equation}
As (\ref{eq:localattr}) is linear in $\tau$, there is only one solution (\ref{eq:tau_ms}).$\blacksquare$

Let's denote the point where the attractor flow of $\Gamma + \gamma$, starting at $t$, hits the locus $\Im \left[ \bar Z(\Gamma;t) Z(\gamma;t)\right]=0$ by $B(t)$.

{\em \noindent \textit{ {\bf Property 2:}  For any $t \not\in AMS(\gamma, \Gamma)$, the distance in the moduli space from $t$ to $B(t)$ along the flow of $\Gamma+\gamma$ is finite.
}}

{\bf Proof:} Using (\ref{eq:attreqns}) one can write an identity:

\begin{eqnarray}
\int d \tau \left( \frac{d l}{d \tau} \right)^2  = \int d \tau {\dot
t}^a g_{a \bar b}  \bar {\dot t}^{\bar b} = \frac{1}{2} \int d \tau
e^{U }\left( -2 \bar {\dot t^{\bar b}}\partial_{\bar b} |Z|   -2
{\dot t^{ a}}\partial_{a} |Z|    \right)  = - \int e^{U} d |Z|,
\end{eqnarray}

\noindent where $dl$ is a line element along the flow, parametrized
by $\tau$. Using the fact that $d |Z| < 0$ along the flow and $e^U
\le 1$, we get

\begin{eqnarray}
\int_0^{\tau_{ms}(t)} d \tau \left( \frac{d l}{d \tau} \right)^2 \le
|Z(\Gamma+\gamma;t)|.
\end{eqnarray}
The parameter $\tau_{ms} $ is finite, since the
only place where it can go to infinity is on the locus
$Z(\Gamma+\gamma;t)=0$, which belongs to $AMS(\gamma,
\Gamma)$. Thus we can conclude that

\begin{equation}
\int_0^{\tau_{ms}} d \tau \left( \frac{d l}{d \tau} \right)^2 <
\infty,
\end{equation}

\noindent which means that the distance is indeed finite

\begin{equation}
    \int_0^{\tau_{ms}} d \tau
\frac{d l}{d \tau} < \infty.
\end{equation}

$\blacksquare$

{\em \noindent \textit{ {\bf Property 3:} The attractor flow of charge $\Gamma$ in the neighborhood of $(A)MS(\gamma, \Gamma)$ always has the direction from stable to unstable side.
}}

{\bf Proof:} Indeed, writing the attractor equation
for charge $\Gamma$ as in (\ref{eq:attreqns}) and taking intersection product of it with
$\gamma$ we get:

\begin{equation}
2 e^{U} \Im \left( \frac{\bar Z(\Gamma;t)}{|Z(\Gamma;t)|} Z(\gamma;t)\right) = - \tau \langle \gamma, \Gamma \rangle + 2 \Im \left( \frac{\bar Z(\Gamma;t_{\infty})}{|Z(\Gamma;t_{\infty})|} Z(\gamma;t_{\infty})\right).
\end{equation}
Dividing both sides by $\langle \gamma, \Gamma \rangle$ we get

\begin{equation}
2 \frac{e^{U}}{|Z(\Gamma;t)|}  \frac{\Im \left( \bar Z(\Gamma;t) Z(\gamma;t) \right)}{\langle \gamma, \Gamma \rangle}  = - (\tau - \tau_{ms}) ,
\end{equation}
which proves the Property after taking into account the definition of stable/unstable side (\ref{eq:Stab-reg}). $\blacksquare$

\begin{figure}[htp]
\centering
\includegraphics[scale=0.4,angle=90,trim=0 0 0 0]{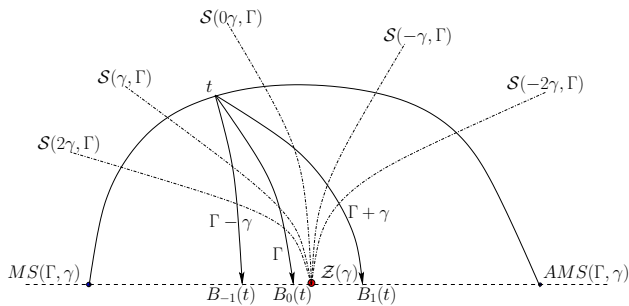}
\caption{Location of conjugation walls $\CS(n \gamma, \Gamma)$ as a function of $n$.  Two attractor flows for charges $\Gamma+\gamma$ and $\Gamma-\gamma$ are shown.  }
\label{fig:BSTwall2}
\end{figure}

\section{Arrangement of conjugation walls near $\CZ(\gamma)$}\label{app:Wall-arrangement}

In figure \ref{fig:BSTwall2} we show the general arrangement of the walls $\CS(p \gamma, \Gamma)$ walls for $p \in \IZ$ in a sufficiently small neighborhood of
 $\CZ(\Gamma)$. Let us give a
 schematic proof that this arrangement of walls is correct. To do
 this we find the gradient vectors, entering the attractor
 equations for charges $\Gamma+ p \gamma$, at the locus
 $\CZ(\gamma)$. For this purpose we can neglect all the moduli
 except the one parameterizing the plane in figure
 \ref{fig:BSTwall2} and can further choose it to be the period of
  $\gamma$, so that $Z_h(\gamma; z) = z$ in a small neighborhood of
 $\CZ(\gamma)$.\footnote{A more complete calculation would involve writing the general attractor equation for the moduli $y^a$, assuming that close to the zero of $\gamma$'s central charge we can write $Z_h(\gamma; z) \simeq \alpha_a y^a:= z$ and then projecting the attractor flow $\dot y^a$ on to the direction $\dot z = \alpha_a \dot y^a$. It gives the same result as we get below.} In this effectively one complex dimensional case the
 attractor equation will look like:
 \begin{equation}\label{eq:attrgradient}
     \dot z = - 2 e^U g^{z \bar z} \partial_{\bar z} | Z(\Gamma+ p \gamma;z)|,
 \end{equation}
 This can be written out as
 \begin{equation}\label{eq:attrgradientp}
     \dot z = -   e^U g^{z \bar z} \partial_{\bar z}K  | Z(\Gamma+ p \gamma;z)|
     - e^U g^{z \bar z} e^{K/2} \sqrt{\frac{Z_h }{\bar Z_h}} \partial_{\bar z}\left( \bar Z_h(\Gamma; \bar z) +  p \bar z \right),
 \end{equation}
and we want to study this flow in the leading approximation near $z=0$ as a function of $p$.

%
%
%

 Making a further
 choice of the phase of $z$ we can assume without loss of generality
 that $\arg Z_h(\Gamma;z)|_{z=0} = \pi$, so that $AMS(\gamma,\Gamma)$ is the positive real axis and
 $MS(\gamma,\Gamma)$ is the negative real axis.
 In the neighborhood of $z=0$ the metric will in general be singular. For example
 the metric will typically behave like $g^{z \bar z} \sim  - \frac{R}{\log \vert z \vert^2}$.
 Nevertheless, we can choose a gauge so that $\partial_{\bar z}K $ is continuous
 and then we can write  the leading approximation
 to the flow equation for $\Gamma + p z$  near $z=0$ in the form
 \begin{equation}
 \frac{dz}{d\tau} =  - K g^{z \bar z} (\Delta - p )
  \end{equation}
 where $K>0$ is constant and $\Delta$ is a complex constant. The stable side is determined by the sign of $\langle \Gamma, \gamma \rangle$ and taking $\langle \Gamma, \gamma \rangle > 0$ the stable side will be on top of figure \ref{fig:BSTwall2}. According to Property 3 from Appendix \ref{app:properties} the attractor flow will cross $(A)MS(\gamma, \Gamma)$ from stable to unstable side which corresponds to $\Im \Delta >0$. Then even if $g^{z \bar z}$ is singular, we know that it is
 positive, and hence  the tangent vectors to the flow have
 $\frac{dy}{d\tau} < 0$ while  $\frac{dx }{d \tau}$ is \emph{positive}
 for $p > \Re\Delta$ and \emph{negative} for $p < \Re\Delta$.
 Thus, in a suffciently small neighborhood of $z=0$ the flows will look
 as shown in Figure \ref{fig:BSTwall2}.

\section{The BST walls on the conifold}\label{app:walls}

In this section we determine the positions of $\CS(\gamma_{m,n}, \Gamma_{m',n'})$ walls in the conifold example.
As discussed on Section \ref{sec:conifold}, we are going to use
large volume approximation\footnote{We use the large volume
periods in what follows. Close to the discriminant locus, the
periods get finite quantum corrections, but one can check that the
effect of these corrections is subleading in $1/\Lambda$.} to
solve (\ref{eq:attreq}) and find the branching point of the flow
$\Gamma_{\tilde m, \tilde n}:=\Gamma_{m',n'}+ \gamma_{m,n}$ from
(\ref{eq:BSTeq}). The starting point of the flow is $t_{\infty} =
z P + L \tilde P e^{i \phi}$ and the large volume
approximation of the holomorphic central charges are given by

\begin{eqnarray}
&& Z^h(\Gamma_{m',n'}; t_{\infty}) = \frac{t_{\infty}^3}{6} - m' z -n' \nonumber\\
&& Z^h(\gamma_{m,n}; t_{\infty}) = - m z -n \nonumber\\
&& Z^h(\Gamma_{\tilde m, \tilde n}; t_{\infty}) = \frac{t_{\infty}^3}{6} - \tilde m z - \tilde n.
\end{eqnarray}
In terms of the harmonic functions $H = -\tau \Gamma + H_{\infty}$, entering (\ref{eq:attreq}), the moduli on the first branch of the attractor tree are given by \cite{Shmakova:1996nz,Bates:2003vx}.

\begin{eqnarray}\label{eq:moduli}
&& t^A = \frac{H^A - \frac{M}{Q^{3/2}} y^A}{H^0} + i \frac{\Sigma}{Q^{3/2}} y^A \nonumber\\
&& D_{ABC} y^B y^C = -2 H^0 H_A + D_{ABC} H^B H^C \nonumber\\
&& Q^{3/2} = 1/3 D_{ABC} y^A y^B y^C \nonumber\\
&& M = H_0 (H^0)^2 + 1/3 D_{ABC} H^A H^B H^C - H^A H_A H^0
\end{eqnarray}
The branching point corresponds to $\tau_{ms} = \langle
\gamma_{m,n}, H_{\infty}\rangle
=\sqrt{\frac{3}{J_{\infty}^3}}\frac{ {\rm Im} \left[\frac{\bar
Z^h(\Gamma_{\tilde m, \tilde n}; t_{\infty})}{|Z^h(\Gamma_{\tilde
m, \tilde n}; t_{\infty})|} Z^h(\gamma_{m,n}; t_{\infty}) \right]
}{\langle \gamma_{m,n}, \Gamma_{\tilde m, \tilde n}\rangle}$, and
the only problem in finding the moduli at the brancing point is
solving the equation  $D_{ABC} y^B y^C = -2 H^0 H_A + D_{ABC} H^B
H^C$ from (\ref{eq:moduli}). In the present case this reduces to

\begin{eqnarray}\label{eq:ys}
&& D_{222} y^2 y^2 + 2 D_{122} y^1 y^2 + D_{112} y^1 y^1 = -2 H^0 H_2 + D_{222} H^2 H^2+ 2 D_{122} H^1 H^2+ D_{112} H^1 H^1 \nonumber\\
&& D_{122} y^2 y^2 + 2 D_{112} y^1 y^2 + D_{111} y^1 y^1 = -2 H^0 H_1 + D_{122} H^2 H^2+ 2 D_{112} H^1 H^2+ D_{111} H^1 H^1
\end{eqnarray}
It is easy to see that $H^1 \sim O(L^0)$ and $H^2 \sim O(L^1)$ and we will look for the solution of (\ref{eq:ys}) of the form

\begin{eqnarray}
&& y^1 = y^1_{(1)} +\frac{1}{L} y^1_{(2)}  + \dots \nonumber\\
&& y^2 = L y^2_{(1)} + y^2_{(2)}  + \dots \nonumber\\
\end{eqnarray}
This allows to solve (\ref{eq:ys}) as an expansion in $\frac{1}{L}$ and gives

\begin{eqnarray}
&& y^2_{(1)} = L  \sqrt{\frac{3}{J_{\infty}^3}}  \sqrt{\sin^2 2 \phi  - \frac{m}{n}\sin \phi \; {\rm Im} \left[ z e^{- 3 i \phi}\right] } \nonumber\\
&& y^1_{(1)} =  \frac{ \sqrt{\frac{3}{J_{\infty}^3}}  \left(  -2 {\rm Im} \left[ e^{- 2 i \phi} (-\frac{m}{n} z -1 ) \right]    {\rm Im} \left[ z e^{- 3 i \phi}\right] \right) }{2 \sqrt{\sin^2 2 \phi  - \frac{m}{n}\sin \phi \; {\rm Im} \left[ z e^{- 3 i \phi}\right] }}.
\end{eqnarray}
Note that triple-intersection numbers completely disappeared from the answer! Now we can write our final expression for the moduli at the branching point :

\begin{eqnarray}\label{eq:branchmoduli}
&& z_{br} = -\frac{n}{m} + \frac{{\rm Im} \left[ e^{- 2 i \phi} (-\frac{m}{n} z -1 ) \right]}{\left(\sin^2 2 \phi  - \frac{m}{n}\sin \phi \; {\rm Im} \left[ z e^{- 3 i \phi}\right] \right)^2} \left( \frac{n}{m} \sin 2 \phi  \left(\sin^2 2 \phi  - \frac{3}{2} \frac{m}{n}\sin \phi \; {\rm Im} \left[ z e^{- 3 i \phi}\right] \right) - \right. \nonumber\\
&&  \left.  -  i \sin \phi {\rm Im} \left[ z e^{- 3 i \phi}\right]  \sqrt{\frac{3}{4} \sin^2 2 \phi  - \frac{m}{n}\sin \phi \; {\rm Im} \left[ z e^{- 3 i \phi}\right] } \right)  \nonumber\\
&& L_{br} = L \frac{\sin \phi}{\sqrt{\sin^2 2 \phi  - \frac{m}{n}\sin \phi \; {\rm Im} \left[ z e^{- 3 i \phi}\right] }}  \nonumber\\
&& \tan \phi_{br} = \sqrt{ \tan^2 \phi + \frac{{\rm Im} [e^{-3i\phi}(-\frac{m}{n}z-1)]}{\sin \phi \cos^2 \phi}}
\end{eqnarray}
The BST wall is determined by the fact that the branching point is on the locus $ Z^h(\gamma_{m,n}; t_{br}) = - m z_{br} -n = 0$.  Using (\ref{eq:branchmoduli}) we find that this happens when

\begin{equation}\label{eq:tildeW1}
\left\{ \begin{array}{l}
{\rm Im} \left[ e^{- 2 i \phi} (-\frac{m}{n} z -1 ) \right] =0 \\
\frac{3}{4} \sin^2 2 \phi  - \frac{m}{n}\sin \phi \; {\rm Im} \left[ z e^{- 3 i \phi}\right] >0.
\end{array}
\right.
\end{equation}
which leads to the final answer (\ref{eq:BSTlocation}).

\section{Enumerating Flow Trees in the FHSV Example}\label{appendix:FHSV}

Let us consider attractor flow trees of the form \eqref{eq:Flow-Trees}
\begin{equation}
    \Gamma_{m,n} \to (\gamma_{m_{1},n_{1}}+(\gamma_{m_{2},n_{2}}+...(\gamma_{m_{L-1},n_{L-1}}+(\Gamma+ \gamma_{m_L,n_L}))...)),
\end{equation}
where $\sum m_i = m$ and $\sum n_i = n$. We make the simplifying assumptions
described in Section \ref{sec:FHSV}. In particular $\tau=i$ and $z$ is arbitrarily small.
We can assume without loss of generality that $n>0$.

If the terminal point is in the lower half-plane (for $z$)
then each successive split drives the flow further into the lower half-plane. On the other hand
$\sum n_i = n$ must be positive, so some of the $n_i$ must be positive, but all the walls $MS_{m_i,n_i}$
with $n_i$ positive are  in the upper half-plane. Therefore,   the initial point must be in the upper half-plane.

Now consider a single split $\Gamma_{m,n} \to \Gamma + \gamma_{m,n}$. The marginal stability
wall $MS_{m,n}$ is always to the right of the BST wall $\CS(\gamma_{m,n},\Gamma)$. If the initial
point is to the right of the BST wall we can construct the tree, otherwise, there is no intersection
with the marginal stability wall.

In general, we cannot construct any flow tree with initial point to the left of the BST
wall $\CS(\gamma_{m,n},\Gamma)$ because the initial branching (which must take place in the
lower half-plane) must necessarily proceed
from unstable to stable region, which is forbidden.

Now, consider a point infinitesimally to the right of the BST wall. When enumerating trees
further to the right we find a subset of these flows.
We first claim that all the $n_i$ are
positive. One can check that if $n_i$ becomes negative, then the conservation of charge and the
stable-to-unstable rule forces the next $n_{i+1}$ to be negative and so on so that the $n$-value
of the core flow grows without bound. Since we explicitly want the terminal core flow to be $\Gamma$ this
cannot happen. Thus, the $n_i$ are all positive and form a partition of $n$, in particular, there are
finitely many choices for the $n_i$.

Next, the flow tree will intersect
a series of marginal stability walls for
$$\gamma_{m_1,n_1}, \gamma_{m_2,n_2}, \dots, \gamma_{m_L,n_L},$$
in that order. Since the flow proceeds to the right (stable to unstable) the walls must be
ordered so that successive walls are in the clockwise direction.
It is easy to check that the wall $MS_{m_{i+1},n_{i+1}}$ is on the clockwise side
of $MS_{m_i,n_i}$ iff
\begin{equation}
m_i n_{i+1} - n_i m_{i+1} >0
\end{equation}
One way to see this is to require the slope   $-n_i/m_i$ to be decreasing. Another way
is to require ${\rm Im} (m_j - i n_j)\overline{(m_{j+1}-i n_{j+1})} >0$. Since $m_i$ can
be positive, zero, or negative, but $n_i$ must be positive it is more convenient to write:
\begin{equation}\label{eq:clockwise-walls}
\frac{m_1}{n_1} > \frac{m_2}{n_2} > \cdots > \frac{m_L}{n_L} .
\end{equation}
Now, if $m-a \leq 0$ then all the successive walls have $m_i<0$. It follows that
$-m_i$ forms a partition of $-m$ and hence there are finitely many choices for the $m_i$.
In that case, the trees are labeled by two partitions of $-m$ and $n$, respectively,
subject to \eqref{eq:clockwise-walls}.

If $m-a >0$ the situation is a little more complicated. The walls of marginal stability divide up
into the first set with $m_i \geq 0 $, which are met first in moving downstream the flow tree
 and the second set with $m_i <0$. For the first set there will be inequalities bounding the
 allowed values of $m_i$. Then the second set must form a partition to saturate total charge
 conjugation. For example, if the first branching happens at $MS_{m_1,n_1}$ with $m_1>0$ then
 requiring that the subsequent flow with tangent vector $-a + m-m_1 - i (n-n_1)$ proceed into
 the unstable region forces
 \begin{equation}
 {\rm Im} (-a + m-m_1 - i (n-n_1))\overline{(m_1-i n_1)} >0 \quad \Rightarrow \quad \frac{m-a}{n}> \frac{m_1}{n_1}
 \end{equation}
 There are therefore finitely many possible values for $m_1$. If the next branching is at $MS_{m_2,n_2}$ with
 $m_2>0$ then we similarly get
 \begin{equation}
 \frac{-a + m - m_1}{n-n_1} > \frac{m_2}{n_2}
 \end{equation}
 giving finitely many values of $m_2$, and so on.

Of course, the walls are only relevant for wall-crossing provided the Hilbert spaces of
halo particles are nonvanishing. Thus  $gcd(m_i,n_i)$ must be $1$ or $2$.

\end{document}